\documentclass[aos,preprint]{imsart}

\usepackage{hyperref, amsmath, amsthm, amssymb, bm}
\usepackage[numbers]{natbib}
\usepackage{graphicx}
\usepackage{enumitem}
\usepackage{accents}
\usepackage{algpseudocode, algorithm}
\usepackage{subcaption}
\usepackage{cleveref}
\usepackage[colorlinks,citecolor=blue,urlcolor=blue]{hyperref}
\usepackage{multirow}
\usepackage{siunitx}
\usepackage{booktabs}

\usepackage{enumitem}

\numberwithin{equation}{section}

\newtheorem{theorem}{Theorem}[section]

\newtheorem{lemma}{Lemma}[section]

\newtheorem{assumption}{Assumption}
\newtheorem{model}{Model}
\crefname{assumption}{Assumption}{Assumptions}
\crefname{model}{Model}{Models}
\crefname{equation}{equation}{equations}
\crefname{appendix}{Appendix}{Appendices}

\newcommand{\thistheoremname}{}
\newtheorem*{genericthm*}{\thistheoremname}
\newenvironment{namedthm*}[1]
{\renewcommand{\thistheoremname}{#1}%
  \begin{genericthm*}}
  {\end{genericthm*}}

\theoremstyle{definition}

\newtheoremstyle{propertystyle}
{3pt} % Space above
{3pt} % Space below
{\it} % Body font
{} % Indent amount
{\bfseries} % Theorem head font
{.} % Punctuation after theorem head
{.5em} % Space after theorem head
{} % Theorem head spec (can be left empty, meaning `normal')
\theoremstyle{propertystyle}

\theoremstyle{remark}

\newcommand\independent{\protect\mathpalette{\protect\independenT}{\perp}}
\def\independenT#1#2{\mathrel{\rlap{$#1#2$}\mkern2mu{#1#2}}}

\newlength{\dhatheight}

\newcommand*\diff{\mathop{}\!\mathrm{d}}

\DeclareMathOperator*{\argmax}{arg\,max}

\usepackage{tikz}

\newcommand*\circled[1]{\tikz[baseline=(char.base)]{
    \node[shape=circle,draw,inner sep=2pt] (char) {#1};}}
\tikzstyle{na} = [baseline=-.5ex]
\usetikzlibrary{arrows,arrows.meta,shapes,shapes.geometric}
\tikzset{circle/.style = {shape=circle,draw,minimum size=1.5em,thick}}
\tikzset{semicircle/.style = {shape=semicircle,draw,minimum size=1em}}
\tikzset{rectangle/.style = {shape=rectangle,draw,minimum size=1.5em}}
\tikzset{dashedrectangle/.style = {shape=rectangle,dashed,draw,minimum size=1.5em}}
\tikzset{dashedcircle/.style = {shape=circle,dashed,draw,minimum size=1.5em}}
\tikzset{edge/.style = {-Latex,> = latex',thick}}
\tikzset{dashededge/.style = {->,dashed,> = latex',thick}}

\tikzstyle{na} = [baseline=-.5ex]
\usetikzlibrary{arrows,shapes}

\usepackage{color}
\definecolor{qz}{RGB}{0,0,0}

\newcommand{\revise}[1]{\textcolor{qz}{#1}}

\begin{document}

\begin{frontmatter}

  \title{Statistical inference in two-sample summary-data Mendelian
    randomization using robust adjusted profile score}

  \runtitle{Mendelian randomization by RAPS}

  % \thankstext{T1}{Working draft, please do not cite.}

  \begin{aug}
    \author{Qingyuan Zhao\corref{}\ead[label=e1]{qyzhao@wharton.upenn.edu}},
    \author{Jingshu Wang\ead[label=e2]{jingshuw@wharton.upenn.edu}},
    \author{Gibran Hemani\ead[label=e3]{g.hemani@bristol.ac.uk}},
    \author{Jack Bowden\ead[label=e4]{jack.bowden@bristol.ac.uk}},
    \and
    \author{Dylan S.\ Small\ead[label=e5]{dsmall@wharton.upenn.edu}}

    \runauthor{Q. Zhao et al.}

    \affiliation{University of Pennsylvania and University of Bristol}

    \address{Department of Statistics,\\ The Wharton School,\\ University
      of Pennsylvania, \\ USA \\ \printead{e1,e2,e5}.}

    \address{MRC Integrative Epidemiology Unit,\\ University of Bristol,\\
      UK\\ \printead{e3,e4}.}

  \end{aug}

  \begin{abstract}
    Mendelian randomization (MR) is a method of exploiting genetic
    variation to unbiasedly estimate a causal effect in presence of
    unmeasured confounding. MR is being
    widely used in epidemiology and other related areas of population
    science. In this paper, we study statistical inference
    in the increasingly popular two-sample summary-data MR design. We show a linear
    model for the observed associations approximately holds in a wide
    variety of settings
    when all the genetic variants satisfy the exclusion restriction
    assumption, or in genetic terms,
    when there is no pleiotropy. In this scenario, we derive a maximum profile
    likelihood estimator with provable consistency and asymptotic
    normality. However, through analyzing real datasets, we
    find strong evidence of both systematic and idiosyncratic
    pleiotropy in MR, echoing the omnigenic model of complex traits
    that is recently proposed in genetics. We model the systematic
    pleiotropy by a random effects
    model, where no genetic variant satisfies the exclusion restriction
    condition exactly. In this case we propose a consistent and
    asymptotically normal estimator
    by adjusting the profile score. We then tackle the idiosyncratic pleiotropy
    by robustifying the adjusted profile score. We demonstrate the
    robustness and efficiency of the proposed methods using several
    simulated and real datasets.
  \end{abstract}



\end{frontmatter}

\section{Introduction}

A common goal in epidemiology is to understand the causal mechanisms
of disease. If it was known that a risk factor causally influenced
an adverse health outcome, effort could be focused to develop an intervention (e.g., a drug or public health intervention) to reduce the risk factor and improve the population's health. In settings where evidence from a randomized
controlled trial is lacking, inferences about causality are made using
observational data. The most common design of observational study
is to control for confounding variables between the exposure and
the outcome. However, this strategy can easily lead to biased
estimates and false conclusions when one or several important confounding variables
are overlooked.

Mendelian randomization (MR) is an alternative study design that
leverages genetic variation to produce an unbiased estimate of the causal
effect even when there is unmeasured confounding. MR is both
old and new. It is a special case  of the
instrumental variable (IV) methods \citep{didelez2007mendelian}, which
date back to the 1920s
\citep{wright1928tariff} and have a long and rich history in
econometrics and statistics. The first MR design was proposed by
\citet{katan1986apoupoprotein} over 3 decades ago and later popularized in
genetic epidemiology by \citet{davey2003mendelian}. As a public health
study design, MR is rapidly gaining popularity from just $5$
publications in 2003 to over $380$ publications in
the year 2016 \citep{mrpopularity}. However,
due to the inherent complexity of genetics (the understanding of which
is rapidly evolving) and the make-up of large international disease
databases being utilized in the analysis, MR has
many unique challenges compared to classical IV analyses in
econometrics and health studies. Therefore, MR does not merely involve
plugging genetic instruments in existing IV methods. In fact,
the unique problem structure has sparked many recent methodological
advancements
\citep{bowden2015mendelian,bowden2016consistent,guo2016confidence,kang2016instrumental,li2017mendelian,tchetgen2017genius,van2017pleiotropy,verbanck2018detection}.

Much of the latest developments in Mendelian randomization has been
propelled by the increasing availability and scale of
genome-wide association studies (GWAS) and other high-throughput genomic
data. A particularly attractive proposal is to automate the causal
inference by using published GWAS data  \citep{burgess2015using}, and a
large database and software platform is currently being developed
\citep{hemani2016mr}. Many existing IV and MR methods
\citep[e.g.][]{guo2016confidence,pacini2016robust,tchetgen2017genius},
though theoretically sound and robust to different kinds of biases,
require having individual-level data. Unfortunately,
due to privacy concerns, the access to individual-level genetic data
is almost always restricted and usually only the GWAS summary
statistics are publicly available. This data structure has sparked a
number of new statistical methods anchored within the framework of meta-analysis
\citep[e.g.][]{bowden2015mendelian,bowden2016consistent,hartwig2017robust}.
They are intuitively simple and can be conveniently used with
GWAS summary data, thus are quickly gaining popularity in practice. However,
the existing summary-data MR methods often make unrealistic simplifying
assumptions and generally lack theoretical support such as statistical
consistency and asymptotic sampling distribution results.

This paper aims to resolve this shortcoming by developing statistical
methods that can be used with summary data, have good theoretical
properties, and are robust to deviations of the usual IV
assumptions. In the rest of the Introduction, we will
introduce a statistical model for GWAS summary data and demonstrate
the MR problem using a real data
example. This example will be repeatedly used in subsequent sections
to motivate and illustrate the statistical methods. We will conclude the
Introduction by discussing the methodological challenges in MR and
outlining our solution.

\subsection{Two-sample MR with summary data}
\label{sec:model-gwas-summary}

We are interested in estimating the causal effect of an
exposure variable $X$ on an outcome variable $Y$. The causal effect is
confounded by unobserved variables, but we have $p$ genetic variants
(single nucleotide polymorphisms, SNPs), $Z_1,Z_2,\dotsc,Z_p$, that are
approximately \emph{valid} instrumental variables (validity of an IV
is defined in \Cref{sec:valid-instr-vari}). These IVs can help us to
obtain unbiased
estimate of the causal effect even when there is unmeasured confounding.
The precise problem considered in this paper is two-sample Mendelian
randomization with summary data, where we observe, for SNP $j = 1,\dotsc,p$,
two associational effects: the SNP-exposure effect $\hat{\gamma}_j$ and the
SNP-outcome effect $\hat{\Gamma}_j$. These estimated effects are
usually computed from two different samples using a simple linear
regression or logistic regression and are or are becoming available in
public domain.

Throughout the paper we assume
\begin{assumption}
  \label{assump:setup}
  For every $j \in \{1,\dotsc,p\}:=[p]$, $\hat{\gamma}_j\sim
  \mathrm{N}(\gamma_j,\sigma_{Xj}^2)$,  $\hat{\Gamma}_j \sim
  \mathrm{N}(\Gamma_j,\sigma_{Yj}^2)$, and the variances $(\sigma_{Xj}^2,\sigma_{Yj}^2)_{j \in [p]}$ are
  known. Furthermore, the $2p$ random variables $(\hat{\gamma}_j)_{j \in
    [p]}$ and $(\hat{\Gamma}_j)_{j \in [p]}$ are mutually independent.
\end{assumption}
The first assumption is quite reasonable as typically there are hundreds
of thousands of samples in modern GWAS, making the normal approximation
very accurate. \revise{We assume the variances of the GWAS marginal
  coefficients are computed very accurately using the individual data
  (as they are typically based on tens of thousands of samples),
but the methods developed in this paper do not utilize individual
data for statistical inference.} The independence between $(\hat{\gamma}_j)_{j \in
  [p]}$ and $(\hat{\Gamma}_j)_{j \in [p]}$ is guaranteed because the
effects are computed from independent samples. The independence across
SNPs is reasonable if we only use uncorrelated SNPs by using a tool
called linkage disequilibrium (LD) clumping
\citep{hemani2016mr,plinksoftware,purcell2007plink}. See
\Cref{sec:stat-model-gwas} for more justifications of the last
assumption.

Our key modeling assumption for summary-data MR is
\begin{namedthm*}{Model for GWAS summary data}
  There exists a real number $\beta_0$ such that
  \begin{equation}
    \label{eq:model-0}
    \Gamma_j \approx \beta_0 \gamma_j~\text{for almost all}~j \in [p].
  \end{equation}
\end{namedthm*}

In \Cref{sec:stat-model-gwas,sec:linear-model-gwas}, we will explain why this model likely holds
for a variety of situations and why the parameter $\beta_0$ may be
interpreted as the causal effect of $X$ on $Y$. However, by investigating a
real data example, we will demonstrate in \Cref{sec:example-continued}
that it is very likely that the strict
equality $\Gamma_j = \beta_0 \gamma_j$ is not true
for some if not most $j$. For now we will proceed with the loose
statement in \eqref{eq:model-0}, but it will be soon
made precise in several ways.

\Cref{assump:setup} and model \eqref{eq:model-0} suggest two different
strategies of estimating $\beta_0$:
\begin{enumerate}
\item Use the Wald ratio
  $\hat{\beta}_j=\hat{\Gamma}_j/\hat{\gamma}_j$
  \citep{wald1940fitting} as each SNP's   individual estimate of
  $\beta_0$, then aggregate the estimates using a robust meta-analysis
  method. Most existing methods for summary-data MR follow this line
  \citep{bowden2015mendelian,bowden2016consistent,hartwig2017robust},
  however the Wald estimator $\hat{\beta}_j$ is heavily biased when
  $\gamma_j$ is small, a phenomenon known as ``weak instrument
  bias''. See \citet{bound1995problems} and \Cref{sec:challenges-mr} below.
\item Treat equation \eqref{eq:model-0} as an errors-in-variables
  regression problem \citep{carroll2006measurement}, where we are
  regressing $\hat{\Gamma}_j$, whose expectation is $\Gamma_j$, on
  $\hat{\gamma}_j$, which can be regarded as a noisy observation of
  the   actual regressor $\gamma_j$. Then we directly
  estimate $\beta_0$ in a robust way. This is the novel approach taken
  in this paper and will be described and tested in detail.
\end{enumerate}

\subsection{A motivating example}
\label{sec:an-example-bmi}

Next we introduce a real data example that will be repeatedly used in
the development of this paper. In this example we are interested in estimating
the causal effect of a person's Body Mass Index (BMI) on Systolic
Blood Pressure (SBP). We obtained publicly available summary data from
three GWAS with non-overlapping samples:
\begin{description}
\item[\texttt{BMI-FEM}:] BMI in females by the Genetic
  Investigation of ANthropometric Traits (GIANT) consortium
  \citep{locke2015genetic} (sample size: 171977, unit: \si{kg/m^2}).
\item[\texttt{BMI-MAL}:] BMI in males in the same study by the GIANT
  consortium (sample size: 152893, unit: \si{kg/m^2}).
\item[\texttt{SBP-UKBB}:] SBP using the United Kingdom BioBank (UKBB)
  data (sample size: 317754, unit: \si{mmHg}).
\end{description}

Using the \texttt{BMI-FEM} dataset and LD
clumping, we selected $25$ SNPs
that are genome-wide significant ($p$-value $\le 5 \times 10^{-8}$)
and uncorrelated ($10000$ kilo base pairs apart and $R^2 \le
0.001$). We then obtained the $25$ SNP-exposure effects
$(\hat{\gamma}_j)_{j=1}^{25}$ and the corresponding standard errors
from \texttt{BMI-MAL} and the SNP-outcome effects
$(\hat{\Gamma}_j)_{j=1}^{25}$ and the corresponding standard errors
from \texttt{SBP-UKBB}. \revise{Later on in the paper we will consider
an expanded set of $160$ SNPs using the selection threshold $p$-value
$\le 10^{-4}$}.

\begin{figure}[t]
  \centering
  \includegraphics[width = \textwidth]{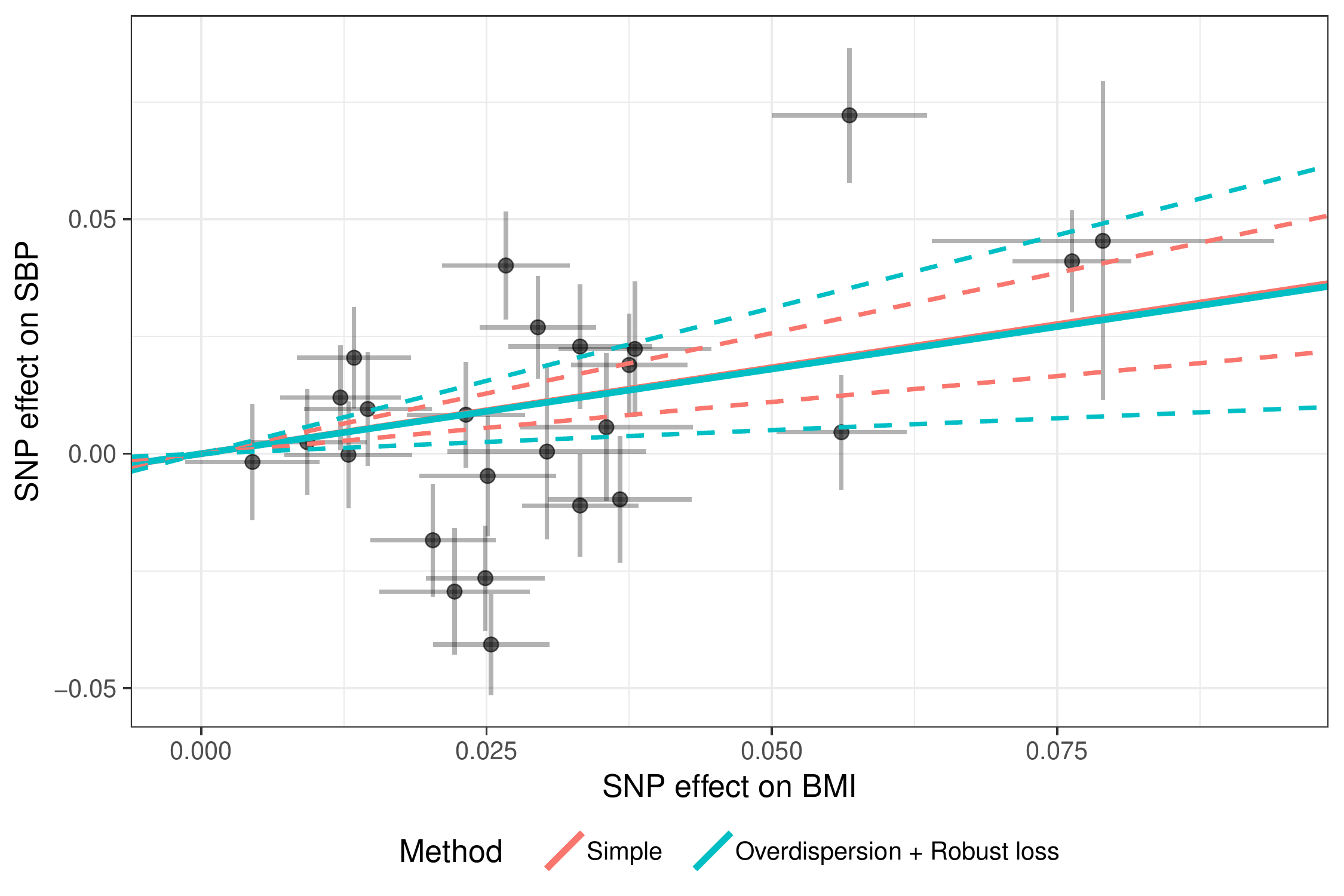}
  \caption{Scatter plot of $\hat{\Gamma}_j$ versus $\hat{\gamma}_j$ in
    the BMI-SBP example. Each point is augmented by the standard error
    of $\hat{\Gamma}_j$ and $\hat{\gamma}_j$ on the vertical and
    horizontal sides. For presentation purposes only, we chose the allele codings so that
    all $\hat{\gamma}_j$ are positive. Solid lines are the regression slope fitted by
    two of our methods. Dashed lines are the $95\%$ confidence interval
    of the slopes. The simple method using unadjusted profile score (PS, described in \Cref{sec:no-plei-prof}) has smaller standard
    error than the more robust method using robust adjusted profile
    score (RAPS, described in \Cref{sec:idiosyncr-plei-robus}),
    because the simple method does not consider genetic pleiotropy. See also
    \Cref{sec:example-continued}.}
  \label{fig:scatter-plot}
\end{figure}

\Cref{fig:scatter-plot} shows the scatter plot of the $25$ pairs of
genetic effects. Since they are measured with error, we added error
bars of one standard error to every point on both sides. The goal of summary-data MR is to
find a straight line through the origin that best fits these
points. The statistical method should also be robust to violations of model
\eqref{eq:model-0} since not all SNPs satisfy the relation $\Gamma_j =
\beta_0 \gamma_j$ exactly. We will come back to this example in
\Cref{sec:example-continued,sec:example-continued-1,sec:example-continued-2}
to illustrate our methods.

\subsection{Statistical Challenges and organization of the paper}
\label{sec:challenges-mr}

Compared to classical IV analyses in econometrics and
health studies, there are many unique challenges in two-sample MR with
summary data:
\begin{enumerate}
\item Measurement error: Both the SNP-exposure and SNP-outcome effects are
  clearly measured with error, but most of the existing methods
  applicable to summary data assume that the sampling error of
  $\hat{\gamma}_j$ is negligible so a weighted linear regression can
  be directly used
  \citep{burgess2013mendelian}.
\item Invalid instruments due to pleiotropy (the phenomenon that one
  SNP can affect seemingly unrelated traits): A SNP $Z_j$ may causally affect
  the outcome $Y$ through other pathways not involving the exposure
  $X$. In this case, the approximate linear model $\Gamma_j \approx
  \beta_0 \gamma_j$ might be entirely wrong for some SNPs.
\item Weak instruments: Including a SNP $j$ with very small $\gamma_j$
  can bias the causal effect estimates (especially when the meta-analysis strategy
  is used). It can also increase the variance of the estimator
  $\hat{\beta}$. See \Cref{sec:selecting-ivs}.
\item Selection bias: To avoid the weak instrument bias, the
  standard practice in MR is to only use the genome-wide significant
  SNPs as instruments (for example, as implemented in the
  \texttt{TwoSampleMR} R package \citep{hemani2016mr}). However, in
  many studies the same dataset is used for both selecting SNPs and
  estimating $\gamma_j$, resulting in substantial selection bias even if the
  selection threshold is very stringent.
\end{enumerate}

Many previous works have considered one or some of these
challenges. \citet{bowden2017improving} proposed a modified Cochran's $Q$
statistic to detect the heterogeneity due to pleiotropy instead of
measurement error in $\hat{\gamma}_j$. Addressing the issue of bias
due to pleiotropy has attracted lots of attention in the
summary-data MR literature
\citep{bowden2015mendelian,bowden2016consistent,hartwig2017robust,li2017mendelian,van2017pleiotropy,verbanck2018detection},
but no solid statistical underpinning has yet been given. Other methods with
more rigorous statistical theory require individual-level
data
\citep{guo2016confidence,pacini2016robust,tchetgen2017genius}. The
weak instrument problem has been thoroughly studied in the econometrics
literature
\citep[e.g.][]{bound1995problems,hansen2008estimation,stock2005asymptotic},
but all of this work operates in the individual-level data setting. Finally,
the selection bias has largely been overlooked in practice; common
wisdom has been that the selection biases the causal effects towards the null (so it
might be less serious) \citep{haycock2016best} and the bias is perhaps small
when a stringent selection criterion is used
(in \Cref{sec:real-data-comparison} we show this is not necessarily
the case).

In this paper we develop a novel approach to overcome all the
aforementioned challenges by adjusting the profile likelihood of the summary
data. The measurement errors of $\hat{\gamma}_j$ and
$\hat{\Gamma}_j$ (challenge 1) are naturally incorporated in computing
the profile score. To tackle invalid IVs (challenge 2), we will
consider three models for the GWAS summary data with increasing
complexity:

\begin{model}[No pleiotropy] \label{model:1}
  The linear model $\Gamma_j = \beta_0 \gamma_j$ is true for every $j \in [p]$.
\end{model}

\begin{model}[Systematic pleiotropy]
  \label{model:2}
  Assume $\alpha_j = \Gamma_j - \beta_0 \gamma_j
  \overset{i.i.d.}{\sim} \mathrm{N}(0,\tau_0^2)$ for $j \in [p]$ and
  some small $\tau_0^2$.
\end{model}

\begin{model}[Systematic and idiosyncratic pleitropy]
  \label{model:3}
  Assume $\alpha_j,~j\in[p]$ are from a contaminated normal
  distribution: most $\alpha_j$ are distributed as $\mathrm{N}(0,\tau_0^2)$
  but some $|\alpha_j|$ may be much larger.
\end{model}

The consideration of these three models is motivated by not only the
theoretical models in \Cref{sec:stat-model-gwas} but also
characteristics observed in real data
(\Cref{sec:example-continued,sec:example-continued-1,sec:example-continued-2})
and recent empirical evidence in genetics
\citep{boyle2017expanded,shi2016contrasting}.

The three models are considered in
\Cref{sec:no-plei-prof,sec:syst-plei-adjust,sec:idiosyncr-plei-robus},
respectively.
We will propose estimators that are provably
consistent and asymptotically normal in \Cref{model:1,model:2}. We
will then derive an estimator that is robust to a small proportion of
outliers in \Cref{model:3}. We believe
\Cref{model:3} best explains the real data and the corresponding
Robust Adjusted Profile Score (RAPS) estimator is the clear winner in
all the empirical examples.

Although weak IVs may bias the individual Wald's ratio
estimator (challenge 3), we will show, both theoretically and
empirically, that including additional weak IVs is usually helpful for our new
estimators when there are already strong IVs or many weak
IVs. Finally, the selection bias (challenge 4) is handled by requiring
use of an independent dataset for IV selection as we have done in
\Cref{sec:an-example-bmi}. \revise{This might not be possible in all
  practical problems, but failing to use a separate dataset for IV
  selection can lead to severe selection bias as illustrated by an
  empirical example in \Cref{sec:real-data-comparison}.}

The rest of the paper is organized as follows. In
\Cref{sec:stat-model-gwas} we give theoretical justifications of the
model \eqref{eq:model-0} for GWAS summary data. Then in
\Cref{sec:no-plei-prof,sec:syst-plei-adjust,sec:idiosyncr-plei-robus}
we describe an adjusted profile score approach of statistical
inference in \Cref{model:1,model:2,model:3}, respectively. The paper
is concluded with simulation examples in \Cref{sec:simulation},
another real data example in \Cref{sec:real-data-comparison} and more
discussion in \Cref{sec:discussion}.

\section{Statistical model for MR}
\label{sec:stat-model-gwas}

In this Section we explain why the \revise{approximate} linear model
\eqref{eq:model-0} for GWAS summary data may hold in many MR
problems. \revise{We will put structural assumptions on the
  original data and show that \eqref{eq:model-0} holds in a variety of
  scenarios. Owing to this heuristic and the wide availability of GWAS
summary datasets, we will focus on statistical inference for
summary-data MR after \Cref{sec:stat-model-gwas}.}
 % A reader may skip this Section for the first time of reading
% if primarily interested in the statistical inference under
% \Cref{model:1,model:2,model:3} using the adjusted profile score.

\subsection{Validity of instrumental variables}
\label{sec:valid-instr-vari}

In order to study the origin of the linear model \eqref{eq:model-0}
for summary data and give a causal interpretation to the parameter
$\beta_0$, we must specify how
the original data $(X,Y,Z_1,\dotsc,Z_p)$ are generated and how the
summary statistics are computed. Consider the
following structural equation model \citep{pearl2009causality} for the
random variables:
\begin{equation}
  \label{eq:npsem}
  \begin{split}
    X &= g(Z_1,\dotsc,Z_p,U,E_X),~\text{and} \\
    Y &= f(X,Z_1,\dotsc,Z_p,U,E_Y), \\
  \end{split}
\end{equation}
where $U$ is the unmeasured confounder, $E_X$ and $E_Y$
are independent random noises, $(E_X,E_Y) \independent
(Z_1,\cdots,Z_p,U)$ and $E_X \independent E_Y$. In two-sample MR, we
observe $n_X$ i.i.d.\ realizations of $(X,Z_1,\dotsc,Z_p)$ and
independently $n_Y$ i.i.d.\ realizations of $(Y,Z_1,\dotsc,Z_p)$. We
shall also assume that the SNPs $Z_1,Z_2,\dotsc,Z_p$ are discrete random
variables supported on $\{0,1,2\}$ and are mutually
independent. To ensure the independence, in practice we only include
SNPs with low pairwise LD score in our model by using standard
genetics software like LD clumping \citep{plinksoftware}.

\begin{figure}[t]
  \centering
  % \begin{subfigure}[c]{0.49\textwidth} \centering
  % \begin{tikzpicture}
  %   % nodes
  %   \node[circle] (z) at (-1.75,0) {$Z_1$};
  %   \node[circle] (x) at (0,0) {$X$};
  %   \node[circle] (y) at (2.5,0) {$Y$};
  %   \node[circle] (c) at (1.25,1.25) {$U$};
  %   % edges
  %   \draw[edge] (z) to node [above] {\small
  %     \color{blue}{\circled{1}}} (x);
  %   \draw[edge] (c) to (x);
  %   \draw[edge] (c) to (y);
  %   \draw [bend right, dashed] (c) to node [above,solid] {\small
  %     \color{blue}{\circled{2}}} (z);
  %   \draw [bend right, dashed] (c) to node [] {\Large $\bm{\times}$} (z);
  %   \draw[edge] (x) to (y);
  %   \draw[edge] [bend right, dashed] (z) to node [below,solid] {\small
  %     \color{blue}{\circled{3}}} (y);
  %   \draw [bend right, dashed] (z) to node [] {\Large $\bm \times$} (y);
  % \end{tikzpicture}
  % \caption{$Z_1$ is a causal variant for $X$.}
  % \end{subfigure} \hfill
  % \begin{subfigure}[c]{0.49\textwidth} \centering
  \begin{tikzpicture}
    % nodes
    \node[circle] (z) at (-1.75,1.5) {$Z_1$};
    \node[circle] (zt) at (-1.75,-1.5) {$\tilde{Z}_1$};
    \node[circle] (x) at (0,0) {$X$};
    \node[circle] (y) at (3.5,0) {$Y$};
    \node[circle] (c) at (1.75,1) {$U$};
    % edges
    \draw[edge, bend right] (z) to (zt);
    \draw[edge, bend left] (zt) to (z);
    \draw[edge] (zt) to node [above] {\small
      \color{blue}{\circled{1}}} (x);
    \draw[edge] (z) to node [below] {\small
      \color{blue}{\circled{1}}} (x);
    \draw[edge] (c) to (x);
    \draw[edge] (c) to (y);
    \draw[edge] [dashed] (c) to node [above,solid] {\small
      \color{blue}{\circled{2}}} (z);
    \draw[edge] [dashed] (c) to node [] {\Large $\bm{\times}$} (z);
    \draw[edge] [dashed] (z) to (c);
    \draw[edge] (x) to (y);
    \draw[edge] [bend right, dashed] (zt) to node [below,solid] {\small
      \color{blue}{\circled{3}}} (y);
    \draw[edge] [out=60, in=90, dashed, edge] (z) to node [above,solid] {\small
      \color{blue}{\circled{3}}} (y);
    \draw [bend right, dashed] (zt) to node [] {\Large $\bm \times$}
    (y);
    \draw [out=60, in=90, dashed, edge] (z) to node [] {\Large $\bm \times$} (y);
  \end{tikzpicture}
  % \caption{$Z_1$ is correlated with $\tilde{Z}_1$ which is a causal
  %   variant for $X$.}
  % \end{subfigure}
  \caption{Causal DAG and the three criteria for valid IV. The
    proposed IV $Z_1$ can either be a causal variant for $X$ or
    correlated with a causal variant \revise{($\tilde{Z}_1$ in the
      figure)}. $Z_1$ must be independent of any
  unmeasured confounder $U$ and cannot have any direct effect on $Y$
  or be correlated with another variant that has direct effect on $Y$.}
  \label{fig:iv-dag}
\end{figure}
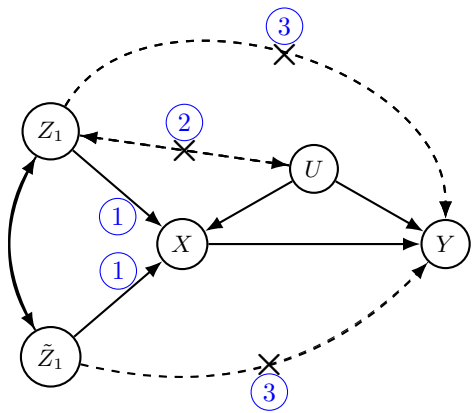

A variable $Z_j$ is called a \emph{valid} IV if it satisfies
the following three criteria:
\begin{enumerate}
\item Relevance: $Z_j$ is associated with the exposure $X$. Notice
  that a SNP that is correlated (in genetics terminology, in LD) with
  the actual causal variant is also considered relevant and does not
  affect the statistical analysis below.
\item Effective random assignment: $Z_j$ is independent of the unmeasured
  confounder $U$.
\item Exclusion restriction: $Z_j$ only affects the outcome $Y$
  through the exposure $X$. In other words, the function $f$ does not
  depend on $Z_j$.
\end{enumerate}
The causal model and the IV conditions are illustrated by a directed
acyclic graph (DAG) with a single instrument $Z_1$ in
\Cref{fig:iv-dag}. Readers who are unfamiliar with this language may find the
tutorial by \citet{baiocchi2014instrumental} helpful.

In Mendelian randomization, the first criterion---relevance---is easily
satisfied by selecting SNPs that are significantly associated with
$X$. \revise{Notice that the genetic instrument does not need to be a
  causal SNP for the exposure. The first criterion is considered satisfied if the SNP
  is correlated with the actual causal SNP
  \citep{hernan2006instruments}. For example, in \Cref{fig:iv-dag},
  $Z_1$ would be considered ``relevant'' even if it is not causal for
  $X$ but it is correlated with $\tilde{Z}_1$.} Aside from the effects of population stratification, the second independence to unmeasured confounder assumption is usually easy to justify
because most of the common confounders in epidemiology are postnatal,
which are independent of genetic variants governed by
Mendel's Second Law of independent assortment
\citep{davey2003mendelian,davey2007clustered}. Empirically, there is
generally a lack of confounding of genetic variants with factors that
confound exposures in conventional observational epidemiological
studies \citep{davey2014mendelian}.

The main concern for Mendelian randomization is the possible
violation of the third exclusion restriction criterion, due to a genetic
phenomenon called pleiotropy
\citep{davey2003mendelian,solovieff2013pleiotropy}, a.k.a.\ the multi-function
of genes. The exclusion restriction assumption does not hold if a SNP
$Z_j$ affects the outcome $Y$ through multiple causal pathways and
some do not involve the exposure $X$. It is also violated if $Z_j$
is correlated with other variants \revise{(such as $\tilde{Z}_1$ in \Cref{fig:iv-dag})} that affect $Y$ through pathways
that does not involve $X$. Pleiotropy is widely prevalent
for complex traits \citep{stearns2010one}.  In fact, a
``universal pleiotropy hypothesis'' developed
by \citet{fisher1930genetical} and \citet{wright1968evolution}
theorizes that every genetic mutation is capable of affecting essentially
all traits. Recent genetics studies have found strong evidence that
there is an extremely large number of causal variants with tiny effect
sizes on many complex traits, which in part motivates our random effects
\Cref{model:2}.

Another important concept is the strength of an IV, defined as its
association with the exposure $X$ and usually measured by the
$F$-statistic of an instrument-exposure regression. Since we assume
all the genetic instruments are independent, the strength of SNP $j$
can be assessed by comparing the statistic
$\hat{\gamma}_j^2/\sigma_{Xj}^2$ with the quantiles of $\chi^2_1$ (or
equivalently $F_{1,\infty}$). When only a few weak instruments are
available (e.g.\ $F$-statistic less than $10$), the usual asymptotic
inference is quite problematic \citep{bound1995problems}. In this
paper, we primarily consider the setting where there is at least one
strong IV or many weak IVs.

\subsection{Linear structural model}
\label{sec:line-struct-model}

We are now ready to derive the linear model \eqref{eq:model-0} for
GWAS summary data. Assuming all the IVs are valid, we start with the
linear structural model where functions $f$ and $g$ in
\eqref{eq:npsem} are linear in their arguments (see also
\citet{bowden2017framework}):
\begin{equation} \label{eq:linear-sem}
  X = \sum_{j=1}^p \gamma_j Z_j + \eta_X U + E_X,~Y = \beta X + \eta_Y U + E_Y.
\end{equation}
In this case, the GWAS summary statistics $(\hat{\gamma}_j)_{j\in[p]}$
and $(\hat{\Gamma}_j)_{j\in[p]}$ are usually computed from simple linear
regressions:
\[
  \hat{\gamma}_j = \frac{\widehat{\mathrm{Cov}}_{n_X}(X,Z_j)}{\widehat{\mathrm{Cov}}_{n_X}(Z_j,Z_j)},~\hat{\Gamma}_j = \frac{\widehat{\mathrm{Cov}}_{n_Y}(Y,Z_j)}{\widehat{\mathrm{Cov}}_{n_Y}(Z_j,Z_j)}.
\]
Here $\widehat{\mathrm{Cov}}_n$ is the sample covariance operator with
$n$ i.i.d.\ samples. Using \eqref{eq:linear-sem}, it is easy to show
that $\hat{\gamma}_j$ and
$\hat{\Gamma}_j$ converge to normal distributions centered at
$\gamma_j$ and $\Gamma_j = \beta \gamma_j$.

However, $\hat{\gamma}_j$ and $\hat{\gamma}_k$ are not exactly
uncorrelated when $j \ne k$ (same for $\hat{\Gamma}_j$ and
$\hat{\Gamma}_k$), even if $Z_j$ and $Z_k$ are independent. After some
simple algebra, one can show that
\[
  \mathrm{Cor}^2(\hat{\gamma}_j,\hat{\gamma}_k) = 4 \cdot
  \frac{\gamma_j^2\mathrm{Var}(Z_j)}{\mathrm{Var}(X) - \gamma_j^2\mathrm{Var}(Z_j)} \frac{\gamma_k^2 \mathrm{Var}(Z_k)}{\mathrm{Var}(X) - \gamma_k^2\mathrm{Var}(Z_k)}.
\]
Notice that $\gamma_j^2
\mathrm{Var}(Z_j)/\mathrm{Var}(X)$ is the proportion of variance of
$X$ explained by $Z_j$. In the genetic context, a single SNP usually has
very small predictability of a complex trait
\citep{boyle2017expanded,ioannidis2006implications,park2010estimation,shi2016contrasting}. Therefore the
correlation between
$\hat{\gamma}_j$ and $\hat{\gamma}_k$ (similarly $\hat{\Gamma}_j$ and
$\hat{\Gamma}_k$) is almost negligible.
In conclusion, the linear model
\eqref{eq:model-0} is approximately true when the phenotypes are believed to
be generated from a linear structural model.

To stick to the main statistical methodology, we postpone
additional justifications of \eqref{eq:model-0} in nonlinear
structural models to \Cref{sec:linear-model-gwas}. In
\Cref{sec:binary-outc-logist}, we will investigate the case where $Y$ is binary and
$\hat{\Gamma}_j$ is obtained via logistic regression, as is very often
the case in applied MR investigations. In
\Cref{sec:mr-identifying-local}, we will show the linearity between
$X$ and $\bm Z$ is also not necessary.

\subsection{Violations of exclusion restriction}
\label{sec:viol-excl-restr}

\Cref{eq:linear-sem} assumes that all the instruments are valid. In
reality, the exclusion restriction assumption is likely violated for
many if not most of the SNPs. To investigate its impact in the model for
summary data, we consider the following modification of the linear
structural model \eqref{eq:linear-sem}:
\begin{equation} \label{eq:linear-sem-alpha}
  X = \sum_{j=1}^p \gamma_j Z_j + \eta_X U + E_X,~Y = \beta X +
  \sum_{j=1}^p \alpha_j Z_j + \eta_Y U + E_Y.
\end{equation}
The difference between \eqref{eq:linear-sem} and
\eqref{eq:linear-sem-alpha} is that the SNPs are now allowed to
directly affect $Y$ and the effect size of SNP $Z_j$ is
$\alpha_j$. In this case, it is not difficult to see that the
regression coefficient $\hat{\Gamma}_j$ estimates $\Gamma_j = \alpha_j
+ \gamma_j \beta$. This inspires our \Cref{model:2,model:3}. In
\Cref{model:2}, we assume the direct effects $\alpha_j$ are normally
distributed random effects. In \Cref{model:3}, we further require the
statistical procedure to be robust against any extraordinarily large direct
effects $\alpha_j$. See \Cref{sec:discussion} for more discussion on
the assumptions on the pleiotropy effects.

\section{No pleiotropy: A profile likelihood approach}
\label{sec:no-plei-prof}

We now consider \Cref{model:1}, the case with no pleiotropy
effects.

\subsection{Derivation of the profile likelihood}
\label{sec:deriv-prof-likel}

A good place to start is writing down the likelihood of GWAS summary
data. Up to some additive constant, the log-likelihood function is
given by
\begin{equation} \label{eq:log-like}
  l(\beta,\gamma_1,\dotsc,\gamma_p) = - \frac{1}{2} \bigg[ \sum_{j=1}^p \frac{(\hat{\gamma}_j -
    \gamma_j)^2}{\sigma_{Xj}^2} + \sum_{j=1}^p \frac{(\hat{\Gamma}_j -
    \gamma_j \beta)^2}{\sigma_{Yj}^2} \bigg].
\end{equation}
Since we are only interested in estimating $\beta_0$, the other
parameters, namely $\bm \gamma := (\gamma_1,\cdots,\gamma_p)$, are
considered nuisance
parameters. There are two ways to proceed from here. One is to view
$\bm \gamma$ as \emph{incidental} parameters
\citep{neyman1948consistent} and try to eliminate them from the
likelihood. The other approach is to assume the sequence
$\gamma_1,\gamma_2,\cdots$ is generated from a fixed unknown
distribution. When $p$ is
large, it is possible to estimate the distribution of $\bm \gamma$ to
improve the efficiency using the second approach
\citep{murphy1996likelihood}.
In this paper we aim to develop a general method for summary-data MR
that can be used regardless of the number of SNPs being used, so we
will take the first approach.

The profile log-likelihood of $\beta$ is given by profiling out $\bm
\gamma$ in \eqref{eq:log-like}:
\begin{equation}
  \label{eq:profile-like}
  l(\beta) = \max_{\bm \gamma} l(\beta, \bm \gamma) = - \frac{1}{2} \sum_{j=1}^p
  \frac{(\hat{\Gamma}_j - \beta
    \hat{\gamma}_j)^2}{\sigma_{Xj}^2 \beta^2 + \sigma_{Yj}^2}.
\end{equation}
The maximum likelihood estimator of $\beta$ is given by $\hat{\beta} =
\argmax_{\beta} l(\beta)$. It is also called a Limited Information
Maximum Likelihood (LIML) estimator in the IV literature, a method due
to \citet{anderson1949estimation} with good consistency and efficiency
properties. See also \citet{pacini2016robust}.

% \begin{remark} \label{rmk:lin-reg}
Equation \eqref{eq:profile-like} can be interpreted as a linear
regression of $\hat{\Gamma}$ on $\hat{\gamma}$, with the intercept of
the regression fixed to zero and the variance of
each observation equaling to $\sigma_{Xj}^2 \beta^2 +
\sigma_{Yj}^2$. There is another
meta-analysis interpretation. Let $\hat{\beta}_j =
\hat{\Gamma}_j/\hat{\gamma}_j$ be the individual
Wald's ratio, then \eqref{eq:profile-like} can be rewritten as
\begin{equation} \label{eq:profile-like-2}
  l(\beta) = -\frac{1}{2} \sum_{j=1}^p \frac{(\hat{\beta}_j -
    \beta)^2}{\sigma_{Xj}^2 \beta^2/\hat{\gamma}_j^2 +
    \sigma_{Yj}^2/\hat{\gamma}_j^2}.
\end{equation}
This expression is also derived by \citet{bowden2017improving} by
defining a generalized version of Cochran's Q statistic to test for
the presence of pleiotropy that takes into account uncertainty in
$\hat{\gamma}_j$.
% \end{remark}

\subsection{Consistency and asymptotic normality}
\label{sec:cons-asympt-norm}

It is well known that the maximum likelihood estimator can
be inconsistent when there are many nuisance parameters in the problem
\citep[e.g.][]{neyman1948consistent}. Nevertheless, due to the connection
with LIML, we expect and will prove below that $\hat{\beta}$ is
consistent and asymptotically normal. However, we will also show that
the profile likelihood \eqref{eq:profile-like} can be information
biased \citep{mccullagh1990simple}, meaning the profile likelihood
ratio test does not generally have a $\chi^2_1$ limiting distribution
under the null.

A major distinction between our asymptotic setting and the classical
errors-in-variables regression setting is that our ``predictors''
$\hat{\gamma}_j,\,j\in[p]$ can be individually weak. This can be seen,
for example, from the linear structural model \eqref{eq:linear-sem}
that
\begin{equation} \label{eq:var-x}
  \mathrm{Var}(X) = \sum_{j=1}^p \gamma_j^2 \mathrm{Var}(Z_j) +
  \eta_X^2\mathrm{Var}(U) + \mathrm{Var}(E_X).
\end{equation}
Note that $Z_j$ takes on the value $0,1,2$ with probability $p_j^2$,
$2 p_j(1 - p_j)$, $(1 - p_j)^2$ where $p_j$ is the allele
frequency of SNP $j$. For simplicity, we assume $p_j$ is bounded away
from $0$ and $1$. In other words, only common genetic variants are
used as IVs. Together with \eqref{eq:var-x}, this implies that, if
$\mathrm{Var}(X)$ exists, $\|\bm \gamma\|_2$ is bounded.
\begin{assumption}[Collective IV strength is bounded] \label{assump:iv-strength}
  $\|\bm \gamma\|_2^2 = O(1)$.
\end{assumption}
% \begin{assumption}[IV strength]
%   \label{assump:iv-strength}
%   $\|\bm \gamma\| = \Theta(1)$, meaning there exist constants
%   $c_{\gamma},c_{\gamma}' > 0$ such that $c_{\gamma} \le \|\bm
%   \gamma\|_2^2 \le c_{\gamma}'$.
% \end{assumption}
% The upper bound is a natural consequence of \eqref{eq:var-x}, if the
% exposure $X$ has finite variance. The lower bound says that the
% collective strength of the genetic instruments is strong, but no
% restriction is put on the individual $\gamma_j$. Since $\|\bm
% \gamma\|$ is bounded,
As a consequence, the average effect size is decreasing to $0$,
\[
  \frac{1}{p} \sum_{j=1}^p |\gamma_j| \le \|\bm \gamma\|_2/\sqrt{p} \to
  0,~\text{when}~p \to \infty.
\]
This is clearly different from the usual linear
regression setting where the ``predictors'' $\hat{\gamma}_j$ are
viewed as random samples from a population. In the one-sample IV literature,
this many weak IV setting ($p \to \infty$) has been considered by
\citet{bekker1994alternative,stock2005asymptotic,hansen2008estimation}
among many others in econometrics.

Another difference between our asymptotic setting and the errors-in-variables regression
is that our measurement errors also converge to $0$ as the sample size
converges to infinity. Recall that $n_X$ is the sample size of
$(X,Z_1,\dotsc,Z_p)$ and $n_Y$ is the sample size of $(Y,
Z_1,\dotsc,Z_p)$.  We assume
\begin{assumption}[Variance of measurement error] \label{assump:variance}
  Let $n = \min(n_X,n_Y)$. There exist constants
  $c_{\sigma},c_{\sigma}'$ such that $c_{\sigma}/n \le \sigma_{Xj}^2
  \le c'_{\sigma}/n$ and $c_{\sigma}/n \le \sigma_{Yj}^2
  \le c'_{\sigma}/n$ for all $j \in [p]$.
\end{assumption}
We write $a = O(b)$ if there exists a constant $c > 0$
such that $|a| \le cb$, and $a = \Theta(b)$ if there exists $c >
0$ such that $c^{-1}b \le |a| \le cb$. In this notation,
\Cref{assump:variance} assumes the known variances $\sigma_{Xj}^2$ and
$\sigma_{Yj}^2$ are $\Theta(1/n)$.

In the linear structural model \eqref{eq:linear-sem},
$\mathrm{Var}(\hat{\gamma}_j) \le
\mathrm{Var}(X)/[\mathrm{Var}(Z_j)/n_X]$. Thus \Cref{assump:variance}
is satisfied when only common variants are used.

We are ready to state our first theoretical result.
\begin{theorem} \label{thm:consistency-simple}
  In \Cref{model:1} and under
  \Cref{assump:setup,assump:iv-strength,assump:variance}, if
  $p/(n^2\|\bm \gamma\|_2^4) \to 0$,
  the maximum likelihood estimator $\hat{\beta}$ is statistically
  consistent, i.e.\ $\hat{\beta} \overset{p}{\to} \beta_0$.
\end{theorem}

A crucial quantity in \Cref{thm:consistency-simple} and the analysis
below is the average strength of the IVs, defined as
\[
  \kappa = \frac{1}{p} \sum_{j=1}^p
  \frac{\gamma_j^2}{\sigma_{Xj}^2} = \Theta(n \|\bm\gamma\|_2^2 / p).
\]
An unbiased estimator of $\kappa$ is the average
$F$-statistic minus $1$,
\[
  \hat{\kappa} = \frac{1}{p} \sum_{j=1}^p \frac{\hat{\gamma}_j^2}{\sigma_{Xj}^2} - 1.
\]
In practice, we require the average $F$-statistic to be large
(say $> 100$) when $p$ is small, or not too small (say $> 3$) when $p$
is large. Thus the condition $p/(n^2 \|\bm\gamma\|_2^4) =
\Theta\big(1/(p\kappa^2)\big) \to 0$ in \Cref{thm:consistency-simple} is usually quite
reasonable. \revise{In particular, since this condition only depends on the average
  instrument strength $\kappa$, the estimator $\hat{\beta}$ remains
  consistent even if a substantial proportion of $\gamma_j = 0$ (for
  example, if the selection step in \Cref{sec:an-example-bmi} using
  \texttt{BMI-FEM} with less stringent $p$-value threshold finds many
  false positives).}

Next we study the asymptotic normality of $\hat{\beta}$. Define the
\emph{profile score} to be the derivative of the profile
log-likelihood:
\begin{equation}
  \label{eq:profile-score}
  \psi(\beta) := - l'(\beta) = \sum_{j=1}^p \frac{(\hat{\Gamma}_j - \beta
    \hat{\gamma}_j)( \hat{\Gamma}_j
    \sigma_{Xj}^2 \beta + \hat{\gamma}_j
    \sigma_{Yj}^2)}{(\sigma_{Xj}^2 \beta^2 + \sigma_{Yj}^2)^2}.
\end{equation}
% For notational simplicity, we multiplied the derive $l'(\beta)$ by
% $-1$ in \eqref{eq:profile-score}.
The maximum
likelihood estimator $\hat{\beta}$ solves the estimating equation
$\psi(\hat{\beta}) = 0$, and we consider the Taylor expansion around the
truth $\beta_0$:
\begin{equation}
  \label{eq:taylor-expansion}
  0 = \psi(\hat{\beta}) = \psi(\beta_0) + \psi'(\beta_0)
  (\hat{\beta} - \beta_0) + \frac{1}{2} \psi''(\tilde{\beta})
  (\hat{\beta} - \beta_0)^2,
\end{equation}
where $\tilde{\beta}$ is between $\hat{\beta}$ and $\beta_0$.
Since $\hat{\beta}$ is statistically consistent, the last term on
the right hand side of \eqref{eq:taylor-expansion} can be proved to be negligible, and
the asymptotic normality of $\hat{\beta}$ can be established by
showing, for some appropriate $V_1$ and $V_2$, $\psi(\beta_0)
\overset{d}{\to} \mathrm{N}(0, V_1)$ and $\psi'(\beta_0)
\overset{p}{\to} -V_2$. When $V_1 = V_2$, the profile likelihood/score
is called \emph{information unbiased} \citep{mccullagh1990simple}.

% We consider the asymptotic regime $n \to \infty$ and $\kappa(n,p,\bm
% \gamma) \to \kappa \in (0, \infty]$. To
% prove $\psi(\beta_0)$ is asymptotically normal, we need an
% additional condition when $\kappa < \infty$:
% \begin{assumption} \label{assump:clt-simple}
%   The number of IVs $p \to \infty$ and for some $\delta > 0$,
%   $p^{-(1+\delta)} \sum_{j=1}^p (n \gamma_j^2 + 1)^{1+\delta} \to 0$.
% \end{assumption}

% This Lyapunov's condition says that the $F$-statistics
% $\gamma_j^2/\sigma_{Xj}^2$ is not too different. For example, if
% $n\gamma_j^2 \equiv \kappa p$, then $p^{-(1+\delta)} \sum_{j=1}^p (n
% \gamma_j^2 + 1)^{1+\delta} = p^{-\delta} \to 0$

\begin{theorem} \label{thm:clt-simple}
  Under the assumptions in \Cref{thm:consistency-simple} and if at least one of the
  following two conditions are true: (1) $p \to \infty$ and $\|\bm
  \gamma\|_3 / \|\bm \gamma\|_2 \to 0$; (2)
  $\kappa \to \infty$; then we have
  \begin{equation}
    \label{eq:clt-simple}
    \frac{V_2}{\sqrt{V_1}}(\hat{\beta} - \beta_0) \overset{d}{\to}
    \mathrm{N}(0, 1),
  \end{equation}
  where
  \begin{equation} \label{eq:V-simple}
    \begin{split}
      V_1 &= \sum_{j=1}^p \frac{\gamma_j^2 \sigma_{Yj}^2 + \Gamma_j^2 \sigma_{Xj}^2 +
        \sigma_{Xj}^2 \sigma_{Yj}^2}{(\sigma_{Xj}^2 \beta_0^2 + \sigma_{Yj}^2)^2},~
      V_2 = \sum_{j=1}^p \frac{\gamma_j^2 \sigma_{Yj}^2 + \Gamma_j^2
        \sigma_{Xj}^2}{(\sigma_{Xj}^2 \beta_0^2 + \sigma_{Yj}^2)^2}.
    \end{split}
  \end{equation}
\end{theorem}

Notice that \Cref{thm:clt-simple} is very general. It can be applied
even in the extreme situation $p$ is fixed and $\kappa \to \infty$ (a
few strong IVs) or $p \to \infty$ and $\kappa \to 0$ (many very weak
IVs). The assumption $\|\bm
\gamma\|_3 / \|\bm \gamma\|_2 \to 0$ is used to verify a Lyapunov's
condition for a central limit theorem. It essentially says the
distribution of IV strengths is not too uneven and this assumption can
be further relaxed.

Using our rate assumption for the variances
(\Cref{assump:variance}), $V_2 = \Theta(n \|\bm \gamma\|_2^2) =
\Theta(p \kappa)$ and $V_1 = V_2
+ \Theta(p)$. This suggests that the profile
likelihood is information unbiased if and only if $\kappa \to
\infty$. In general, the amount of information bias depends on
the instrument strength $\kappa$. As an example, suppose $\beta_0 = 0$
and $\sigma_{Yj}^2 \equiv \sigma_{Y1}^2$. Then by
\eqref{eq:clt-simple} and \eqref{eq:V-simple}, $
\mathrm{Var}(\hat{\beta}) \approx V_1/V_2^2 =
(1 + {\kappa^{-1}})/V_2$. Alternatively, if we make the simplifying
assumption that $\sigma_{Yj}^2 / \sigma_{Xj}^2$ does not depend on
$j$, it is straightforward to show that
\[
  \mathrm{Var}(\hat{\beta}) \propto \frac{1 + \kappa^{-1}}{p \kappa}.
\]
This approximation can be used as a rule of thumb to select the
optimal number of IVs.

In order to obtain standard error of $\hat{\beta}$, we must estimate
$V_1$ and $V_2$ using the GWAS summary data. We propose to replace
$\gamma_j^2$ and $\Gamma_j^2$ in \eqref{eq:V-simple} by their unbiased
sample estimates, $\hat{\gamma}_j^2 - \sigma_{Xj}^2$ and
$\hat{\Gamma}_j^2 - \sigma_{Yj}^2$:
\[
  \begin{split}
    \hat{V}_1 &= \sum_{j=1}^p \frac{(\hat{\gamma}_j^2 - \sigma_{Xj}^2)
      \sigma_{Yj}^2 + (\hat{\Gamma}_j^2 - \sigma_{Yj}^2) \sigma_{Xj}^2 +
      \sigma_{Xj}^2 \sigma_{Yj}^2}{(\sigma_{Xj}^2
      \hat{\beta}^2 + \sigma_{Yj}^2)^2}, \\
    \hat{V}_2 &= \sum_{j=1}^p \frac{(\hat{\gamma}_j^2 - \sigma_{Xj}^2)
      \sigma_{Yj}^2 + (\hat{\Gamma}_j^2 - \sigma_{Yj}^2)
      \sigma_{Xj}^2}{(\sigma_{Xj}^2 \hat{\beta}^2 +
      \sigma_{Yj}^2)^2}. \\
  \end{split}
\]

\begin{theorem} \label{thm:consistent-variance-simple}
  Under the same assumptions in \Cref{thm:clt-simple}, we have $\hat{V}_1 =
  V_1 (1 + o_p(1))$, $\hat{V}_2 = V_2 (1 + o_p(1))$, and
  \begin{equation}
    \label{eq:pivot-simple}
    \frac{\hat{V}_2}{\sqrt{\hat{V}_1}}(\hat{\beta} - \beta_0) \overset{d}{\to}
    \mathrm{N}(0, 1)~\mathrm{as}~n \to \infty.
  \end{equation}
\end{theorem}

\subsection{\revise{Weak IV bias}}
\label{sec:weak-iv-bias}

\revise{
As mentioned in \Cref{sec:challenges-mr}, many existing statistical
methods for summary-data MR ignore the measurement error in
$\hat{\gamma}_j$. We briefly describe the amount of bias this
may incur for the inverse variance weighted (IVW) estimator
\citep{burgess2013mendelian}. The IVW estimator is equivalent to the
maximum likelihood estimator \eqref{eq:profile-like} assuming
$\sigma_{Xj}^2 = 0$, which has an explicit expression and can be
approximated by:
\begin{equation} \label{eq:ivw-bias-approx}
\hat{\beta}_{\mathrm{IVW}} = \frac{\sum_{j=1}^p \hat{\Gamma}_j
  \hat{\gamma}_j}{\sum_{j=1}^p \hat{\gamma}_j^2} \approx \frac{\mathbb{E}\big[\sum_{j=1}^p \hat{\Gamma}_j
  \hat{\gamma}_j\big]}{\mathbb{E}\big[\sum_{j=1}^p
  \hat{\gamma}_j^2\big]} = \frac{\beta \|\bm \gamma\|^2}{\|\bm
  \gamma\|^2 + \sum_{j=1}^p \sigma_{Xj}^2} \approx \frac{\beta}{1 + (1/\kappa)}.
\end{equation}
Thus the amount of bias for the IVW estimator crucially depends on the
average IV strength $\kappa$. In comparison, our consistency result
(\Cref{thm:consistency-simple}) only requires $\kappa \gg 1/\sqrt{p}$.
}

\subsection{Practical issues}
\label{sec:practical-issues}

Next we discuss several practical implications of the theoretical
results above.

\subsubsection{Influence of a single IV}
\label{sec:influence-single-iv}

Under the assumptions in \Cref{thm:clt-simple},
\eqref{eq:taylor-expansion} and \eqref{eq:profile-score} lead to the
following asymptotically linear form of $\hat{\beta}$:
\[
  \hat{\beta} = \frac{1 + o_p(1)}{V_2} \sum_{j=1}^p \frac{(\hat{\Gamma}_j - \beta_0
    \hat{\gamma}_j)( \hat{\Gamma}_j
    \sigma_{Xj}^2 \beta_0 + \hat{\gamma}_j
    \sigma_{Yj}^2)}{(\sigma_{Xj}^2 \beta_0^2 + \sigma_{Yj}^2)^2}.
\]
The above equation characterizes the influence of a single IV on the
estimator $\hat{\beta}$ \citep{hampel1974influence}. Intuitively, the
IV $Z_j$ has large influence if it is strong or it has large
residual $\hat{\Gamma}_j - \beta_0 \hat{\gamma}_j$. Alternatively, we can measure the influence of a single IV by
computing the leave-one-out estimator $\hat{\beta}_{-j}$ that maximizes
the profile likelihood with all the SNPs except $Z_j$. In practice, it
is desirable to limit the influence of each SNP to make the estimator
robust against idiosyncratic pleiotropy (\Cref{model:3}). This problem
will be considered in \Cref{sec:idiosyncr-plei-robus}.

\subsubsection{Selecting IVs}
\label{sec:selecting-ivs}

The formulas \eqref{eq:clt-simple} and \eqref{eq:V-simple} suggest that using
extremely weak
instruments may deteriorate the efficiency. Consider the following example
in which we have a new instrument $Z_{p+1}$ that is independent of $X$, so
$\gamma_{p+1}=0$. When adding $Z_{p+1}$ to the analysis, $V_1$
increases but $V_2$ remains the same, thus the variance of $\hat{\beta}$
becomes larger. Generally, this suggests that we should screen out
extremely weak IVs to improve efficiency. To avoid selection bias, we
recommend to use two independent GWAS datasets in practice, one to
screen out weak IVs and perform LD clumping and one to estimate the
SNP-exposure effects $\gamma_j$ unbiasedly.

\subsubsection{Residual quantile-quantile plot}
\label{sec:quant-quant-plot}

One way to check the modeling assumptions in
\Cref{assump:setup,model:1} is the residual Quantile-Quantile (Q-Q) plot, which
plots the quantiles of standardized residuals
\[
  \hat{t}_j = \frac{\hat{\Gamma}_j - \hat{\beta}
    \hat{\gamma}_j}{\sqrt{\hat{\beta}^2\sigma_{Xj}^2 + \sigma_{Yj}^2}}
\]
against the quantiles of the standard normal distribution. This is
reasonable because when
$\hat{\beta} = \beta_0$, $\hat{t}_j \sim \mathrm{N}(0,1)$ under
\Cref{assump:setup,model:1}. The Q-Q plot is helpful at identifying
IVs that do not satisfy the linear relation $\Gamma_j =
\beta_0\gamma_j$, most likely due to genetic pleiotropy.

Besides the residual Q-Q plot, other diagnostic tools can be found in
related works. \citet{bowden2017improving} considered using each SNP's
contribution to the generalized Q statistic to assess
whether it is an outlier. \citet{bowden2017improving2} proposed a
radial plot $\hat{\beta}_j \sqrt{w}_j$ versus $\sqrt{w}_j$, where $w_j$ is the
``weight'' of the $j$-th SNP in \eqref{eq:profile-like-2}. Since these
diagnostic methods are based on the Wald ratio estimates $\hat{\beta}_j$,
they can suffer from the weak instrument bias.

\subsection{Example (continued)}
\label{sec:example-continued}

We conclude this Section by applying the profile likelihood or Profile
Score (PS) estimator
in the BMI-SBP example in \Cref{sec:an-example-bmi}. Here
we used $160$ SNPs that have $p$-values $\le 10^{-4}$
in the \texttt{BMI-FEM} dataset. The PS point estimate is $0.601$
with standard error $0.054$.

\begin{figure}[t]
  \centering
  \includegraphics[width = \textwidth]{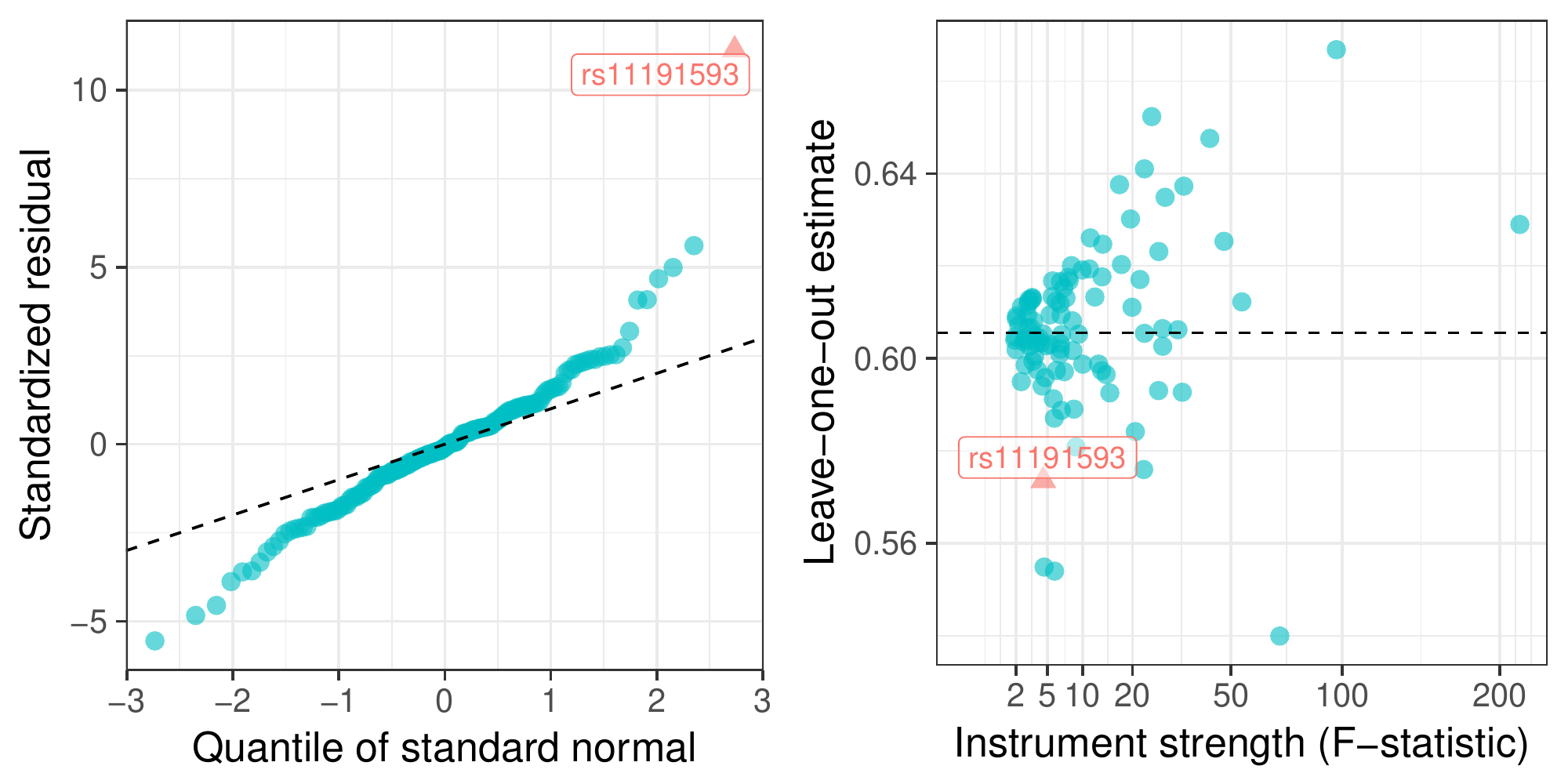}
  \caption{Diagnostic plots of the Profile Score (PS) estimator. Left
    panel is a Q-Q plot of the standardized residuals against standard
    normal. Right panel is the leave-one-out estimates against
    instrument strength.}
  \label{fig:ps}
\end{figure}

\Cref{fig:ps} shows the Q-Q plot and the leave-one-out estimates
discussed in \Cref{sec:practical-issues}. The Q-Q plot clearly
indicates the linear model \Cref{model:1} is not appropriate to
describe the summary data. Although the standardized residuals are
roughly normally distributed, their standard deviations are apparently larger
than $1$. This
motivates the random pleiotropy effects assumption in \Cref{model:2}
which will be considered next.

\section{Systematic pleiotropy: Adjusted profile score}
\label{sec:syst-plei-adjust}

\subsection{Failure of the profile likelihood}
\label{sec:fail-prof-likel}

Next we consider \Cref{model:2}, where the deviation from the linear
relation $\Gamma_j = \beta_0\gamma_j$ is described by a random effects
model $\alpha_j = \Gamma_j - \beta_0 \gamma_j \sim \mathrm{N}(0,
\tau_0^2)$. The normality assumption is motivated by \Cref{fig:ps}
and does not appear to be very consequential in the simulation studies.
In this model, the
variance of $\hat{\Gamma}$ is essentially inflated by an unknown additive constant $\tau_0^2$:
\[
  \hat{\gamma}_j \sim \mathrm{N}(\gamma_j, \sigma_{Xj}^2),~
  \hat{\Gamma}_j \sim \mathrm{N}(\gamma_j \beta_0, \sigma_{Yj}^2 + \tau_0^2),~j \in [p].
\]

Similar to \Cref{sec:deriv-prof-likel}, the profile
log-likelihood of $(\beta,\tau^2)$ is given by
\begin{equation*}
  \label{eq:profile-like-overdispersed}
  l(\beta,\tau^2) = - \frac{1}{2} \sum_{j=1}^p
  \frac{(\hat{\Gamma}_j - \beta
    \hat{\gamma}_j)^2}{\sigma_{Xj}^2 \beta^2 + \sigma_{Yj}^2 + \tau^2}
  + \log (\sigma_{Yj}^2 + \tau^2),
\end{equation*}
and the corresponding profile score equations are
\begin{equation*}
  \frac{\partial}{\partial \beta} l(\beta,\tau^2) = 0,~\frac{\partial}{\partial \tau^2} l(\beta,\tau^2) = 0.
\end{equation*}
It is straightforward to verify that the first estimating equation is unbiased, i.e.\ it has
expectation $0$ at $(\beta_0,\tau_0^2)$. However, the other
profile score is
\begin{equation} \label{eq:naive-ps}
  \frac{\partial}{\partial \tau^2} l(\beta,\tau^2) = \frac{1}{2} \sum_{j=1}^p
  \frac{(\hat{\Gamma}_j - \beta
    \hat{\gamma}_j)^2}{(\sigma_{Xj}^2
    \beta^2 + \sigma_{Yj}^2 + \tau^2)^2} - \frac{1}{\sigma_{Yj}^2 + \tau^2}.
\end{equation}
It is easy to see that its expectation is not
equal to $0$ at the true value $(\beta, \tau^2) =
(\beta_0, \tau_0^2)$. This means the profile score is biased in
\Cref{model:2}, thus the corresponding
maximum likelihood estimator is not statistically consistent.

\subsection{Adjusted profile score}
\label{sec:adjust-prof-score}

The failure of maximizing the profile likelihood should not be surprising,
because it is well known that maximum likelihood estimator can be
biased when there are many nuisance parameters
\citep{neyman1948consistent}. There are many proposals to modify the
profile likelihood, see, for example,
\citet{barndorff1983formula,cox1987parameter}. Here we take the
approach of \citet{mccullagh1990simple} that directly modifies the
profile score so it has mean $0$ at the true value. The \emph{Adjusted
  Profile Score (APS)} is given by $\bm \psi(\beta,\tau^2) =
(\psi_1(\beta,\tau^2), \psi_2(\beta,\tau^2))$, where
\begin{align}
  \psi_1(\beta,\tau^2) &= - \frac{\partial}{\partial \beta} l(\beta,\tau^2) = \sum_{j=1}^p \frac{(\hat{\Gamma}_j - \beta
                         \hat{\gamma}_j)(\hat{\Gamma}_j
                         \sigma_{Xj}^2 \beta + \hat{\gamma}_j (\sigma_{Yj}^2 + \tau^2))}{(\sigma_{Xj}^2 \beta^2 + \sigma_{Yj}^2 + \tau^2)^2}, \label{eq:aps-1} \\
  \psi_2(\beta,\tau^2) &= \sum_{j=1}^p \sigma_{Xj}^2
                         \frac{(\hat{\Gamma}_j - \beta
                         \hat{\gamma}_j)^2 - (\sigma_{Xj}^2
                         \beta^2 + \sigma_{Yj}^2 + \tau^2)}{(\sigma_{Xj}^2
                         \beta^2 + \sigma_{Yj}^2 + \tau^2)^2}. \label{eq:aps-2}
\end{align}
Compared to \eqref{eq:naive-ps}, we replaced $(\sigma_{Yj}^2 +
\tau^2)^{-1}$ by $(\sigma_{Xj}^2 \beta^2 + \sigma_{Yj}^2 + \tau^2)^{-1}$,
so each summand in \eqref{eq:aps-2} has mean $0$ at
$(\beta_0,\tau_0^2)$. We also weighted the IVs by $\sigma_{Xj}^2$ in
\eqref{eq:aps-2}, which is useful in the proof of statistical
consistency.

Notice that both the denominators and numerators in $\psi_1$ and
$\psi_2$ are polynomials of $\beta$ and $\tau^2$. However, the
denominators are of higher degrees. This implies that the APS estimating
equations always have diverging solutions:
$\bm{\psi}(\beta,\tau^2) \to \bm{0}$ if $\beta \to
\pm \infty$ or $\tau^2 \to \infty$. We define the APS estimator
$(\hat{\beta},\hat{\tau}^2)$ to be the non-trivial finite solution to $\bm
\psi(\beta,\tau^2) = \bm 0$ if it exists.

\subsection{Consistency and asymptotic normality}
\label{sec:cons-asympt-norm-1}

Because of the diverging solutions of the APS equations, we
need to impose some compactness constraints on the parameter space to study
the asymptotic property of $(\hat{\beta},\hat{\tau}^2)$:
\begin{assumption} \label{assump:compactness}
  $(\beta_0,p\tau_0^2)$ is in the interior of a bounded set $\mathcal{B} \subset \mathbb{R} \times
  \mathbb{R}^{+}$.
\end{assumption}
The overdispersion parameter $\tau_0^2$ is scaled up in
\Cref{assump:compactness} by $p$. This is motivated by
the linear structural model \eqref{eq:linear-sem-alpha}, where $\sum_{j=1}^2 \tau_0^2\mathrm{Var}(Z_j)
= \Theta(p \tau_0^2)$ is the variance of $Y$ explained by the direct
effects of $\bm Z$. Thus it is reasonable to treat $p \tau_0^2$ as a
constant.

We also assume, in addition to \Cref{assump:iv-strength}, that the variance
of $X$ explained by the IVs is non-diminishing:
\begin{assumption} \label{assump:iv-strength-2}
  $\|\bm\gamma\|_2 = \Theta(1)$.
\end{assumption}

\begin{theorem} \label{thm:consistency-overdispersed}
  In \Cref{model:2} and suppose
  \Cref{assump:setup,assump:iv-strength-2,assump:variance,assump:compactness}
  hold, $p \to \infty$ and $p/n^2 \to
  0$. Then with probability going to $1$ there exists a solution
  of the APS equation such that $(\hat{\beta},p \hat{\tau}^2)$ is in $\mathcal{B}$. Furthermore, all
  solutions in $\mathcal{B}$ are statistically consistent, i.e.\ $\hat{\beta}
  \overset{p}{\to} \beta_0$ and $p\hat{\tau}^2 - p\tau_0^2
  \overset{p}{\to} 0$.
\end{theorem}

Next we consider the asymptotic distribution of the APS estimator.

\begin{theorem} \label{thm:clt-overdispersed}
  In \Cref{model:2} and under the assumptions in
  \Cref{thm:consistency-overdispersed}, if additionally
  $p = \Theta(n)$ and $\|\bm \gamma\|_3 / \|\bm \gamma\|_2 \to 0$, then
  \begin{equation} \label{eq:clt-overdispersed}
    \big(\tilde{\bm{V}}_2^{-1} \tilde{\bm{V}}_1
    \tilde{\bm{V}}_2^{-T}\big)^{1/2}
    \begin{pmatrix}
      \hat{\beta} - \beta_0 \\
      \hat{\tau}^2 - \tau_0^2
    \end{pmatrix}
    \overset{d}{\to} \mathrm{N}(\bm{0}, \bm{I}_2),
  \end{equation}
  where
  \[
    \tilde{\bm{V}}_1 = % \mathrm{Var}(\bm{\psi}(\beta_0,\tau_0^2)) =
    \sum_{j=1}^p \frac{1}{(\sigma_{Xj}^2 \beta_0^2 + \sigma_{Yj}^2 +
      \tau_0^2)^2}
    \begin{pmatrix}
      (\gamma_j^2 + \sigma_{Xj}^2)
      (\sigma_{Yj}^2 + \tau_0^2) + \Gamma_j^2 \sigma_{Xj}^2 & 0 \\
      0 & 2 (\sigma_{Xj}^2)^2 \\
    \end{pmatrix},
  \]
  \[
    \tilde{\bm{V}}_2 = % \mathbb{E}\Big[
    % \big(\frac{\partial}{\partial\beta}~\frac{\partial}{\partial\tau^2}\big)
    % \begin{pmatrix}
    %   \psi_1(\beta_0,\tau_0^2) \\
    %   \psi_2(\beta_0,\tau_0^2)
    % \end{pmatrix}
    % \Big] =
    \sum_{j=1}^p
    \frac{1}{(\sigma_{Xj}^2 \beta_0^2 + \sigma_{Yj}^2 +
      \tau_0^2)^2}
    \begin{pmatrix}
      \gamma_j^2 (\sigma_{Yj}^2 + \tau_0^2) + \Gamma_j^2 \sigma_{Xj}^2 & \sigma_{Xj}^2 \beta_0 \\
      0 & \sigma_{Xj}^2 \\
    \end{pmatrix}.
  \]
\end{theorem}

Similar to \Cref{thm:consistent-variance-simple}, the information matrices $\tilde{\bm V}_1$ and $\tilde{\bm V}_2$ can
be estimated by substituting $\gamma_j^2$ by
$\hat{\gamma}_j^2 - \sigma_{Xj}^2$ and $\Gamma_j^2$ by $\hat{\Gamma}_j^2 -
\sigma_{Yj}^2 - \hat{\tau}^2$. We omit the details for brevity.

\subsection{Example (continued)}
\label{sec:example-continued-1}

We apply the APS estimator to the BMI-SBP example. Using the same 160 SNPs in
\Cref{sec:example-continued}, the APS point estimate is $\hat{\beta} =
0.301$ (standard error $0.158$) and $\hat{\tau}^2 = 9.2 \times
10^{-4}$ (standard error $1.7 \times 10^{-4}$). Notice that the APS
point estimate of $\beta$ is much
smaller than the PS point estimate. One possible explanation of this phenomenon
is that the PS estimator tends to use a larger $\beta$ to compensate for the
overdispersion in \Cref{model:2} (the variance of $\hat{\Gamma}_j -
\beta \hat{\gamma}_j$ is $\beta^2\sigma_{Xj}^2 + \sigma_{Yj}^2$ in
\Cref{model:1} and $\beta^2 \sigma_{Xj}^2 + \sigma_{Yj}^2 + \tau_0^2$
in \Cref{model:2}).

\begin{figure}[t]
  \centering
  \includegraphics[width = \textwidth]{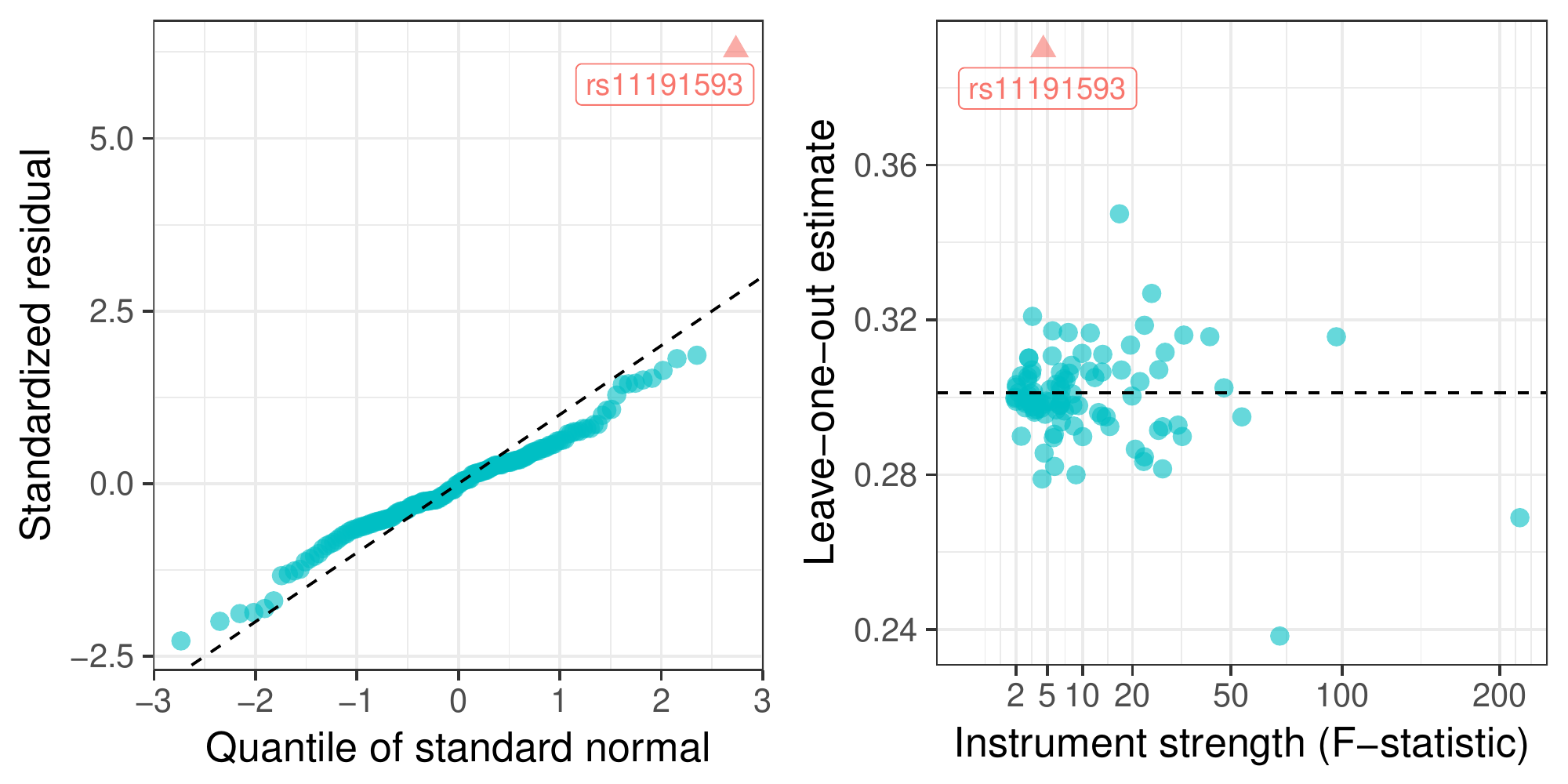}
  \caption{Diagnostic plots of the Adjusted Profile Score (APS)
    estimator. Left panel is a Q-Q plot of the standardized residuals
    against standard normal. Right panel is the leave-one-out
    estimates against instrument strength.}
  \label{fig:aps}
\end{figure}

\Cref{fig:aps} shows the diagnostic plots of the APS
estimator. Compared to the PS estimator in
\Cref{sec:example-continued}, the overdispersion issue is much more
benign. However, there is an outlier which corresponds to the SNP
\texttt{rs11191593}. It heavily biases the APS estimate too: when
excluding this SNP, the APS point estimate changes from $0.301$ to
almost $0.4$ in the right panel of \Cref{fig:aps}. The outlier might also
inflate $\hat{\tau}^2$ so the Q-Q plot looks a little
underdispersed. These observations motivate the consideration of a robust
modification of the APS in the next Section.

\section{Idiosyncratic pleiotropy: Robustness to outliers}
\label{sec:idiosyncr-plei-robus}

Next we consider \Cref{model:3} with idiosyncratic pleiotropy. As mentioned in
\Cref{sec:influence-single-iv}, a single IV can have
unbounded influence on the PS (and APS) estimators. When the
IV $Z_j$ has other strong causal pathways, its pleiotropy parameter
$\alpha_j$ can be much larger than what is predicted by the random
effects model $\alpha_j \sim \mathrm{N}(0, \tau_0^2)$, leading to a biased
estimate of the causal effect as illustrated in
\Cref{sec:example-continued-1}. In this Section, we propose a general method to robustify the APS
to limit the influence of outliers such as SNP \texttt{rs11191593} in
the example.

\subsection{Robustify the adjusted profile score}
\label{sec:robust-prof-score}

Our approach is an application of the robust regression techniques
pioneered by \citet{huber1964robust}. As mentioned in
\Cref{sec:deriv-prof-likel}, the profile likelihood
\eqref{eq:profile-like} can
be viewed as a linear regression of $\hat{\Gamma}_j$ on
$\hat{\gamma}_j$ using the $l_2$-loss. To limit the influence of a
single IV, we consider changing the $l_2$-loss to a robust loss
function. Two celebrated examples are the Huber loss
\[
  \rho_{\mathrm{huber}}(r;k) =
  \begin{cases}
    r^2/2, & \mathrm{if}~|r| \le k, \\
    k (|r| - k/2), & \mathrm{otherwise}, \\
  \end{cases}
\]
and Tukey's biweight loss
\[
  \rho_{\mathrm{tukey}}(r;k) =
  \begin{cases}
    1 - (1 - (r/k)^2)^3, & \mathrm{if}~|r| \le k, \\
    1, & \mathrm{otherwise}. \\
  \end{cases}
\]
This heuristic motivates the following modification of the profile
log-likelihood when $\tau_0^2 = 0$:
\begin{equation}
  \label{eq:robust-profile-loglike}
  l_{\rho}(\beta) := - \sum_{j=1}^p \rho\bigg(\frac{\hat{\Gamma}_j - \beta
    \hat{\gamma}_j}{\sqrt{\sigma_{Xj}^2 \beta^2 + \sigma_{Yj}^2}}\bigg)
\end{equation}
It is easy to see that $l_{\rho}(\beta)$ reduces to the regular
profile log-likelihood \eqref{eq:profile-like} if $\rho(r) =
r^2/2$.

When $\tau_0^2 > 0$, we cannot directly use the profile score
$(\partial/\partial \tau^2) l(\beta,\tau^2)$ as discussed in
\Cref{sec:fail-prof-likel}. This issue can be resolved using the APS
approach in \Cref{sec:adjust-prof-score} by using $\psi_2$ in
\eqref{eq:aps-2}. However, a single IV can still have unbounded influence in
$\psi_2$. We must further robustify $\psi_2$, which is analogous to
estimating a scale parameter robustly.

Next we briefly review the robust M-estimation of scale parameter. Consider
repeated measurements of a scale family with density
$f_0(r/\sigma)/\sigma$. Then a general way of robust estimation of
$\sigma$ is to solve the following estimating equation \citep[Section 2.5]{maronna2006robust}
\[
  \hat{\mathbb{E}}[(R/\sigma) \cdot \rho'(R/\sigma)] = \delta,
\]
where $\hat{\mathbb{E}}$ stands for the empirical average and $\delta = \mathbb{E}[R \cdot \rho'(R)]$ for $R \sim f_0$.%  and
% some robust loss function $\rho(\cdot)$.

Based on the above discussion, we propose the following \emph{Robust
  Adjusted Profile Score (RAPS)} estimator of $\beta$. Denote
\[
  t_j(\beta,\tau^2) = \frac{\hat{\Gamma}_j - \beta
    \hat{\gamma}_j}{\sqrt{\sigma_{Xj}^2 \beta^2 + \sigma_{Yj}^2 + \tau^2}}.
\]
Then the RAPS $\bm \psi^{(\rho)} = (\psi_1^{(\rho)}, \psi_2^{(\rho)})$ is
given by
\begin{align}
  \psi_1^{(\rho)}(\beta,\tau^2) &= \sum_{j=1}^p \rho'(t_j(\beta,\tau^2))
                                  u_j(\beta,\tau^2), \label{eq:raps-1} \\
  \psi_2^{(\rho)}(\beta,\tau^2) &= \sum_{j=1}^p \sigma_{Xj}^2
                                  \frac{t_j(\beta,\tau^2) \cdot \rho'(t_j(\beta,\tau^2)) -
                                  \delta}{\sigma_{Xj}^2 \beta^2 + \sigma_{Yj}^2 + \tau^2}, \label{eq:raps-2}
\end{align}
where $\rho'(\cdot)$ is the derivative of $\rho(\cdot)$,
$u_j(\beta,\tau^2)= - (\partial/\partial \beta) t_j(\beta,\tau^2)$
and $\delta = \mathbb{E}[R \cdot \rho'(R)]$ for $R \sim
\mathrm{N}(0,1)$. Notice that $\bm \psi^{(\rho)}$ reduces to the
non-robust APS $\bm \psi$ in \eqref{eq:aps-1} and \eqref{eq:aps-2}
when $\rho(r) = r^2/2$ is the squared error loss. Finally, the RAPS
estimator $(\hat{\beta},\hat{\tau}^2)$ is given by the non-trivial
finite solution of $\bm \psi^{(\rho)}(\beta, \tau^2) = \bm 0$.

\subsection{Asymptotics}
\label{sec:asymptotics}

Because the RAPS estimator is the solution of a system of nonlinear
equations, its asymptotic behavior is very difficult to analyze. For
instance, it is difficult to establish statistical consistency because
there could be multiple roots for the RAPS equations in the
population level. Thus $\beta$ might not be globally identified. We
can, nevertheless, verify the local identifiability
\citep{rothenberg1971identification}:
\begin{theorem}[Local identification of RAPS] \label{prop:raps-local-id}
  In \Cref{model:2}, $\mathbb{E}[\bm\psi^{(\rho)}(\beta_0,\tau_0^2)] = \bm
  0$ and $\mathrm{E}[\nabla \bm \psi^{(\rho)}]$ has full
  rank.
\end{theorem}

In practice, we find that the RAPS estimating equation usually only
has one finite solution. To study the asymptotic normality of the RAPS
estimator, we will assume $(\hat{\beta},p\hat{\tau}^2)$ is consistent
under \Cref{model:2}. We further impose the following smoothness
condition on the robust loss function $\rho$:

\begin{assumption} \label{assump:rho}
  The first three derivatives of $\rho(\cdot)$ exist and are bounded.
\end{assumption}

\begin{theorem} \label{thm:clt-raps}
  In \Cref{model:2} and under the assumptions in
  \Cref{thm:clt-overdispersed}, if additionally we assume
  \begin{enumerate}
  \item the RAPS estimator is consistent: $\hat{\beta} - \beta_0 \overset{p}{\to} 0$, $p(\hat{\tau}^2 -
    \tau_0^2) \overset{p}{\to} 0$,
  \item \Cref{assump:rho} holds, and
  \item $\|\bm \gamma\|_3^3/\|\bm \gamma\|_2^3 = O(p^{-1/2})$,
  \end{enumerate}
  then
  \begin{equation} \label{eq:clt-raps}
    \big((\tilde{\bm{V}}^{(\rho)}_2)^{-1} \tilde{\bm{V}}_1^{(\rho)}
    (\tilde{\bm{V}}_2^{(\rho)})^{-T}\big)^{1/2}
    \begin{pmatrix}
      \hat{\beta} - \beta_0 \\
      \hat{\tau}^2 - \tau_0^2
    \end{pmatrix}
    \overset{d}{\to} \mathrm{N}(\bm{0}, \bm{I}_2),
  \end{equation}
  where
  \[
    \begin{split}
      \tilde{\bm V}_1^{(\rho)} &=
      \begin{pmatrix}
        c_1 (\tilde{\bm V}_1)_{11} & 0 \\
        0 & c_2 (\tilde{\bm V}_1)_{22} \\
      \end{pmatrix}, \\
      \tilde{\bm V}_2^{(\rho)} &=
      \begin{pmatrix}
        \delta (\tilde{\bm V}_2)_{11} & \delta (\tilde{\bm V}_2)_{12} \\
        0 & [(\delta + c_3)/2] (\tilde{\bm V}_2)_{22} \\
      \end{pmatrix},
    \end{split}
  \]
  and the constants are: for $R \sim \mathrm{N}(0, 1)$, $c_1 = \mathbb{E}[\rho'(R)^2]$, $c_2 =
  \mathrm{Var}(R \rho'(R))/2$, $c_3 =
  \mathbb{E}[R^2 \rho''(R)]$.
\end{theorem}

It is easy to verify that when $\rho(r) = r^2/2$, $\delta = c_1 = c_2
= c_3 = 1$, so $\tilde{\bm V}_1^{(\rho)}$ and $\tilde{\bm
  V}_2^{(\rho)}$ reduce to $\tilde{\bm V}_1$ and $\tilde{\bm V}_2$. In
other words, the asymptotic variance formula in \Cref{thm:clt-raps} is
consistent with the one in \Cref{thm:clt-overdispersed}. However,
additional technical assumptions are needed in \Cref{thm:clt-raps} to
bound the higher-order terms in the Taylor expansion.

\subsection{Example (continued)}
\label{sec:example-continued-2}

As before, we illustrate the RAPS estimator using the BMI-SBP
example. Using the Huber loss with $k = 1.345$ (corresponding to 95\%
asymptotic efficiency in the simple location problem), the point estimate is
$\hat{\beta} = 0.378$ (standard error $0.121$), $\hat{\tau}^2 = 4.7
\times 10^{-4}$ (standard error $1.0 \times 10^{-4}$). Using the
Tukey loss with $k = 4.685$ (also corresponding to 95\% asymptotic
efficiency in the simple location problem), the point estimate is
$\hat{\beta} = 0.402$ (standard error $0.106$), $\hat{\tau}^2 = 3.4
\times 10^{-4}$ (standard error $7.8 \times 10^{-5}$).

\begin{figure}[p]
  \centering
  \begin{subfigure}[b]{\textwidth}
    \includegraphics[width = \textwidth]{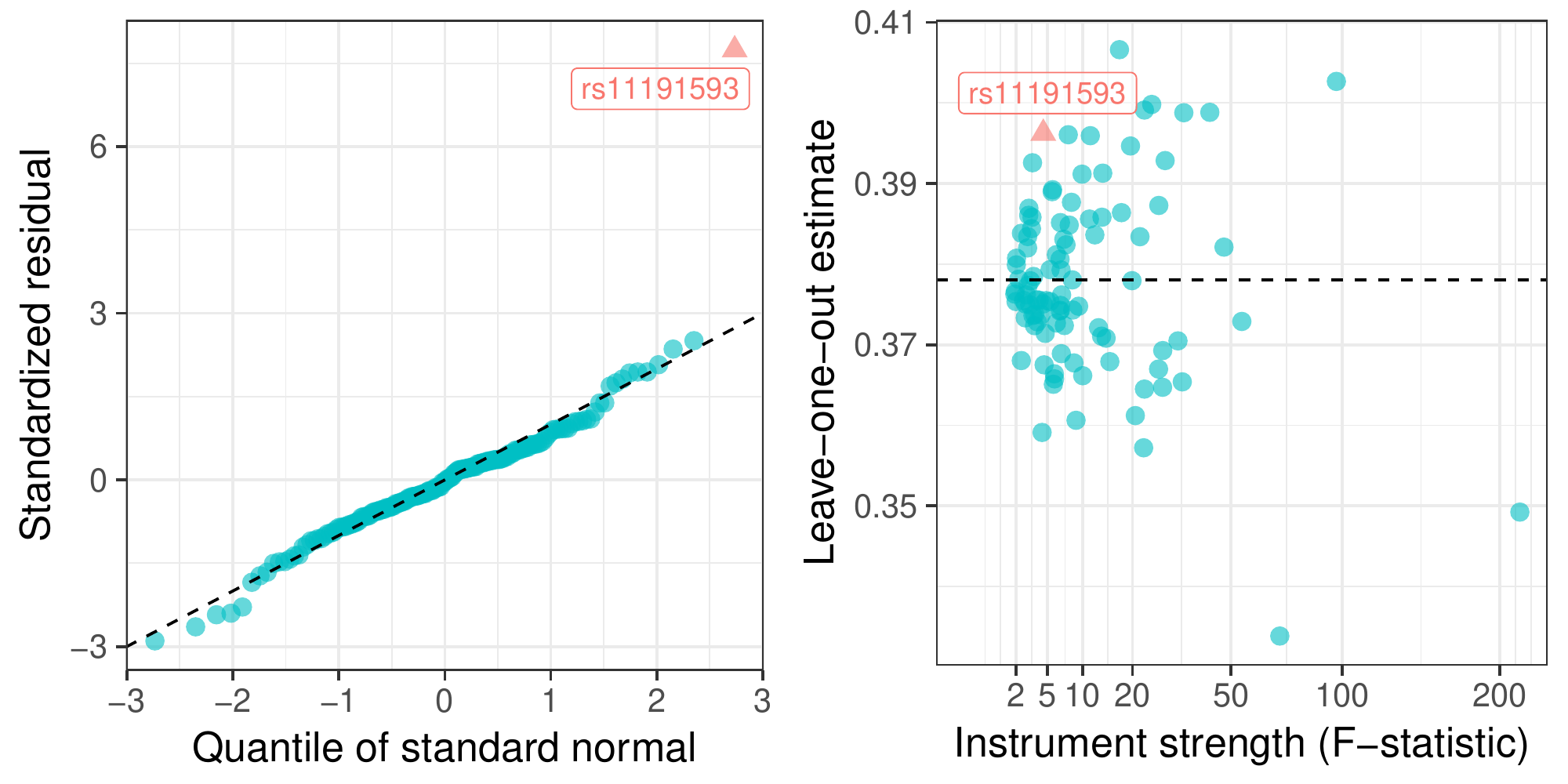}
    \caption{RAPS using the Huber loss.} \label{fig:huber}
  \end{subfigure}
  \begin{subfigure}[b]{\textwidth}
    \includegraphics[width = \textwidth]{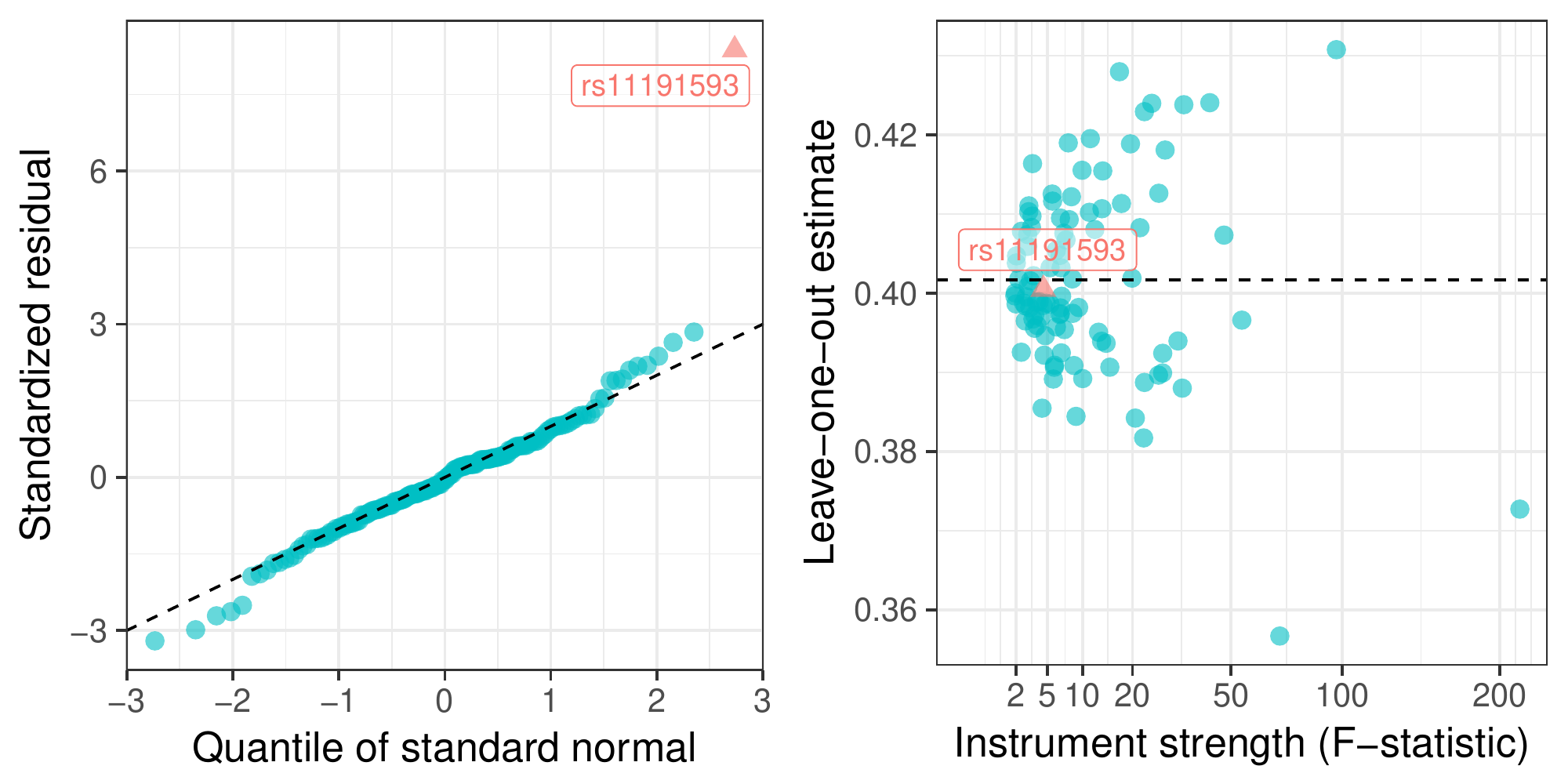}
    \caption{RAPS using the Tukey loss.} \label{fig:tukey}
  \end{subfigure}
  \caption{Diagnostic plots of the Robust Adjusted Profile Score (RAPS)
    estimator. Left panels are Q-Q plots of the standardized residuals
    against standard normal. Right panels are the leave-one-out
    estimates against instrument strength.}
  \label{fig:raps}
\end{figure}

\Cref{fig:raps} shows the diagnostic plots of the two RAPS
estimators. Compared to \Cref{fig:aps}, the robust loss functions
limit the influence of the outlier (SNP \texttt{rs11191593}), and the
resulting $\hat{\beta}$ becomes larger. In
\Cref{fig:tukey}, the outlier's influence is essentially zero because
the Tukey loss function is redescending. This shows the robustness of
our RAPS estimator to the idiosyncratic pleiotropy.

\section{Simulation}
\label{sec:simulation}

Throughout the paper all of our theoretical results are asymptotic. We
usually require both the sample size $n$ and the number of IVs $p$ to
go to infinity (except for \Cref{thm:clt-simple} where finite $p$ is
allowed). We now assess if the asymptotic approximations are
reasonably accurate in practical situations, where $p$ may range from
tens to hundreds.

\subsection{\revise{Simulating summary data directly from \Cref{assump:setup}}}
\label{sec:simulating-from-cref}

To this end, we first created simulated summary-data MR datasets that
mimic the BMI-SBP example in \Cref{sec:an-example-bmi}. In
particular, we considered two scenarios: $p=25$, which corresponds to using the
selection threshold $5 \times 10^{-8}$ as described in
\Cref{sec:an-example-bmi}, and $p=160$, which corresponds to using the
threshold $1 \times 10^{-4}$ as in
\Cref{sec:example-continued,sec:example-continued-1,sec:example-continued-2}. The
model parameters are chosen as follows: the variances of the
measurement error, $\{(\sigma_{Xj}^2,\sigma_{Yj}^2)\}_{j \in [p]}$, are
the same as those in the BMI-SBP dataset. The true \revise{marginal} IV-exposure effects,
$\{\gamma_j\}_{j\in[p]}$, are chosen to be the observed effects in the
BMI-SBP dataset, and $\hat{\gamma}_j$ is generated according to
\Cref{assump:setup} by $\hat{\gamma}_j \overset{ind.}{\sim}
\mathrm{N}(\gamma_j,\sigma_{Xj}^2)$. The true \revise{marginal} IV-outcome effects,
$\{\Gamma_j\}_{j\in[p]}$, are generated in six different ways with
$\beta_0 = 0.4$:
\begin{enumerate}
\item $\Gamma_j = \gamma_j \beta_0$;
\item $\Gamma_j = \gamma_j \beta_0 + \alpha_j$, $\alpha_j
  \overset{i.i.d.}{\sim} \mathrm{N}(0,\tau_0^2)$, where $\tau_0 = 2 \cdot
  (1/p) \sum_{j=1}^p \sigma_{Yj}$;
\item $\Gamma_j$ is generated according to setup $2$ above, except
  that $\alpha_1$ has mean $5 \cdot \tau_0$ (the IVs are sorted so
  that the first IV has the largest $|\gamma_j|/\sigma_{Xj}$).
\item $\Gamma_j = \gamma_j \beta_0 + \alpha_j$, $\alpha_j
  \overset{i.i.d.}{\sim} \tau_0 \cdot \mathrm{Lap}(1)$, where
  $\mathrm{Lap}(1)$ is the Laplace (double exponential) distribution with rate
  $1$.
\item $\Gamma_j = \gamma_j \beta_0 + \alpha_j$, $\alpha_j =
  |\gamma_j|/(p^{-1}\sum_{j=1}^p |\gamma_j|) \cdot
  \mathrm{N}(0,\tau_0^2)$.
\item $\Gamma_j$ is generated according to setup $2$ above, except
  that for 10\% randomly selected IVs, their direct effects $\alpha_j$
  have mean $5 \cdot \tau_0$.
\end{enumerate}
The first three setups correspond to \Cref{model:1,model:2,model:3},
respectively, and the last three setups violate our modeling
assumptions and are used to assess the robustness of the procedures.
Finally, $\hat{\Gamma}_j$ is generated according to
\Cref{assump:setup} by $\hat{\Gamma}_j \overset{ind.}{\sim}
\mathrm{N}(\Gamma_j,\sigma_{Yj}^2)$.

We applied six methods to the simulated data (10,000 replications in
each setting). The first three are existing methods to benchmark our performance: the inverse variance weighting (IVW) estimator
\citep{burgess2013mendelian}, MR-Egger regression
\citep{bowden2015mendelian}, and the weighted median estimator
\citep{bowden2016consistent}. The next three methods are proposed in
this paper: the profile score (PS) estimator in
\Cref{sec:no-plei-prof}, the adjusted profile score (APS) estimator
in \Cref{sec:syst-plei-adjust}, and the robust adjusted profile score
(RAPS) estimator in \Cref{sec:idiosyncr-plei-robus} with Tukey's loss
function ($k=4.685$).

\begin{table}[htbp]
  \centering
  \caption{Simulation results for $p = 25$. The summary statistics reported are:
    bias divided by $\beta_0$, root-median-square error (RMSE)
    divided by $\beta_0$, length of the confidence interval (CI)
    divided by $\beta_0$, and the coverage rate of the CI (nominal rate is
    95\%), all in \%.}
  \label{tab:sim1}
  \begin{tabular}{ll|rrrr}
    \toprule
    Setup & Method & Bias \% & RMSE \% & CI Len.\ \% &
                                                       \multicolumn{1}{c}{Cover.\ \%} \\
    \midrule
    1 & IVW  & $\phantom{00}-2.9$ & $\phantom{0}12.7$ & $\phantom{0}73.8$ & $95.4$ \\
          & Egger  & $\phantom{00}-7.4$ & $\phantom{0}24.4$ & $142.3$ & $95.3$ \\
          & W. Median  & $\phantom{00}-5.2$ & $\phantom{0}17.0$ & $105.5$ & $96.5$ \\
          & PS  & $\phantom{00}-0.1$ & $\phantom{0}12.7$ & $\phantom{0}74.9$ & $95.1$ \\
          & APS  & $\phantom{00}-0.4$ & $\phantom{0}12.7$ & $\phantom{0}76.8$ & $96.0$ \\
          & RAPS  & $\phantom{00}-0.4$ & $\phantom{0}13.0$ & $\phantom{0}79.0$
                                                     & $96.1$ \\
    \midrule
    2 & IVW  & $\phantom{00}-3.0$ & $\phantom{0}29.3$ & $167.9$ & $93.3$ \\
          & Egger  & $\phantom{00}-8.2$ & $\phantom{0}59.7$ & $319.2$ & $92.1$ \\
          & W. Median  & $\phantom{0}-12.8$ & $\phantom{0}39.9$ & $121.4$ & $70.6$ \\
          & PS  & $\phantom{0}\phantom{-}14.7$ & $\phantom{0}36.1$ & $\phantom{0}71.4$ & $49.2$ \\
          & APS  & $\phantom{00}-0.2$ & $\phantom{0}28.8$ & $165.4$ & $93.4$ \\
          & RAPS  & $\phantom{00}-0.1$ & $\phantom{0}30.1$ & $170.2$ & $93.1$
    \\
    \midrule
    3 & IVW  & $-115.5$ & $115.2$ & $225.6$ & $48.1$ \\
          & Egger  & $-264.2$ & $262.8$ & $409.1$ & $25.5$ \\
          & W. Median  & $\phantom{0}-80.7$ & $\phantom{0}79.5$ & $151.4$ & $47.3$ \\
          & PS  & $-122.3$ & $121.3$ & $\phantom{0}66.1$ & $\phantom{0}6.9$ \\
          & APS  & $\phantom{0}-86.2$ & $\phantom{0}85.6$ & $207.0$ & $65.0$ \\
          & RAPS  & $\phantom{0}-11.6$ & $\phantom{0}40.6$ & $168.7$ & $84.3$
    \\
    \midrule
    4 & IVW  & $\phantom{00}-5.1$ & $\phantom{0}25.1$ & $159.5$ & $96.0$ \\
          & Egger  & $\phantom{0}-54.5$ & $\phantom{0}58.8$ & $300.9$ & $90.0$ \\
          & W. Median  & $\phantom{0}-22.5$ & $\phantom{0}26.0$ & $113.2$ & $83.8$ \\
          & PS  & $\phantom{0}\phantom{-}13.4$ & $\phantom{0}31.2$ & $\phantom{0}71.7$ & $55.9$ \\
          & APS  & $\phantom{00}\phantom{-}4.0$ & $\phantom{0}25.6$ & $158.4$ & $96.1$ \\
          & RAPS  & $\phantom{00}\phantom{-}2.6$ & $\phantom{0}20.3$ & $117.5$
                                                     & $93.3$ \\
    \midrule
    5 & IVW  & $\phantom{00}-2.4$ & $\phantom{0}48.2$ & $169.7$ & $76.3$ \\
          & Egger  & $\phantom{00}-8.2$ & $\phantom{0}98.0$ & $321.0$ & $72.9$ \\
          & W. Median  & $\phantom{0}-24.4$ & $\phantom{0}60.4$ & $136.7$ & $56.0$ \\
          & PS  & $\phantom{0}\phantom{-}15.8$ & $\phantom{0}57.2$ & $\phantom{0}71.6$ & $33.0$ \\
          & APS  & $\phantom{00}\phantom{-}0.9$ & $\phantom{0}46.8$ & $183.0$ & $81.1$ \\
          & RAPS  & $\phantom{00}\phantom{-}1.5$ & $\phantom{0}44.9$ & $169.0$
                                                     & $78.3$ \\
    \midrule
    6 & IVW  & $\phantom{00}-8.1$ & $\phantom{0}64.2$ & $382.8$ & $94.8$ \\
          & Egger  & $-102.2$ & $134.8$ & $723.7$ & $90.7$ \\
          & W. Median  & $\phantom{0}-30.8$ & $\phantom{0}50.3$ & $130.6$ & $63.1$ \\
          & PS  & $\phantom{-}200.2$ & $309.6$ & $\phantom{0}82.1$ & $\phantom{0}4.1$ \\
          & APS  & $\phantom{0}\phantom{-}13.7$ & $\phantom{0}62.1$ & $327.1$ & $92.8$ \\
          & RAPS  & $\phantom{0}\phantom{-}12.3$ & $\phantom{0}50.3$ & $298.2$ & $85.4$ \\
    \bottomrule
  \end{tabular}
\end{table}

\begin{table}[htbp]
  \centering
  \caption{Simulation results for $p = 160$. The summary statistics reported are:
    bias divided by $\beta_0$, root-median-square error (RMSE)
    divided by $\beta_0$, length of the confidence interval (CI)
    divided by $\beta_0$, and the coverage rate of the CI (nominal rate is
    95\%), all in \%.}
  \label{tab:sim2}
  \begin{tabular}{ll|rrrr}
    \toprule
    Setup & Method & Bias \% & RMSE \% & CI Len.\ \% &
                                                       \multicolumn{1}{c}{Cover.\ \%} \\
    \midrule
    1 & IVW  & $\phantom{000}-11.1$ & $\phantom{00}12.2$ & $\phantom{00}51.0$ & $87.0$ \\
          & Egger  & $\phantom{000}-10.1$ & $\phantom{00}15.2$ & $\phantom{00}79.9$ & $92.6$ \\
          & W. Median  & $\phantom{000}-12.6$ & $\phantom{00}15.6$ & $\phantom{00}84.3$ & $93.9$ \\
          & PS  & $\phantom{0000}\phantom{-}0.1$ & $\phantom{000}9.6$ & $\phantom{00}57.0$ & $95.2$ \\
          & APS  & $\phantom{0000}-0.4$ & $\phantom{000}9.5$ & $\phantom{00}58.3$ & $95.8$ \\
          & RAPS  & $\phantom{0000}-0.5$ & $\phantom{000}9.8$ &
                                                                $\phantom{00}59.9$ & $95.8$ \\
    \midrule
    2 & IVW  & $\phantom{000}-11.6$ & $\phantom{00}23.2$ & $\phantom{0}122.5$ & $92.6$ \\
          & Egger  & $\phantom{000}-10.8$ & $\phantom{00}34.9$ & $\phantom{0}191.5$ & $93.6$ \\
          & W. Median  & $\phantom{000}-25.7$ & $\phantom{00}34.3$ & $\phantom{0}105.5$ & $68.9$ \\
          & PS  & $\phantom{00}\phantom{-}119.2$ & $\phantom{0}119.8$ & $\phantom{00}51.0$ & $\phantom{0}6.2$ \\
          & APS  & $\phantom{0000}-0.4$ & $\phantom{00}23.0$ & $\phantom{0}134.8$ & $95.1$ \\
          & RAPS  & $\phantom{0000}-0.4$ & $\phantom{00}23.8$ &
                                                                $\phantom{0}138.7$ & $95.1$ \\
    \midrule
    3 & IVW  & $\phantom{000}-70.1$ & $\phantom{00}69.9$ & $\phantom{0}131.3$ & $44.7$ \\
          & Egger  & $\phantom{00}-125.5$ & $\phantom{0}125.6$ & $\phantom{0}203.8$ & $32.3$ \\
          & W. Median  & $\phantom{000}-65.0$ & $\phantom{00}65.0$ & $\phantom{0}111.5$ & $41.5$ \\
          & PS  & $\phantom{0000}\phantom{-}4.1$ & $\phantom{00}77.9$ & $\phantom{00}44.6$ & $15.5$ \\
          & APS  & $\phantom{000}-47.9$ & $\phantom{00}48.3$ & $\phantom{0}139.3$ & $73.2$ \\
          & RAPS  & $\phantom{0000}-3.9$ & $\phantom{00}27.4$ &
                                                                $\phantom{0}137.9$ & $90.6$ \\
    \midrule
    4 & IVW  & $\phantom{000}-11.9$ & $\phantom{00}20.5$ & $\phantom{0}121.5$ & $94.7$ \\
          & Egger  & $\phantom{000}-13.6$ & $\phantom{00}31.5$ & $\phantom{0}189.5$ & $94.7$ \\
          & W. Median  & $\phantom{000}-24.1$ & $\phantom{00}24.8$ & $\phantom{00}93.9$ & $80.2$ \\
          & PS  & $\phantom{00}\phantom{-}134.7$ & $\phantom{0}114.3$ & $\phantom{00}51.4$ & $\phantom{0}7.1$ \\
          & APS  & $\phantom{0000}\phantom{-}4.8$ & $\phantom{00}20.8$ & $\phantom{0}133.6$ & $96.5$ \\
          & RAPS  & $\phantom{0000}\phantom{-}4.3$ & $\phantom{00}16.1$ &
                                                                          $\phantom{00}91.3$ & $93.6$ \\
    \midrule
    5 & IVW  & $\phantom{000}-11.0$ & $\phantom{00}53.9$ & $\phantom{0}139.7$ & $62.2$ \\
          & Egger  & $\phantom{0000}-9.8$ & $\phantom{00}92.5$ & $\phantom{0}217.7$ & $56.9$ \\
          & W. Median  & $\phantom{000}-56.0$ & $\phantom{00}63.7$ & $\phantom{0}125.2$ & $49.3$ \\
          & PS  & $\phantom{00}-819.8$ & $\phantom{0}244.0$ & $\phantom{00}57.8$ & $\phantom{0}4.7$ \\
          & APS  & $\phantom{0000}-0.3$ & $\phantom{00}55.3$ & $\phantom{0}170.7$ & $71.6$ \\
          & RAPS  & $\phantom{0000}\phantom{-}1.5$ & $\phantom{00}48.6$ &
                                                                          $\phantom{0}120.4$ & $59.8$ \\
    \midrule
    6 & IVW  & $\phantom{000}-12.7$ & $\phantom{00}47.2$ & $\phantom{0}278.8$ & $95.0$ \\
          & Egger  & $\phantom{000}-16.4$ & $\phantom{00}74.2$ & $\phantom{0}435.3$ & $94.9$ \\
          & W. Median  & $\phantom{000}-34.9$ & $\phantom{00}43.6$ & $\phantom{0}115.2$ & $63.1$ \\
          & PS  & $\phantom{-}>999.9$ & $>999.9$ & $>999.9$ & $12.8$ \\
          & APS  & $\phantom{000}\phantom{-}13.6$ & $\phantom{00}50.2$ & $\phantom{0}291.2$ & $95.2$ \\
          & RAPS  & $\phantom{000}\phantom{-}10.8$ & $\phantom{00}42.7$ & $\phantom{0}258.4$ & $91.2$ \\
    \bottomrule
  \end{tabular}
\end{table}

The simulation results are reported in \Cref{tab:sim1} for $p = 25$
and \Cref{tab:sim2} for $p = 160$. Here is a summary of the results:
\begin{enumerate}
\item  In setup 1, the PS estimator has the smallest root-median square error
  (RMSE) and the shortest confidence interval (CI)
  with the desired coverage rate. The IVW estimator performs very well
  when $p = 25$ but has considerable bias and less than nominal coverage
  when $p = 160$. The APS and RAPS estimators have slightly longer CI
  than PS. The MR-Egger and weighted median estimators are less accurate
  than the other methods.
\item In setup 2, the PS estimator, as well as the weighted
  median, have substantial bias and perform poorly. The APS estimator
  is overall the best with very small bias and desired coverage, followed very closely by RAPS. The IVW and MR-Egger
  estimators also perform quite well, though their relative biases are
  more than 10\% when $p = 160$.
\item In setup 3, all estimators besides RAPS have very large
  bias and poor CI coverage. The RMSE of the RAPS estimator is
  slightly larger than the RMSE in \Cref{model:2}, and the coverage of
  RAPS is slightly below the nominal rate.
\item In setup 4, the direct effects $\alpha_j$ are distributed as
  Laplace instead of normal. The RAPS estimator
  has the smallest bias and RMSE, though the coverage is slightly
  below the nominal level.
\item In setup 5, the variance of $\alpha_j$ is proportional to
  $|\gamma_j|$. In this case APS and RAPS are approximately unbiased but the
  coverage is significantly lower than 95\%.
\item In setup 6, 10\% of the IVs have very large but roughly balanced
  pleiotropy effects $\alpha_j$. All estimators are biased in this case. The RAPS
  estimator has the smallest RMSE but the CI coverage is slightly
  below 95\%. The IVW and APS estimators have slightly larger RMSE and
  the CI has the desired coverage rate.
\end{enumerate}

\revise{
Finally, we briefly remark on the bias of IVW and other existing
estimators. In \ref{sec:weak-iv-bias} we have derived that the IVW
estimator is biased towards $0$ and the relative bias is approximately
$1/\kappa$. The average instrument strength $\kappa$ in the two
settings are $\kappa = 33.1$ ($p = 25$) and $\kappa = 9.1$
($p=160$). The simulation results for setup 1 in
\Cref{tab:sim1,tab:sim2} almost exactly match the prediction from our
approximation formula \eqref{eq:ivw-bias-approx}.
}

Overall, the RAPS estimator is the clear winner in this simulation: when there is no
idiosyncratic outlier (setups 1 and 2), it behaves almost as well as the best
performer; when there is an idiosyncratic outlier (setup 3), it
still has very small bias and close-to-nominal coverage; when our
modeling assumptions are not satisfied (setups 4, 5, 6), it still has the
smallest bias and RMSE, though the CI may fail to cover $\beta_0$ at
the nominal rate.

\subsection{\revise{Simulating from real genotypes}}
\label{sec:simulating-from-real}

\revise{
As pointed out by an anonymous reviewer, the marginal GWAS
coefficients might not perfectly follow the distributional assumptions
in \Cref{assump:setup}. In fact, in \Cref{sec:line-struct-model} we
already showed that even in linear structural models the marginal
coefficients have small but non-zero covariances. As a proof of
concept, we perform another simulation study using real genotypes from
the 1000 Genomes Project \citep{10002015global}.
}

\revise{
In total, the 1000 Genomes Project phase 1 dataset contains the
genotypes of 1092 individuals. We simulated the exposure $X$ and
outcome $Y$ according to the linear structural equation model
\eqref{eq:linear-sem-alpha} using the entire 10th chromosome as
$Z$ (containing $1,882,663$ genetic variants). $100$ random entries of $\bm
\gamma$ are set to be non-zero and follow the Laplace distribution
with rate $1$. The unmeasured confounder $U$ is simulated from the
standard normal distribution and the parameters were set to $\eta_X
= 3, \eta_Y = 5$. The noise variables were simulated from $E_X \sim
\mathrm{N}(0,3^2)$ and $E_Y \sim \mathrm{N}(0,5^2)$. The direct
effects $\bm \alpha$ had $p_{\alpha}$ random non-zero entries that were
simulated from the Laplace distribution with rate $r_{\alpha}$. In
total we considered five settings:
\begin{enumerate}
\item $\beta = 0$, $p_{\alpha} = 0$;
\item $\beta = 0$, $p_{\alpha} = 200$, $r_{\alpha} = 0.5$;
\item $\beta = 1$, $p_{\alpha} = 0$;
\item $\beta = 1$, $p_{\alpha} = 200$, $r_{\alpha} = 0.5$;
\item $\beta = 1$, $p_{\alpha} = 200$, $r_{\alpha} = 1.5$;
% \item $\beta = 1$, $p_{\alpha} = 1000$, $r_{\alpha} = 0.5$;
\end{enumerate}
}

\revise{
In this dataset, $368,977$ variants have minor allele frequency
greater $5\%$ and are considered as potential instrumental variables.
We used $292$, $400$ and $400$ individuals (random partition) as the
selection, exposure and outcome data and obtained GWAS summary data by
running marginal linear regressions. We simulated $Y$ using one of the
five settings described above. After LD clumping ($p$-value $\le
5 \times 10^{-3}$), $121$ independent variants were selected as
IVs, and we applied existing and our methods to these $121$ SNPs. To
provide a more comprehensive comparison, we also applied two classical
IV estimator, two-stage least squares (2SLS) and limited information
maximum likelihood (LIML), to the outcome sample of $400$
individuals. For the LIML estimator we computed the standard error
using the ``many weak IV asymptotics''
\citep{hansen2008estimation}. Note that 2SLS and LIML
cannot be computed using just the GWAS summary data and they assume
all the IVs are valid.
}

\revise{
We used $5,000$ replications to
obtain the same performance metrics in
\Cref{sec:simulating-from-cref}, which are reported in
\Cref{tab:sim-real}. Overall, our estimators (in particular, APS and
RAPS) are unbiased and maintain the nominal CI coverage rate in all 5
settings. The three existing estimators---IVW, MR-Egger, and weighted
median---are heavily biased towards $0$ when $\beta \ne 0$. Also,
notice that their RMSE and CI length are (abnormally) smaller than the
RMSE and CI length of the ``oracle'' LIML estimator that uses individual
genotypes. The 2SLS estimator is heavily biased due to weak instruments.
}

\revise{
Although the simulation results in \Cref{tab:sim-real} are
encouraging, we want to point out that the sample size and simulation
parameters we used might be quite different from actual MR
studies. The pleiotropy models (parametrized by $p_{\alpha}$ and
$r_{\alpha}$) being tested here are also limited. Nonetheless, this
simulation shows that using the statistical framework developed in
this paper, it is possible to obtain summary-data MR estimators that
perform almost as well as the ``oracle'' LIML estimator that uses
individual data.
}

\begin{table}
\caption{\revise{Results for the numerical simulation using real
  genotypes. The performance metrics reported are: bias (median
  $\hat{\beta}$ minus $\beta$), root-median-square error (RMSE),
  median length of the confidence interval (CI), and the coverage rate
  of the CI (nominal rate is 95\%).}}
\label{tab:sim-real}
\color{black}{
\begin{tabular}{ll|cccc}
\toprule
Setup & Method & Bias & RMSE & CI Len. & Coverage \% \\
\midrule
1 & IVW  & $\phantom{-}0.00$ & $0.08$ & $0.42$ & $93.1$ \\
 & Egger  & $\phantom{-}0.00$ & $0.11$ & $0.62$ & $95.1$ \\
 & W.\ Median  & $\phantom{-}0.00$ & $0.12$ & $0.74$ & $96.8$ \\
 & PS  & $\phantom{-}0.01$ & $0.26$ & $1.42$ & $92.9$ \\
 & APS  & $\phantom{-}0.01$ & $0.23$ & $1.61$ & $98.9$ \\
 & RAPS  & $\phantom{-}0.00$ & $0.23$ & $1.76$ & $98.2$ \\
 & 2SLS  & $-0.46$ & $0.46$ & $0.41$ & $\phantom{0}0.9$ \\
 & LIML  & $\phantom{-}0.00$ & $0.26$ & $1.40$ & $94.5$ \\
\midrule
2 & IVW  & $-0.02$ & $0.08$ & $0.45$ & $94.0$ \\
 & Egger  & $-0.02$ & $0.11$ & $0.65$ & $95.4$ \\
 & W.\ Median  & $-0.04$ & $0.12$ & $0.78$ & $97.2$ \\
 & PS  & $-0.06$ & $0.29$ & $1.42$ & $89.2$ \\
 & APS  & $-0.05$ & $0.25$ & $1.67$ & $98.6$ \\
 & RAPS  & $-0.05$ & $0.25$ & $1.82$ & $97.5$ \\
 & 2SLS  & $-0.47$ & $0.47$ & $0.43$ & $\phantom{0}1.1$ \\
 & LIML  & $\phantom{-}0.02$ & $0.28$ & $1.56$ & $95.4$ \\
\midrule
3 & IVW  & $-0.63$ & $0.63$ & $0.43$ & $\phantom{0}0.1$ \\
 & Egger  & $-0.45$ & $0.45$ & $0.61$ & $21.1$ \\
 & W.\ Median  & $-0.64$ & $0.64$ & $0.76$ & $\phantom{0}8.7$ \\
 & PS  & $\phantom{-}0.08$ & $0.22$ & $1.35$ & $96.9$ \\
 & APS  & $\phantom{-}0.02$ & $0.22$ & $1.78$ & $97.6$ \\
 & RAPS  & $\phantom{-}0.01$ & $0.22$ & $1.87$ & $93.1$ \\
 & 2SLS  & $-0.46$ & $0.46$ & $0.41$ & $\phantom{0}1.2$ \\
 & LIML  & $-0.01$ & $0.26$ & $1.41$ & $94.8$ \\
\midrule
4 & IVW  & $-0.65$ & $0.65$ & $0.46$ & $\phantom{0}0.2$ \\
 & Egger  & $-0.47$ & $0.47$ & $0.65$ & $22.4$ \\
 & W.\ Median  & $-0.61$ & $0.61$ & $0.79$ & $13.6$ \\
 & PS  & $\phantom{-}0.13$ & $0.26$ & $1.39$ & $95.1$ \\
 & APS  & $\phantom{-}0.01$ & $0.25$ & $1.86$ & $96.6$ \\
 & RAPS  & $-0.01$ & $0.24$ & $1.95$ & $92.2$ \\
 & 2SLS  & $-0.46$ & $0.46$ & $0.43$ & $\phantom{0}1.4$ \\
 & LIML  & $\phantom{-}0.03$ & $0.28$ & $1.57$ & $95.4$ \\
\midrule
5 & IVW  & $-0.68$ & $0.68$ & $0.62$ & $\phantom{0}0.9$ \\
 & Egger  & $-0.50$ & $0.50$ & $0.90$ & $40.4$ \\
 & W.\ Median  & $-0.44$ & $0.44$ & $0.97$ & $57.0$ \\
 & PS  & $\phantom{-}0.41$ & $0.49$ & $1.72$ & $87.3$ \\
 & APS  & $\phantom{-}0.01$ & $0.37$ & $2.40$ & $96.8$ \\
 & RAPS  & $-0.04$ & $0.33$ & $2.48$ & $94.8$ \\
 & 2SLS  & $-0.47$ & $0.47$ & $0.57$ & $10.4$ \\
 & LIML  & $\phantom{-}0.23$ & $0.51$ & $2.93$ & $97.9$ \\
% 6 & IVW  & $-0.68$ & $0.68$ & $0.54$ & $\phantom{0}0.2$ \\
%  & Egger  & $-0.51$ & $0.51$ & $0.79$ & $26.0$ \\
%  & W.\ Median  & $-0.51$ & $0.51$ & $0.89$ & $37.6$ \\
%  & PS  & $\phantom{-}0.15$ & $0.30$ & $1.61$ & $95.2$ \\
%  & APS  & $-0.01$ & $0.29$ & $2.13$ & $97.1$ \\
%  & RAPS  & $-0.03$ & $0.28$ & $2.25$ & $95.5$ \\
%  & 2SLS  & $-0.50$ & $0.50$ & $0.52$ & $\phantom{0}3.3$ \\
%  & LIML  & $\phantom{-}0.07$ & $0.38$ & $2.23$ & $97.2$ \\
\bottomrule
\end{tabular}
}
\end{table}

\section{Comparison in real data examples}
\label{sec:real-data-comparison}

\subsection{\revise{In the BMI-SBP example}}
\label{sec:bmi-sbp-example}

\revise{
\Cref{tab:summary-bmi-sbp} briefly summarize the results using
different estimators in this and previous papers for the
BMI-SBP example introduced in \Cref{sec:an-example-bmi}. Since the
ground truth is unknown, we do not know which estimate is closer to
the truth. Nevertheless, we can still make two observations. First,
the point estimates varied considerably between the methods, so the
choice of estimator may make a difference in practice. Second,
the PS, IVW, and MR-Egger point estimates changed substantially
when all $160$ SNPs were used instead of just the $25$ strongest ones,
whereas the RAPS estimators and the weighted median were more stable.
}

\begin{table}[t]
  \centering
  \caption{\revise{Comparison of results in the BMI-SBP example.}}
  \label{tab:summary-bmi-sbp} \color{black}
  \begin{tabular}{l|cc|cc}
    \toprule
    \multirow{2}{*}{Method} & \multicolumn{2}{c|}{$p=25$} &
    \multicolumn{2}{c}{$p=160$} \\
    & $\hat{\beta}$ & SE & $\hat{\beta}$ & SE \\
    \midrule
    PS & 0.367 & 0.075 & 0.601 & 0.054 \\
    APS & 0.364 & 0.133 & 0.301 & 0.158 \\
    RAPS (Huber) & 0.354 & 0.131 & 0.378 & 0.121 \\
    RAPS (Tukey) & 0.361 & 0.133 & 0.402 & 0.106 \\
    \midrule
    IVW & 0.332 & 0.140 & 0.514 & 0.102 \\
    MR-Egger & 0.647 & 0.283 & 0.472 & 0.176 \\
    Weighted median & 0.516 & 0.125 & 0.514 & 0.102 \\
    \bottomrule
  \end{tabular}
\end{table}

\subsection{\revise{An illustration of weak IV bias and selection bias}}
\label{sec:real-data-revis}

Finally, we consider another real data validation example, which shall
be referred to as the BMI-BMI example. In this example, both the
``exposure'' and the ``outcome'' are BMI. Although there is no
``causal'' effect of BMI on itself,
\Cref{model:1} for GWAS summary data should technically hold with
$\beta_0 = 1$. Therefore, this is a rare scenario where we know the
truth in real data. Since there are many SNPs that are only
weakly associated with BMI, this example also offers a good opportunity to
probe the issue of weak instrument bias and the efficiency gain by
including many weak IVs. \revise{The downside is that this example
  does not test the methods' robustness to pleiotropy because the
  exposure and outcome are the same trait.}

We obtained three GWAS datasets for this example:
\begin{description}
\item[\texttt{BMI-GIANT}:] full dataset from the GIANT consortium
  \citep{locke2015genetic} (i.e.\ combining \texttt{BMI-FEM} and
  \texttt{BMI-MAL}), used to select SNPs.
\item[\texttt{BMI-UKBB-1}:] half of the UKBB data, used as the ``exposure''.
\item[\texttt{BMI-UKBB-2}:] another half of UKBB data, used as the ``outcome''.
\end{description}

We applied in total six methods. Four have been previously developed:
besides the three estimators considered in \Cref{sec:simulation}, we
also included the weighted mode estimator of
\citet{hartwig2017robust}. We use the implementation in the
\texttt{TwoSampleMR} software package \citep{hemani2016mr} for the existing
methods. The last two methods were the PS and RAPS estimators
developed in this
paper (APS performs similarly to PS and RAPS and is omitted).

The results are reported in \Cref{tab:bmi-bmi}. Overall, the PS and RAPS
estimators provided very accurate estimate of $\beta_0 = 1$. PS has
the smallest standard error because there is no pleiotropy at all in
this example. When there is pleiotropy (as expected in most real
studies), PS can perform poorly as demonstrated in
\Cref{sec:simulation}. All the existing methods are biased especially
when there are many weak IVs.

In \Cref{tab:bmi-bmi-2} we illustrate the danger of selection bias. In
this example we discard the \texttt{BMI-GIANT} dataset and use
\texttt{BMI-UKBB-1} for both selection and inference (estimating
$\gamma_j$). The estimators are biased towards $0$ in almost all cases,
even if we only use the genome-wide significant $p$-value
threshold $10^{-9}$ or
$10^{-8}$. This is because the assumption $\hat{\gamma}_j \sim
\mathrm{N}(\gamma_j,\sigma_{Xj}^2)$ is violated. In fact, due to
selection bias, the selected $\hat{\gamma}_j$ are stochastically
larger than their mean $\gamma_j$ (if $\gamma_j > 0$). Compared with
other methods, the MR-Egger regression seems to be less affected by
the selection bias.

\begin{table}[t]
  \caption{Results of the BMI-BMI example. The true $\beta_0$ should
    be $1$. We considered $8$ selection thresholds $p_{\mathrm{sel}}$
    from $1 \times
    10^{-9}$ to $1 \times 10^{-2}$. The mean and median of the
    $F$-statistics $\hat{\gamma}_j^2/\sigma_{Xj}^2$ are reported. In
    each setting, we report the point estimate and the standard error
    of all the methods.}
  \label{tab:bmi-bmi}
  \begin{tabular}{lcc|ccc}
    \toprule
    $p_{\mathrm{sel}}$ & \# SNPs & Mean $F$ & IVW & W.\ Median &
                                                                 \multicolumn{1}{c}{W.\ Mode} \\
    \midrule
    1e-9  & $\phantom{0}48$ & $78.6$ & 0.983 (0.026) & 0.945 (0.039) & 0.941 (0.042) \\
    1e-8  & $\phantom{0}58$ & $69.2$ & 0.983 (0.024) & 0.945 (0.039) & 0.939 (0.044) \\
    1e-7  & $\phantom{0}84$ & $55.0$ & 0.988 (0.024) & 0.945 (0.036) & 0.933 (0.041) \\
    1e-6  & $126$ & $44.1$ & 0.986 (0.022) & 0.944 (0.034) & 0.931 (0.038) \\
    1e-5  & $186$ & $34.3$ & 0.986 (0.019) & 0.943 (0.033) & 0.928 (0.039) \\
    1e-4  & $287$ & $26.1$ & 0.981 (0.017) & 0.941 (0.031) & 0.929 (0.035) \\
    1e-3  & $474$ & $18.8$ & 0.955 (0.015) & 0.903 (0.027) & 0.917 (0.231) \\
    1e-2  & $812$ & $12.7$ & 0.928 (0.014) & 0.879 (0.023) & 0.739 (7.130) \\
    \midrule
    $p_{\mathrm{sel}}$ & \# SNPs & Median $F$ & Egger & PS &
                                                             \multicolumn{1}{c}{RAPS} \\
    \midrule
    1e-9  & $\phantom{0}48$ & $51.8$ & 0.926 (0.055) & 0.999 (0.023) & 0.998 (0.026) \\
    1e-8  & $\phantom{0}58$ & $42.0$ & 0.928 (0.050) & 0.999 (0.023) & 0.998 (0.025) \\
    1e-7  & $\phantom{0}84$ & $32.1$ & 0.905 (0.048) & 1.012 (0.021) & 1.004 (0.025) \\
    1e-6  & $126$ & $27.4$ & 0.881 (0.043) & 1.017 (0.019) & 1.009 (0.023) \\
    1e-5  & $186$ & $21.0$ & 0.874 (0.036) & 1.020 (0.018) & 1.013 (0.020) \\
    1e-4  & $287$ & $15.8$ & 0.921 (0.031) & 1.023 (0.017) & 1.018 (0.018) \\
    1e-3  & $474$ & $10.8$ & 0.913 (0.027) & 1.010 (0.016) & 1.006 (0.016) \\
    1e-2  & $812$ & $5.6$ & 0.909 (0.022) & 1.010 (0.015) & 1.005 (0.015) \\
    \bottomrule
  \end{tabular}
\end{table}

\begin{table}[t]
  \caption{Illustration of selection bias. The same
    \texttt{BMI-UKBB-1} dataset is used for both selecting SNPs and
    estimating the SNP-exposure effects $\gamma_j$. All estimators are
    biased (true $\beta_0 = 1$) due to not accounting for selection bias.}
  \label{tab:bmi-bmi-2}
  \begin{tabular}{lcc|ccc}
    \toprule
    $p_{\mathrm{sel}}$ & \# SNPs & Mean $F$ & IVW & W.\ Median &
                                                                 \multicolumn{1}{c}{W.\ Mode} \\
    \midrule
    1e-9  & $\phantom{0}110$ & $68.63$ & 0.851 (0.02) & 0.83 (0.025) & 0.896 (0.046) \\
    1e-8  & $\phantom{0}168$ & $57.00$ & 0.823 (0.017) & 0.8 (0.022) & 0.885 (0.053) \\
    1e-7  & $\phantom{0}228$ & $50.08$ & 0.799 (0.016) & 0.768 (0.019) & 0.886 (0.058) \\
    1e-6  & $\phantom{0}305$ & $43.92$ & 0.761 (0.015) & 0.736 (0.019) & 0.865 (0.079) \\
    1e-5  & $\phantom{0}443$ & $36.98$ & 0.721 (0.013) & 0.667 (0.016) & 0.824 (0.12) \\
    1e-4  & $\phantom{0}652$ & $30.68$ & 0.678 (0.012) & 0.616 (0.015) & 0.593 (0.122) \\
    1e-3  & $\phantom{0}929$ & $25.36$ & 0.629 (0.011) & 0.57 (0.014) & 0.576 (0.096) \\
    1e-2  & $1289$ & $20.70$ & 0.592 (0.01) & 0.528 (0.013) & 0.554 (0.093) \\
    \midrule
    $p_{\mathrm{sel}}$ & \# SNPs & Median $F$ & Egger & PS &
                                                             \multicolumn{1}{c}{RAPS} \\
    \midrule
    1e-9  & $\phantom{0}110$ & $49.20$ & 1.071 (0.051) & 0.871 (0.015) & 0.862 (0.021) \\
    1e-8  & $\phantom{0}168$ & $41.12$ & 1.018 (0.046) & 0.848 (0.014) & 0.831 (0.018) \\
    1e-7  & $\phantom{0}228$ & $37.12$ & 1.016 (0.043) & 0.824 (0.012) & 0.803 (0.016) \\
    1e-6  & $\phantom{0}305$ & $33.68$ & 1.006 (0.041) & 0.793 (0.011) & 0.763 (0.016) \\
    1e-5  & $\phantom{0}443$ & $28.74$ & 0.957 (0.037) & 0.762 (0.01) & 0.716 (0.015) \\
    1e-4  & $\phantom{0}652$ & $23.23$ & 0.89 (0.033) & 0.724 (0.009) & 0.66 (0.014) \\
    1e-3  & $\phantom{0}929$ & $19.12$ & 0.823 (0.03) & 0.687 (0.008) & 0.594 (0.013) \\
    1e-2  & $1289$ & $15.26$ & 0.749 (0.025) & 0.657 (0.008) & 0.541 (0.012) \\
    \bottomrule
  \end{tabular}
\end{table}

\section{Discussion}
\label{sec:discussion}

In this paper we have proposed a systematic approach for two-sample
summary-data Mendelian randomization based on modifying the profile
score function. By considering increasingly more complex models, we
arrived at the Robust Adjusted Profile Score (RAPS) estimator which is
robust to both systematic and idiosyncratic pleiotropy and performed
excellently in all the numerical examples. Thus we recommend to
routinely use the RAPS estimator in practice, especially if the exposure
and the outcome are both complex traits.

Our theoretical and empirical results advocate for a new design of
two-sample MR. Instead of using just a few strong SNPs (those with
large $|\hat{\gamma_j}|/\sigma_{Xj}$), we find that adding many
(potentially hundreds of) weak SNPs usually substantially decreases the
variance of the estimator. This is not feasible with existing methods
for MR because they usually require the instruments to be
strong. An additional advantage of using many weak instruments is that
outliers in the sense of \Cref{model:3} are
more easily detected, so the results are generally more robust
to pleiotropy. There is one caveat: selection bias is
more significant for weaker instruments, so a sample-splitting design (such as
the one in \Cref{sec:an-example-bmi}) should be used.

In \Cref{model:2,model:3}, we have assumed that the pleiotropy effects
are completely independent and normally or nearly normally
distributed. We view this assumption as
an approximate modeling assumption rather than the precise data
generating mechanism. It is motivated by the real data
(\Cref{sec:example-continued}) and seems to fit the data very well (\Cref{sec:example-continued-2}). It is a special instance of the INstrument Strength
Independent of Direct Effect (INSIDE) assumption
\citep{bowden2017framework} that is common in the summary-data MR
literature. Apart from normality, two other implicit but key
assumptions we made are:
\begin{enumerate}
\item The pleiotropy effects $\alpha_j$ are additive rather than
  multiplicative (the variance of $\alpha_j$ is proportional to
  $\sigma_{Yj}$) \citep{bowden2015mendelian}. Multiplicative random
  effects model are easier to fit especially if the measurement error
  in $\hat{\gamma}_j$ is ignored, however it is quite unrealistic
  because $\alpha_j$ is a population quantity and thus is unlikely to be
  dependent on a sample quantity (for example, $\sigma_{Yj}$ may vary
  due to missing data). In contrast, the additive model is well motivated by the
  linear structural model in \ref{sec:viol-excl-restr}.
\item The pleiotropy effects $\alpha_j$ have mean $0$. In comparison,
  the MR-Egger regression \citep{bowden2015mendelian} assumes
  $\alpha_j$ has an unknown mean $\mu$ and refers to the case $\mu \ne
  0$ as ``directional pleiotropy''. We have not seen strong evidence of
  ``directional pleiotropy'' in real datasets, and, more importantly,
  assuming $\mu \ne 0$ implies that there is a ``special'' allele
  coding so that $\alpha_j \sim \mathrm{N}(\mu,\tau^2)$. It is thus
  impossible to obtain estimators of $\beta$ that are invariant to
  allele recoding without completely reformulating the MR-Egger
  model. For further details see \citet{bowden2017improving2}.
\end{enumerate}

There are many technical challenges in the development of this
paper. Due to the nature of the many weak IV problem, the asymptotics
we considered are quite different from the classical measurement error
literature. In \Cref{sec:no-plei-prof} we showed the profile likelihood
is information biased when there are many weak IVs, and in
\Cref{sec:fail-prof-likel} we showed the profile likelihood is biased
when there is overdispersion caused by systematic pleiotropy. This
issue is solved by adjusting the profile score, but the proof of the
consistency of the APS estimator is nontrivial. Consistency of
the the RAPS estimator is even more
challenging and still open because the estimating equations may have
multiple roots, although we found its practical performance is usually
quite benign. A possible solution is to initialize by another robust
and consistent estimator (similar to the MM-estimation in robust
regression, see \citet{yohai1987high}). However, we are not aware of
any other provably robust and consistent estimator in our setting, and
deriving such estimator is beyond the scope of this paper.

\section*{Software and reproducibility}

\texttt{R} code for the methods proposed in this paper can be found in
the package \texttt{mr.raps} that is publicly available at
\url{https://github.com/qingyuanzhao/mr.raps} and can be directly
called from \texttt{TwoSampleMR}. Numerical
examples can be reproduced by running examples in the \texttt{R}
package.

% \begin{supplement}[id=suppA]
% \sname{Supplement to ``Statistical inference in two-sample summary-data Mendelian randomization using robust adjusted profile score''}\label{suppA}
% % \stitle{Supplement}
% \slink[url]{add link later}
% \sdatatype{.pdf}
% \sdescription{In this supplement we provide additional justifications
%   of the linear model for GWAS summary data and detailed proof for the
% theoretical results.}
% \end{supplement}

\bibliographystyle{imsart-nameyear}
\bibliography{ref}

\appendix

\section{Linear model for GWAS summary data}
\label{sec:linear-model-gwas}

In this Appendix we give additional justifications of the linear model
\eqref{eq:model-0} for GWAS summary data. We will show
\eqref{eq:model-0} is very likely to hold in very general situations, much
beyond the linear structural model considered in \Cref{sec:stat-model-gwas}.

\subsection{Binary outcome and logistic model}
\label{sec:binary-outc-logist}

When the outcome $Y$ is binary, the linear structural model
\eqref{eq:linear-sem} is no longer appropriate. Instead, we consider the
following logistic model of $Y$ (let $H(t) = 1 / (1 + e^{-t})$ be the
logistic link function):
\begin{equation} \label{eq:logistic-sem}
  X = \sum_{j=1}^p \gamma_j Z_j + \eta_X U +
  E_X,~Y \sim \mathrm{Bernoulli}\big(H(\beta X + \eta_Y
  U)\big).
\end{equation}

Next we derive an approximation of the coefficient $\Gamma_j$ when
we run a logistic regression of $Y$ on $Z_j$. By
\eqref{eq:logistic-sem}, we have
\[
\mathbb{P}(Y=1|Z_j=z_j) = \mathbb{E} \big[H\big(\beta\gamma_j z_j +
E' \big)\big],
\]
where $E' = \beta\sum_{k\ne j}\gamma_kZ_k + \beta \eta_X U +
\beta E_X + \eta_Y U$. If we assume $E'
\sim \mathrm{N}(\mu,\sigma_j^2)$, then
\[
\mathbb{P}(Y=1|Z_j=z_j) = \int_{-\infty}^{\infty} H\big(\mu + \beta\gamma_j z_j +
\sigma_j e\big) \phi(e) \diff e.
\]
Note that $\sigma_j^2 \approx \sigma^2 = \mathrm{Var}(\beta X + \eta_Y
U)$ when $\gamma_j$ is small.

To proceed further we introduce a well-known probit approximation of
logistic function \citep{carroll2006measurement}:
\[
H(t) \approx \Phi(t/1.7).
\]
By using the following Gaussian integral identity,
\[
\int_{-\infty}^{\infty} \Phi(a + bx) \phi(x) \diff x = \Phi\left( \frac{a}{\sqrt{1+b^2}} \right),
\]
we obtain
\[
\begin{split}
  \mathbb{P}(Y=1|Z_j=z_j) &\approx \int_{-\infty}^{\infty} \Phi\left(\frac{\mu + \beta\gamma_j z_j +\sigma e}{1.7}\right) \phi(e) \diff e \\
  &= \Phi \left( \frac{\mu + \beta \gamma_j z_j}{1.7 \sqrt{1 +
        (\sigma/1.7)^2}} \right) \\
  &\approx H \left( \frac{\mu + \beta \gamma_j z_j}{\sqrt{1 +
        (\sigma/1.7)^2}} \right).
\end{split}
\]
Therefore $\Gamma_j \approx \beta \gamma_j / \sqrt{1 +
  (\sigma/1.7)^2}$. In other words, model \eqref{eq:model-0} is
approximately correct with $\beta_0 = \beta / \sqrt{1 +
  (\sigma/1.7)^2}$. The attenuation bias is due to the
non-collapsibility of odds ratio
\citep{greenland1999confounding}. Notice that although we assumed $E'$
is normally distributed in our calculation, this approximation is
quite accurate for many other distributions
\citep[Section 4.8.2]{carroll2006measurement}. A similar result can be
found in \citet{vansteelandt2011instrumental} who also discussed the
general intepretation of causal odds ratios.

\subsection{General situation: a local argument}
\label{sec:mr-identifying-local}

The linear model \eqref{eq:model-0} may actually hold in much
broader situations than the linear and logistic models considered
above. The main reason is that for most SNPs, the influence on a
complex trait $X$ is usually minuscule
\citep{boyle2017expanded,ioannidis2006implications,park2010estimation,shi2016contrasting}. Let's consider a
continuous exposure $X$ and the quantity $\mathbb{E}[h(Y)|Z_j=1] -
\mathbb{E}[h(Y)|Z_j=0]$ for
some function $h$ of interest. Assuming appropriate differentiability
and using the shorthand notation $X(z_1) = g(z_1,Z_2,\dotsc,Z_p,U,E_X)$,
we have,
\[
\begin{split}
  &\mathbb{E}[h(Y)|Z_1=1] - \mathbb{E}[h(Y)|Z_1=0] \\
  =&
  \mathbb{E}\big[h\big(f(X(1),U,E_Y)\big) -
  h\big(f(X(0),U,E_Y)\big)\big] \\
  \approx& \mathbb{E}[h'(f^{(1)}(X,U,E_Y)) \cdot (X(1) - X(0))],
\end{split}
\]
where $h'$ is th derivative of $h$ and $f^{(1)}$ is the partial derivative of $f$ with respect to its first
argument. In this approximation we have used the assumption that $X(1)
- X(0)$ is small, i.e.\ the exposure $X$ is not changed by a single instrument $Z_1$ by much.

In many epidemiological problems, the causal effect of $X$ on the
outcome $Y$ is also very small compared to the variance of
$Y$. Therefore, when it is reasonable to assume that the variability
of the term $f^{(1)}(X,U,E_Y)$ is mostly driven by the noise variable
$E_Y$ which is independent of $X(0)$ and $X(1)$, we have
\[
\mathbb{E}[h(Y)|Z_1=1] - \mathbb{E}[h(Y)|Z_1=0] \approx
\mathbb{E}[h'(f^{(1)}(X,U,E_Y))] \cdot \mathbb{E}[X(1) - X(0)].
\]
The left hand side of the above equation may be regarded as a general
version of $\Gamma_1$ and $\mathrm{E}[X(1)-X(0)]$ a general version of
$\gamma_1$. Thus we arrive at the approximation $\Gamma_1 \approx
\beta_0 \gamma_1$ for $\beta_0 =
\mathbb{E}[h'(f^{(1)}(X,U,E_Y))]$. This may be interpreted as the
average of ``local'' causal effect: let $h$ be the identity function
and write the potential outcome $Y(x) = f(x,U,E_Y)$, then
\[
\beta_0 \approx \lim_{\Delta x \to 0} \frac{\mathbb{E}[Y(X + \Delta x) - Y(X)]}{\Delta x},
\]
where the expectation is taken jointly over $X$, $U$, and $E_Y$.

The above local argument reflects a meta-analysis interpretation of
MR \citep{bowden2015mendelian}: each SNP can be viewed as randomized experiment that
changes the exposure $X$ by just a little. Because all the changes are
relatively small compared to the variability of $X$ and $Y$, the
relationship between $\gamma_j$ and $\Gamma_j$ is almost linear.
This is why we expect the approximate linear relation
\eqref{eq:model-0} may hold in many problems beyond those discussed in
\Cref{sec:line-struct-model,sec:binary-outc-logist}.

%%% Local Variables:
%%% mode: latex
%%% TeX-master: t
%%% End:

\newpage

\section{Proofs}
\label{sec:proofs}

\subsection{Proof of \Cref{thm:consistency-simple}}
Notice that by \Cref{assump:iv-strength}, $p/(n^2\|\bm \gamma\|_2^2) \to
0$ implies that $n \to \infty$.
Let $e_j = \hat{\Gamma}_j -
\Gamma_j$ and $\epsilon_j = \hat{\gamma}_j - \gamma_j$. After some
algebra, we have
\[
  l(\beta) % &= \sum_{j=1}^p\frac{(\hat{\Gamma}_j - \beta
  % \hat{\gamma}_j)^2}{\sigma_{Xj}^2 \beta^2 + \sigma_{Yj}^2}
  = -\frac{1}{2} \sum_{j=1}^p \frac{\gamma_j^2(\beta_0 - \beta)^2 + (e_j - \beta
    \epsilon_j)^2 + 2 \gamma_j (\beta_0 - \beta) (e_j - \beta
    \epsilon_j)}{\sigma_{Xj}^2 \beta^2 + \sigma_{Yj}^2}.
  % &= (\beta_0 - \beta)^2 \left[ \sum_{j=1}^p \frac{\gamma_j^2}{\sigma_{Xj}^2 \beta^2 + \sigma_{Yj}^2} \right] +
  % \chi^2_p + O_p(\sqrt{n} |\beta_0 - \beta| \cdot \|\gamma\|_2) \\
  % &= (\beta_0 - \beta)^2 \left[ \sum_{j=1}^p \frac{\gamma_j^2}{\sigma_{Xj}^2 \beta^2 + \sigma_{Yj}^2} \right] + p +
  % O_p\big(\sqrt{p} + \sqrt{n}|\beta_0 - \beta|\big).
\]
Notice that $e_j - \beta \epsilon_j \sim \mathrm{N}(0, \sigma_{Xj}^2
\beta^2 + \sigma_{Yj}^2)$. By \Cref{assump:variance}, suppose
$\sigma_{Xj}^2 \ge c_{\sigma}/n$ and $\sigma_{Yj}^2 \ge c_{\sigma}/n$
for all $j \in [p]$. Using the
elementary inequality $2/(a+b) \ge \min(1/a,1/b)$ for $a,b>0$, %  $\gamma_j^2/(\sigma_{j2}^2 +
% \sigma_{Xj}^2 \beta^2) \ge \min
% \big(\gamma_j^2/\sigma_{j2}^2,
% \gamma_j^2/(\sigma_{Xj}^2 \beta^2)\big)/2$,
we obtain
\[
  \begin{split}
    &-2 l(\beta)\\ =& (\beta_0 - \beta)^2 \bigg[ \sum_{j=1}^p
    \frac{\gamma_j^2}{\sigma_{Xj}^2 \beta^2 + \sigma_{Yj}^2} \bigg] + p +
    O_p\big(\sqrt{p} + \sqrt{n}\|\bm \gamma\|\cdot|\beta_0 - \beta|\big) \\
    % &\ge (\beta_0 - \beta)^2 \left[ \sum_{j=1}^p \frac{1}{2} \min
    %   \Big(\frac{\gamma_j^2}{\sigma_{j2}^2},
    %   \frac{\gamma_j^2}{\sigma_{Xj}^2 \beta^2}\Big) \right] + p +
    % O_p\big(\sqrt{p} + \sqrt{n}|\beta_0 - \beta| \big) \\
    \ge& \frac{1}{2} (\beta_0 - \beta)^2 \min \bigg( \sum_{j=1}^p
    \frac{\gamma_j^2}{\sigma_{Yj}^2}, \sum_{j=1}^p
    \frac{\gamma_j^2}{\sigma_{Xj}^2 \beta^2} \bigg) + p +
    O_p\big(\sqrt{p} + \sqrt{n} \|\bm \gamma\|\cdot|\beta_0 - \beta| \big) \\
    % \ge& \frac{1}{2} (\beta_0 - \beta)^2 \cdot \Omega(n) \min \Big(1, \frac{1}{\beta^2}\Big) + p +
    % O_p\big(\sqrt{p} + \sqrt{n}|\beta_0 - \beta| \big) \\
    \ge& \frac{n \|\bm \gamma\|_2^2}{2 c_{\sigma}} \min \Big((\beta_0 - \beta)^2, \frac{(\beta_0 -
      \beta)^2}{\beta^2}\Big) + p +
    O_p\big(\sqrt{p} + \sqrt{n}\|\bm\gamma\|\cdot|\beta_0 - \beta| \big) \\
  \end{split}
\]

Consider the case $\beta_0 > 0$. By taking derivative, it is easy to verify that
$f(\beta) = (\beta_0 - \beta)^2 / \beta^2$ is decreasing in $\beta$
when $0 < \beta < \beta_0$ and increasing in $\beta$ when $\beta < 0$
or $\beta > \beta_0$. Since $f(\beta) \to 1$ as $|\beta| \to \infty$, for any $\epsilon > 0$ there exists
constant $C(\beta_0,\epsilon) > 0$ such that
$
\inf_{|\beta - \beta_0|\ge \epsilon}(\beta_0 -
\beta)^2/\beta^2 \ge C(\beta_0,\epsilon)$.
Similarly, we can show this is also true for $\beta_0 < 0$ and
$\beta_0 = 0$. Also, notice that the last term $O_p(\sqrt{n} |\beta_0
- \beta|)$ is negligible compared to the first term when $|\beta -
\beta_0| \ge \epsilon$. Let $C'(\beta_0,\epsilon) =
\min(\epsilon^2,C(\beta_0,\epsilon)) > 0$. We have
\[
  \inf_{|\beta - \beta_0|\ge \epsilon} -2 l(\beta) \ge
  (1 + o_p(1)) C'(\beta_0,\epsilon) \frac{n \|\bm \gamma\|_2^2}{2
    c_{\sigma}} + p + O_p\big(\sqrt{p}\big).
\]
Finally, by comparing this to $-2l(\beta_0) = p + O_p(\sqrt{p})$, we have
\[
  \begin{split}
    &\mathbb{P}\Big(l(\beta_0) > \sup_{|\beta - \beta_0|\ge
      \epsilon}l(\beta)\Big)
    = \mathbb{P}\Big( O_p(\sqrt{p}) \le (1 + o_p(1)) C'(\beta_0,\epsilon) \frac{n \|\bm \gamma\|_2^2}{2
      c_{\sigma}} + O_p\big(\sqrt{p}\big) \Big).
  \end{split}
\]
When $p \ll n^2\|\bm\gamma\|^4$, it is easy to see that this probability converges to $1$.

\subsection{Proof of \Cref{thm:clt-simple}}

By \eqref{eq:taylor-expansion} and the consistency of $\hat{\beta}$, we have
\[
  \begin{split}
    \hat{\beta} - \beta_0 &= \frac{-\psi(\beta_0)}{\psi'(\beta_0) +
      (1/2) \psi''(\tilde{\beta}) (\hat{\beta} - \beta_0)}\\
    &= \frac{ V_2}{\sqrt{V_1}} \cdot \frac{-\psi(\beta_0) / \sqrt{V_1}}{[\psi'(\beta_0) +
      o_p(\psi''(\tilde{\beta}))] / V_2}
  \end{split}
\]

The central limit theorem \Cref{eq:clt-simple} is immediately proven
using Slutsky's lemma after showing the following three lemmas:
\begin{lemma} \label{lem:1}
  $(1/\sqrt{V_1}) \psi(\beta_0) \overset{d}{\to}     \mathrm{N}(0, 1)$.
\end{lemma}
\begin{lemma} \label{lem:2}
  $(- 1/V_2) \psi'(\beta_0) \overset{p}{\to} 1$.
\end{lemma}
\begin{lemma} \label{lem:3}
  For a neighborhood $\mathcal{N}$ of $\beta_0$, $\sup_{\beta \in
    \mathcal{N}} (1/V_2) \psi''(\beta) = O_p(1)$.
\end{lemma}

Next we prove the three lemmas. For the first lemma, let
$\psi_j(\beta)$ be the $j$-th summand in \eqref{eq:profile-score}, so
$\psi(\beta) = \sum_{j=1}^p \psi_j(\beta)$. It is easy to show that
\begin{equation} \label{eq:psi-j-simple}
  \psi_j(\beta_0) = \frac{(e_j - \beta_0
    \epsilon_j)[\gamma_j (\sigma_{Yj}^2 + \sigma_{Xj}^2 \beta_0^2) +
    e_j \sigma_{Xj}^2 \beta_0 + \epsilon_j \sigma_{Yj}^2]}{(\sigma_{Yj}^2 +
    \sigma_{Xj}^2 \beta_0^2)^2}.
\end{equation}

The expectation of $\psi_j(\beta_0)$ is
\[
  \mathbb{E}[\psi_j(\beta_0)] = \frac{\mathbb{E}[e_j^2 \sigma_{Xj}^2 \beta_0 -
    \epsilon_j^2 \sigma_{Yj}^2 \beta_0]}{(\sigma_{Yj}^2 +
    \sigma_{Xj}^2 \beta_0^2)^2} = 0.
\]
This shows that $\mathbb{E}[\psi(\beta_0)] = 0$. The second moment of
$\psi_j(\beta_0)$ is given by
\[
  \mathbb{E}[\psi_j(\beta_0)^2] = A_j + B_j,
\]
where
\[
  A_j = \mathbb{E}\bigg[\frac{(e_j - \beta_0
    \epsilon_j)^2 \gamma_j^2 (\sigma_{Yj}^2 + \sigma_{Xj}^2 \beta_0^2)^2}{(\sigma_{Yj}^2 +
    \sigma_{Xj}^2 \beta_0^2)^4}\bigg] = \frac{\gamma_j^2}{\sigma_{Yj}^2 +
    \sigma_{Xj}^2 \beta_0^2},
\]
and
\[
  \begin{split}
    B_j &= \mathbb{E} \bigg[ \frac{(e_j - \beta_0
      \epsilon_j)^2 (e_j \sigma_{Xj}^2 \beta_0 + \epsilon_j
      \sigma_{Yj}^2)^2}{(\sigma_{Yj}^2 +
      \sigma_{Xj}^2 \beta_0^2)^4} \bigg] \\
    &= \mathbb{E} \bigg[\frac{e_j^4 \sigma_{Xj}^4 \beta_0^2 + \epsilon_j^4
      \sigma_{Yj}^4 \beta_0^2 + e_j^2 \epsilon_j^2 (\sigma_{Yj}^4 - 4
      \beta_0 \sigma_{Xj}^2 \beta_0 \sigma_{Yj}^2 + \beta_0^2
      \sigma_{Xj}^4 \beta_0^2)}{(\sigma_{Yj}^2 + \sigma_{Xj}^2
      \beta_0^2)^4}\bigg] \\
    &= \frac{3 \sigma_{Yj}^4 \sigma_{Xj}^4 \beta_0^2 + 3 \sigma_{Xj}^4
      \sigma_{Yj}^4 \beta_0^2 + \sigma_{Yj}^2 \sigma_{Xj}^2 (\sigma_{Yj}^4 - 4
      \beta_0 \sigma_{Xj}^2 \beta_0 \sigma_{Yj}^2 + \beta_0^2
      \sigma_{Xj}^4 \beta_0^2)}{(\sigma_{Yj}^2 + \sigma_{Xj}^2
      \beta_0^2)^4} \\
    &= \frac{\sigma_{Yj}^2 \sigma_{Xj}^2 (\sigma_{Yj}^4 + 2
      \beta_0^2 \sigma_{Xj}^2 \sigma_{Yj}^2 + \beta_0^4
      \sigma_{Xj}^4)}{(\sigma_{Yj}^2 + \sigma_{Xj}^2
      \beta_0^2)^4} \\
    &= \frac{\sigma_{Xj}^2 \sigma_{Yj}^2}{(\sigma_{Yj}^2 + \sigma_{Xj}^2
      \beta_0^2)^2}.
  \end{split}
\]
In summary,
\[
  \begin{split}
    \mathbb{E}[\psi_j(\beta_0)^2] % &=
    % \frac{\mathbb{E}[(e_j - \beta_0
    % \epsilon_j)^2]\gamma_j^2}{(\sigma_{Yj}^2 +
    % \sigma_{Xj}^2 \beta_0^2)^2} + \frac{\mathbb{E}[ (e_j - \beta_0
    % \epsilon_j)^2 (e_j \sigma_{Xj}^2 \beta_0 + \epsilon_j \sigma_{Yj}^2)^2]}{(\sigma_{Yj}^2 +
    % \sigma_{Xj}^2 \beta_0^2)^4} \\
    &= \frac{\gamma_j^2 \sigma_{Yj}^2 +
      \Gamma_j^2 \sigma_{Xj}^2 + \sigma_{Xj}^2
      \sigma_{Yj}^2}{(\sigma_{Yj}^2 +
      \sigma_{Xj}^2 \beta_0^2)^2}.
  \end{split}
\]
Notice that by \Cref{assump:variance}, $\mathbb{E}[\psi_j(\beta_0)^2] = \Theta (n
\gamma_j^2 + 1)$.

To prove \Cref{lem:1}, we consider two scenarios:

{\noindent \it Scenario 1: $p\to\infty$.} In this case, we hope to use
central limit theorem to show
\[
  \frac{1}{\sqrt{V_1}} \psi(\beta_0) = \frac{1}{\sqrt{V_1}} \sum_{j=1}^p \psi_j(\beta_0)
  \to \mathrm{N}(0, 1).
\]
Next we check Lyapunov's condition by computing the third moment of
$\psi_j(\beta_0)$. Notice that
\[
  \begin{split}
    \mathbb{E}[|\psi_j(\beta_0)|^3] &= \frac{\mathbb{E}\big\{\big|(e_j - \beta_0
      \epsilon_j)^3[\gamma_j (\sigma_{Yj}^2 + \sigma_{Xj}^2 \beta_0^2) +
      e_j \sigma_{Xj}^2 \beta_0 + \epsilon_j \sigma_{Yj}^2]^3\big|\big\}}{(\sigma_{Yj}^2 +
      \sigma_{Xj}^2 \beta_0^2)^6} \\
    &= \frac{\mathbb{E}[|C_0 + C_1\gamma_j + C_2\gamma_j^2 + C_3 \gamma_j^3|]}{(\sigma_{Yj}^2+
      \sigma_{Xj}^2 \beta_0^2)^6} \\
    &\le \frac{\mathbb{E}[|C_0|] + \gamma_j\mathbb{E}[|C_1|] + \gamma_j^2\mathbb{E}[|C_2|] + \gamma_j^3\mathbb{E}[|C_3|]]}{(\sigma_{Yj}^2+
      \sigma_{Xj}^2 \beta_0^2)^6}
  \end{split}
\]
We omit the detailed expressions for $C_0$, $C_1$, $C_2$ and $C_3$ but
note that there exists constant $C(\beta_0) > 0$ such that
$\mathbb{E}[|C_i|] \le C(\beta_0) (1/n)^{6-i/2}$ for
$i=0,1,2,3$. Therefore
\[
  \sum_{j=1}^p\mathbb{E}[|\psi_j(\beta_0)|^3] = O(p + \sqrt{n} \|\bm
  \gamma\|_1 + n \|\bm \gamma\|_2^2 + n^{3/2} \|\bm \gamma\|_3^3).
\]
By the Cauchy-Schwarz inequality, $\sqrt{n} \|\bm\gamma\|_1 \le
\sqrt{n} \sqrt{p} \|\bm\gamma\|_2 \le (p +
n\|\bm\gamma\|_2^2)/2$. Using the assumption $\|\bm \gamma\|_3 / \|\bm
\gamma\|_2 \to 0$, it is easy to show the Lyapunov condition
\[
  \frac{\sum_{j=1}^p\mathbb{E}[|\psi_j(\beta_0)|^3]}{\big\{\sum_{j=1}^p\mathbb{E}[\psi_j(\beta_0)^2]\big\}^{3/2}}
  = O\bigg(\frac{p + n \|\bm \gamma\|_2^2 + n^{3/2} \|\bm
    \gamma\|_3^3}{(p + n \|\bm \gamma\|_2^2)^{3/2}}\bigg) \to 0.
\]

{\noindent \it Scenario 2: $p$ is finite.} By the
assumption in the Theorem statement, $\kappa = n
\|\bm\gamma\|_2^2/p \to \infty$. We can rewrite \eqref{eq:psi-j-simple}
to obtain
\[
  \begin{split}
    \psi(\beta_0) &= \sum_{j=1}^p \frac{(e_j - \beta_0
      \epsilon_j)\gamma_j}{\sigma_{Yj}^2 +
      \sigma_{Xj}^2 \beta_0^2} + \sum_{j=1}^p \frac{(e_j - \beta_0
      \epsilon_j)[e_j \sigma_{Xj}^2 \beta_0 +
      \epsilon_j \sigma_{Yj}^2]}{(\sigma_{Yj}^2 +
      \sigma_{Xj}^2 \beta_0^2)^2}.
  \end{split}
\]
The first term on the right hand side is distributed as
$\mathrm{N}(0,V_2)$ and $V_2 = \Theta(n \|\bm \gamma\|_2^2)$. The second
term has variance $O(p)$ and is thus ignorable compared to the first
term. Therefore, $(1/\sqrt{V_2})
\psi(\beta_0) \overset{d}{\to} \mathrm{N}(0, 1)$. Since $p/n \to 0$, it is
easy to show that $V_1 / V_2 \to 1$. By Slutsky's lemma, $(1/\sqrt{V_1})
\psi(\beta_0) \overset{d}{\to} \mathrm{N}(0, 1)$.

Now we turn to \Cref{lem:2}. It suffices to prove
$\mathrm{E}[\psi'(\beta_0)] = V_2$ and
$\mathrm{Var}(\psi'(\beta_0)/V_2) \to 0$. Next we compute the first
two moments of $\psi'(\beta_0)$. By differentiating
\eqref{eq:profile-score}, we get
\[
  \psi'(\beta_0) = \sum_{j=1}^p \frac{A_j + B_j}{(\sigma_{Yj}^2 +
    \sigma_{Xj}^2 \beta_0^2)^4},
\]
where
\[
  \begin{split}
    A_j &= (\hat{\Gamma}_j^2
    \sigma_{Xj}^2-\hat{\gamma}_j^2 \sigma_{Yj}^2 - 2 \beta_0
    \hat{\gamma}_j \hat{\Gamma}_j \sigma_{Xj}^2)(\sigma_{Yj}^2 +
    \sigma_{Xj}^2 \beta_0^2)^2, \\
    B_j &= - (\hat{\Gamma}_j - \beta_0
    \hat{\gamma}_j)(\hat{\gamma}_j \sigma_{Yj}^2 + \hat{\Gamma}_j
    \sigma_{Xj}^2 \beta_0) \cdot 2 (\sigma_{Yj}^2 +
    \sigma_{Xj}^2 \beta_0^2) \cdot 2 \sigma_{Xj}^2 \beta_0.
  \end{split}
\]
The expected values of these two terms are
\[
  \begin{split}
    \mathbb{E}[A_j] &= \big[(\Gamma_j^2 + \sigma_{Yj}^2)
    \sigma_{Xj}^2-(\gamma_j^2 + \sigma_{Xj}^2) \sigma_{Yj}^2 - 2 \beta_0
    \gamma_j \Gamma_j \sigma_{Xj}^2\big](\sigma_{Yj}^2 +
    \sigma_{Xj}^2 \beta_0^2)^2 \\
    &= - (\gamma_j^2 \sigma_{Yj}^2 + \Gamma_j^2 \sigma_{Xj}^2) (\sigma_{Yj}^2 +
    \sigma_{Xj}^2 \beta_0^2)^2,
  \end{split}
\]
\[
  \begin{split}
    \mathbb{E}[B_j] &= - \big[ (\Gamma_j^2 + \sigma_{Yj}^2) \sigma_{Xj}^2
    \beta_0 - (\gamma_j^2 + \sigma_{Xj}^2) \sigma_{Yj}^2 \beta_0 +
    \Gamma_j \gamma_j(\sigma_{Yj}^2 - \sigma_{Xj}^2 \beta_0^2) \big]
    \cdot \\
    &\quad \cdot 2(\sigma_{Yj}^2 +
    \sigma_{Xj}^2 \beta_0^2) \cdot 2 \sigma_{Xj}^2 \beta_0 \\
    &= -\big[(\Gamma_j^2 \sigma_{Xj}^2 - \gamma_j^2 \sigma_{Yj}^2) \beta_0
    + \Gamma_j \gamma_j(\sigma_{Yj}^2 - \sigma_{Xj}^2 \beta_0^2) \big]
    \cdot 4(\sigma_{Yj}^2 +
    \sigma_{Xj}^2 \beta_0^2) \sigma_{Xj}^2 \beta_0 \\
    &= 0.
  \end{split}
\]
Therefore $\mathrm{E}[\psi'(\beta_0)] = - V_2$.

For the variance of $\psi'(\beta_0)$, consider any $\beta$ in a
neighborhood $\mathcal{N}$ of $\beta$.
Our argument is based on the key observation that $\psi_j(\beta)$ is a
homogeneous quadratic polynomial of
$(\tilde{\gamma}_j, \tilde{e}_j, \tilde{\epsilon}_j) = (\sqrt{n}
\gamma_j, \sqrt{n}e_j,\sqrt{n}\epsilon_j)$:
\small
\[
  \psi_j(\beta) = \frac{((\beta_0 - \beta)  \tilde{\gamma}_j + \tilde{e}_j -
    \beta \tilde{\epsilon}_j) [(n\sigma_{Xj}^2 \beta \beta_0 + n
    \sigma_{Yj}^2)\tilde{ \gamma}_j + (n \sigma_{Xj}^2 \beta) \tilde{e}_j +
    (n \sigma_{Yj}^2) \tilde\epsilon_j]}{(n \sigma_{Xj}^2 \beta^2 + n \sigma_{Yj}^2)^2}.
\]
\normalsize
Therefore its
derivative, $\psi'_j(\beta)$, remains to be a homogeneous quadratic polynomial of
$(\tilde{\gamma}_j, \tilde{e}_j, \tilde{\epsilon}_j)$. This observation
suggests that $\mathbb{E}[\psi_j'(\beta_0)]$ is a quadratic function
of $\tilde{\gamma}_j$. Notice that the
any term in $\psi'_j(\beta)$ that has odd degree of $\tilde{\gamma}_j$
must have expectation equal to $0$, because it must have odd degree in either
$\tilde{e}_j$ or $\tilde{\epsilon}_j$. A simple calculation then yields
$\mathbb{E}[\psi_j'(\beta)]] = \Theta(\tilde{\gamma}_j^2)$, so for $\beta \in \mathcal{N}$,
\[
  \mathbb{E}[\psi'(\beta)] = \Theta(n \|\bm\gamma\|_2^2).
\]
Similarly, the variance $\mathrm{Var}[\psi_j'(\beta)^2]$ is also a quadratic
polynomial of $\tilde{\gamma}_j$ (because the $\tilde{\gamma}_j^2$ term in
$\psi_j'(\beta)$ is non-random). Thus for $\beta \in \mathcal{N}$,
\[
  \mathrm{Var}(\psi'(\beta)) = O(n \|\bm \gamma\|_2^2 + p).
\]
Using the assumption $p/(n^2 \|\bm \gamma\|^4) \to 0$, it is then easy
to see that
\[
\mathrm{Var}(\psi'(\beta_0)) \ll
\big(\mathrm{E}[\psi'(\beta_0)]\big)^2.
\]
This concludes the proof of \Cref{lem:2}.

The above argument for $\psi'(\beta_0)$ can also be applied to
$\psi''(\beta)$ for any $\beta$ in a neighborhood $\mathcal{N}$ of $\beta_0$, so $\mathrm{Var}(\psi''(\beta)) = O(n \|\bm
\gamma\|_2^2 + p) = o(V_2^2)$. Since $\psi(\beta)$ is smooth in
$\beta$, this proves the \Cref{lem:3}.

\subsection{Proof of \Cref{thm:consistent-variance-simple}}
\label{sec:proof-crefthm:c-vari}

By \Cref{thm:clt-simple}, $\hat{\beta} - \beta_0 =
O_p(1/\sqrt{n})$. Thus
\[
  \sigma_{Yj}^2 + \sigma_{Xj}^2 \hat{\beta}^2 = (\sigma_{Yj}^2 +
  \sigma_{Xj}^2 \beta_0^2) (1 + o_p(1/n)).
\]
This implies that
\[
  \hat{V}_1 = (1 + o_p(1)) \cdot \sum_{j=1}^p \frac{(\hat{\gamma}_j^2 - \sigma_{Xj}^2)
    \sigma_{Yj}^2 + (\hat{\Gamma}_j^2 - \sigma_{Yj}^2) \sigma_{Xj}^2 +
    \sigma_{Xj}^2 \sigma_{Yj}^2}{(\sigma_{Yj}^2 + \sigma_{Xj}^2
    \beta_0^2)^2}.
\]
It is easy to show that the summation on the right hand side has mean
$V_1$. Similar to the proof of \Cref{thm:clt-simple}, the variance
of this term is $O(p)$. Note that $V_1 = \Theta(n\|\bm \gamma\|_2^2 +
p) = \Theta(p\cdot(1 + \kappa))$, using the assumption in
\Cref{thm:clt-simple} that $p \to \infty$ or $\kappa \to \infty$, $\hat{V}_1 = (1 + o_p(1)) V_1$. Similarly, $\hat{V}_2 =
(1 + o_p(1)) V_2$. Equation \eqref{eq:pivot-simple} follows
immediately from Slutsky's lemma.

\subsection{Proof of \Cref{thm:consistency-overdispersed}}
To prove consistency of $(\hat{\beta},\hat{\tau}_0^2)$, we
need to study the asymptotic behavior of the adjusted profile
score. Let $\psi_{1j}(\beta,\tau^2)$ and $\psi_{2j}(\beta,\tau^2)$ be
the $j$-th
term in the summation in \eqref{eq:aps-1} and \eqref{eq:aps-2}, so
$\psi_i(\beta,\tau^2) =
\sum_{j=1}^p \psi_{ij}(\beta,\tau^2),~i=1,2$. We first consider the
expectation of $\psi_{1j}$ and $\psi_{2j}$: \small
\begin{equation*} \label{eq:psi1j}
  \begin{split}
    &\mathbb{E}[\psi_{1j}(\beta,\tau^2)] \\
    =& \mathbb{E} \bigg\{ \frac{[(e_j - \beta \epsilon_j) + (\beta_0 -
      \beta) \gamma_j][\gamma_j (\sigma_{Xj}^2
      \beta \beta_0 + \sigma_{Yj}^2 + \tau^2) + \epsilon_j
      (\sigma_{Yj}^2 + \tau^2) + e_j \sigma_{Xj}^2 \beta
      ]}{(\sigma_{Xj}^2 \beta^2 + \sigma_{Yj}^2 + \tau^2)^2} \bigg\} \\
    =& \frac{(\tau^2 + \sigma_{Yj}^2 + \sigma_{Xj}^2
      \beta \beta_0) (\beta_0 -
      \beta) \gamma_j^2 + \sigma_{Xj}^2 \beta (\tau_0^2 - \tau^2)}{(\tau^2 + \sigma_{Yj}^2 +
      \sigma_{Xj}^2 \beta^2)^2},
  \end{split}
\end{equation*} \normalsize
and
\begin{equation}
  \label{eq:psi2-expansion}
  \begin{split}
    \mathbb{E}[\psi_{2j}(\beta,\tau^2)] &= \mathbb{E}\bigg\{\sigma_{Xj}^2\frac{[(e_j - \beta \epsilon_j) + (\beta_0 -
      \beta) \gamma_j]^2 - (\tau^2 + \sigma_{Yj}^2 + \sigma_{Xj}^2
      \beta^2)}{(\tau^2 + \sigma_{Yj}^2 + \sigma_{Xj}^2
      \beta^2)^2}\bigg\} \\
    &= \frac{\sigma_{Xj}^2(\beta_0 - \beta)^2 \gamma_j^2 + \sigma_{Xj}^2(\tau_0^2 - \tau^2)}{(\tau^2 + \sigma_{Yj}^2 + \sigma_{Xj}^2
      \beta^2)^2}.
  \end{split}
\end{equation}

Now consider the following contrast of the two estimating equations:
\[
  \tilde{\psi}(\beta,\tau^2) = \psi_1(\beta,\tau^2) - \beta
  \psi_2(\beta, \tau^2).
\]
It is straightforward to verify that
\[
  \begin{split}
    \mathbb{E}[\tilde{\psi}(\beta,\tau^2)]
    =& \sum_{j=1}^p\frac{(\beta_0 - \beta) \gamma_j^2 (\tau^2 + \sigma_{Yj}^2 +
      \sigma_{Xj}^2 \beta \beta_0 - \beta \sigma_{Xj}^2 (\beta_0 - \beta))}{(\tau^2 + \sigma_{Yj}^2 + \sigma_{Xj}^2
      \beta^2)^2} \\
    =& \sum_{j=1}^p \frac{(\beta_0 - \beta) \gamma_j^2}{\tau^2 +
      \sigma_{Yj}^2 + \sigma_{Xj}^2 \beta^2}.
  \end{split}
\]
In other words, $\mathbb{E}[\tilde{\psi}(\beta,\tau^2)] = 0$ if and
only if $\beta = \beta_0$.

Next we bound the variance of $\tilde{\psi}(\beta,\tau^2)$ over
$\mathcal{B}$. Because of \Cref{assump:variance,assump:compactness}, $\mathrm{Var}(\epsilon_j) = \Theta(1/n)$ and
$\mathrm{Var}(e_j) = \Theta(1/n + 1/p)$. % An important observation is
% \begin{equation} \label{eq:ob}
%   p (\tau_n^2)^2 = O(p / n^2 + 1/n + 1/p) \to 0.
% \end{equation}
Using the inequality $\mathrm{Var}(X+Y) \le 2
[\mathrm{Var}(X) + \mathrm{Var}(Y)]$ repeatedly, we have
\[
  \mathrm{Var(\tilde{\psi}(\beta,\tau^2))} = O\big(\mathrm{Var}(\psi_1(\beta,\tau^2))\big) + O\big(
  \mathrm{Var}(\psi_{2}(\beta,\tau^2))\big),
\]
and, after some algebra,
\[
  \begin{split}
    \mathrm{Var}(\psi_1(\beta,\tau^2)) &= O((n + p) \|\bm \gamma\|_2^2 +
    p) = o(n^2),~\text{and} \\
    % =& O\Big(\sum_{j=1}^p\frac{n^{-1} \gamma_j^2 n^{-2} + n^{-4} n^{-1}
    %   (\tau_n^2)^2 + (\tau_n^2)^2 n^{-2} + \gamma_j^2
    %   [n^{-1} (\tau_n^2)^2 + \tau_{0n}^2 n^{-2}]}{(\tau_n^2)^4}\Big) \\
    % =& O\Big(\frac{\|\bm \gamma\|_2^2 \tau_{0n}^2 (\tau_n^2)^2 + (p/n)
    %   \tau_{0n}^2 (\tau_n^2)^2 + \|\bm \gamma\|_2^2 [n^{-1} (\tau_n^2)^2 +
    %   \tau_{0n}^2 n^{-2}]}{(\tau_n^2)^4} \Big) \\
    % =& O\Big(\frac{\|\bm \gamma\|_2^2 \tau_{0n}^2 (\tau_n^2)^2 + (p/n)
    %   \tau_{0n}^2 (\tau_n^2)^2}{(\tau_n^2)^4} \Big), \\
    \mathrm{Var}(\psi_2(\beta,\tau^2)) &= O((n + p) \|\bm \gamma\|_2^2 +
    p) = o(n^2).
    % & O\Big( \sum_{j=1}^p \frac{1}{n^2} \frac{(\tau_n^2)^2 + \gamma_j^2
    %   \tau_n^2}{(\tau_n^2)^4} \Big) = O\Big(\frac{p (\tau_n^2)^2 + \|\bm \gamma\|_2^2 \tau_n^2}{n^2(\tau_n^2)^4}\Big). \\
  \end{split}
\]
% Using the assumptions $\tau_0^2 = O(1/p)$, $\|\bm\gamma\|_2 =
% \Theta(1)$ and $p/n^2 \to 0$, we have
% \[
%   \mathrm{Var}(\psi_1(\beta,\tau^2)) = O \Big( \frac{(\|\bm \gamma\|_2^2 + p/n)
%     (1/p + 1/n) }{(\tau_n^2)^2} \Big) = o \Big( \frac{1}{(\tau_n^2)^2} \Big),
% \]
% \[
%   \mathrm{Var}(\psi_2(\beta,\tau^2)) = O\Big(\frac{p}{n^2
%     (\tau_n^2)^2}\Big) = o \Big( \frac{1}{(\tau_n^2)^2} \Big).
% \]
To summarize, we have shown that
\begin{align}
    \tilde{\psi}(\beta,\tau^2) &= (\beta_0 - \beta) \cdot
    \Theta(n) + o_p( n ), \label{eq:psi-tilde-final} \\
  \psi_2(\beta,\tau^2) &= (\beta_0 - \beta)^2 \Theta (n) + (p\tau_0^2 - p\tau) \Theta (n) +
  o_p(n). \label{eq:psi-2-final}
\end{align}

Consider a box $\mathcal{B}' = [-C_1,C_1] \times [0, C_2]$ that contains
$\mathcal{B}$. Using \eqref{eq:psi-tilde-final} and
\eqref{eq:psi-2-final}, if $C_1$ and $C_2/C_1$ are
sufficiently large, all the following events have probabilities going to $1$:
\[
  \begin{split}
    \sup_{|p\tau^2| \le C_2} \tilde{\psi}(C_1,\tau^2) \le 0,~
    &\inf_{|p\tau^2| \le C_2} \tilde{\psi}(-C_1,\tau^2) \ge 0,\\
    \sup_{|\beta| \le C_1} \psi_2(\beta,C_2/p) \le 0,~
    &\inf_{|\beta| \le C_1} \psi_2(\beta,0) \ge 0.\\
  \end{split}
\]
If all the events are true, by continuity of $\tilde{\psi}$ and
$\psi_2$ and the Poincar\'{e}-Miranda theorem, there exists
$(\hat{\beta},p\hat{\tau}^2) \in \mathcal{B}'$ such that
\[
  \tilde{\psi}(\hat{\beta},\hat{\tau}^2) =
  \psi_2(\hat{\beta},\hat{\tau}^2) = 0.
\]
Using \eqref{eq:psi-tilde-final} and \eqref{eq:psi-2-final}, it is then
straightforward to show $\hat{\beta} \overset{p}{\to} \beta_0$ and $(p
\hat{\tau}^2 - p\tau_0^2) \overset{p}{\to} 0$. As a consequence,
$(\hat{\beta},p\hat{\tau}^2) \in \mathcal{B}$ with probability going
to $1$, thus concluding our proof.

% Since $\tilde{\psi}(\hat{\beta},\hat{\tau}^2) = 0$, this shows that
% $\hat{\beta} - \beta_0 \overset{p}{\to} 0$.

% Using the consistency of $\hat{\beta}$ and $p (\tau_n^2)^2 = O( p /
% n^2 + 1/n + 1/p) \to 0$, we can show that
% \[
%   \begin{split}
%     \psi_2(\hat{\beta},\tau^2) &= \sum_{j=1}^p \frac{\sigma_{Xj}^2(\tau_0^2 - \tau^2)}{(\tau^2 + \sigma_{Yj}^2 + \sigma_{Xj}^2
%     \hat{\beta}^2)^2} + o_p\Big(\frac{\|\bm\gamma\|_2^2}{n
%     (\tau_n^2)^2}\Big) + O_p\Big(\frac{\sqrt{p}}{n\tau_n^2}\Big) \\
%     &=
%     p(\tau_0^2 - \tau^2) \Omega\Big(\frac{1}{n (\tau_n^2)^2} \Big) +
%     o_p\Big(\frac{1}{n (\tau_n^2)^2}\Big).
%   \end{split}
% \]
% Therefore $p(\tau_0^2 - \hat{\tau}^2) \overset{p}{\to} 0$.

\subsection{Proof of \Cref{thm:clt-overdispersed}}
\label{sec:proof-crefthm:clt-ov}

We begin with proving $\tilde{\bm V}_1$ and $\tilde{\bm V}_2$ are the
corresponding moments of $\bm \psi (\beta_0,\tau_0^2)$ and $\nabla \bm
\psi (\beta_0,\tau_0^2)$.

\begin{lemma} \label{lem:4}
  $\mathrm{Var}(\bm
  \psi(\beta_0,\tau_0^2)) = \tilde{\bm V}_1$, $\mathbb{E}[\nabla \bm
  \psi(\beta_0,\tau_0^2)] = \tilde{\bm V}_2$.
\end{lemma}
\begin{proof}[Proof of \Cref{lem:4}]
  In \Cref{sec:adjust-prof-score} we have already
  shown that $\mathbb{E}[\bm \psi(\beta_0,\tau_0^2)] = \bm 0$. The variance of
  $\psi_1(\beta_0,\tau_0^2)$ and the expectation of $(\partial/\partial
  \beta)\psi_1(\beta_0,\tau_0^2)$ can be obtained from the proof of
  \Cref{thm:clt-simple} by replacing $\sigma_{Yj}^2$ with $\sigma_{Yj}^2
  + \tau_0^2$.

  Next we compute the other moments. Let
  $\psi_{1j}(\beta,\tau^2)$ and $\psi_{2j}(\beta,\tau^2)$ be the $j$-th
  summand in \eqref{eq:aps-1} and \eqref{eq:aps-2}, so
  $\psi_i(\beta,\tau^2) = \sum_{j=1}^p \psi_{ij}(\beta,\tau^2)$ for $i =
  1,2$. Because $\hat{\Gamma}_j - \beta_0 \hat{\gamma}_j \sim
  \mathrm{N}(0, \beta_0^2\sigma_{Xj}^2 + \Sigma_{Yj}^2 + \tau_0^2)$,
  it is easy to see that
  \[
    \mathrm{Var}(\psi_{2j}(\beta_0,\tau_0^2)) =
    \frac{2(\sigma_{Xj}^2)^2}{(\sigma_{Xj}^2 \beta_0^2 + \sigma_{Yj}^2 + \tau_0^2)^2}
  \]
  The covariance of $\psi_{1j}$ and $\psi_{2j}$ is given by \small
  \[
    \begin{split}
      &\mathrm{Cov}(\psi_{1j}(\beta_0,\tau_0^2),\psi_{2j}(\beta_0, \tau_0^2)) \\
      =&
      \mathbb{E}\Big[ \frac{(\hat{\Gamma}_j - \beta_0
        \hat{\gamma}_j)(\hat{\gamma}_j (\tau_0^2+ \sigma_{Yj}^2) + \hat{\Gamma}_j
        \sigma_{Xj}^2 \beta_0) \sigma_{Xj}^2 [(\hat{\Gamma}_j - \beta_0
        \hat{\gamma}_j)^2 - (\sigma_{Xj}^2
        \beta_0^2 + \sigma_{Yj}^2 + \tau_0^2)]}{(\sigma_{Xj}^2\beta_0^2 + \sigma_{Yj}^2 + \tau_0^2)^4} \Big] \\
      =& \sigma_{Xj}^2 \mathbb{E}\Big[ \frac{(\hat{\Gamma}_j - \beta_0
        \hat{\gamma}_j)^3(\hat{\gamma}_j (\sigma_{Yj}^2 + \tau_0^2) + \hat{\Gamma}_j
        \sigma_{Xj}^2 \beta_0)}{(\sigma_{Xj}^2\beta_0^2 + \sigma_{Yj}^2 + \tau_0^2)^4} \Big] \\
      =& \sigma_{Xj}^2 \mathbb{E}\Big[ \frac{(e_j - \beta_0 \epsilon_j)^3[\gamma_j (\sigma_{Xj}^2
        \beta_0^2 + \sigma_{Yj}^2 + \tau_0^2) + \epsilon_j (\sigma_{Yj}^2
        + \tau_0^2) + e_j
        \sigma_{Xj}^2 \beta_0]}{(\sigma_{Xj}^2\beta_0^2 + \sigma_{Yj}^2 + \tau_0^2)^4} \Big] \\
      =& \sigma_{Xj}^2 \mathbb{E}\Big[ \frac{(e_j - \beta_0
        \epsilon_j)^3[\epsilon_j (\sigma_{Yj}^2 + \tau_0^2) + e_j
        \sigma_{Xj}^2 \beta_0]}{(\sigma_{Xj}^2\beta_0^2 + \sigma_{Yj}^2 + \tau_0^2)^4} \Big] \\
      =& \sigma_{Xj}^2 \mathbb{E}\Big[ \frac{e_j^4 \sigma_{Xj}^2 \beta_0 + e_j^2
        \epsilon_j^2[-3\beta_0(\sigma_{Yj}^2 + \tau_0^2) + 3 \beta_0^2
        \sigma_{Xj}^2 \beta_0] - \epsilon_j^4\beta_0^3(\sigma_{Yj}^2 + \tau_0^2)}{(\sigma_{Xj}^2\beta_0^2 + \sigma_{Yj}^2 + \tau_0^2)^4} \Big] \\
      % =& \frac{3 (\tau_0^2+ \sigma_{Yj}^2)^2 \sigma_{Xj}^2 \beta_0 + (\tau_0^2
      % + \sigma_{Yj}^2) \sigma_{Xj}^2[-3\beta_0(\tau_0^2+ \sigma_{Yj}^2) + 3 \beta_0^3
      % \sigma_{Xj}^2] - 3 \sigma_{Xj}^4\beta_0^3(\tau_0^2+ \sigma_{Yj}^2)}{(\tau_0^2+ \sigma_{Yj}^2 +
      % \sigma_{Xj}^2 \beta_0^2)^4} \\
      =& 0.
    \end{split}
  \]
  \normalsize
  Thus $\mathrm{Var}(\bm \psi(\beta_0,\tau_0^2)) =
  \tilde{\bm{V}}_1$. Next we consider the expectation of $\nabla \bm
  \psi(\beta_0,\tau_0^2)$:
  \[
    \begin{split}
      &\mathbb{E}\Big[ \frac{\partial}{\partial \tau^2} \psi_{1j}(\beta_0^2,\tau_0^2)
      \Big] \\
      =& \mathbb{E}\Big[\frac{(\hat{\Gamma}_j - \beta_0 \hat{\gamma}_j) \hat{\gamma}_j (\sigma_{Xj}^2\beta_0^2 + \sigma_{Yj}^2 + \tau_0^2)^2}{(\sigma_{Xj}^2\beta_0^2 + \sigma_{Yj}^2 + \tau_0^2)^4} \Big] - \\
      &-\mathbb{E}\Big[\frac{(\hat{\Gamma}_j - \beta_0
        \hat{\gamma}_j)(\hat{\gamma}_j (\sigma_{Yj}^2 + \tau_0^2) + \hat{\Gamma}_j
        \sigma_{Xj}^2 \beta_0) \cdot 2 (\sigma_{Xj}^2\beta_0^2 + \sigma_{Yj}^2 + \tau_0^2)}{(\sigma_{Xj}^2\beta_0^2 + \sigma_{Yj}^2 + \tau_0^2)^4} \Big] \\
      =& \mathbb{E} \Big[\frac{(\hat{\Gamma}_j - \beta_0 \hat{\gamma}_j) [\hat{\gamma}_j
        (\sigma_{Xj}^2 \beta_0^2 - \sigma_{Yj}^2 - \tau_0^2) - 2 \hat{\Gamma}_j
        \sigma_{Xj}^2 \beta_0]}{(\sigma_{Xj}^2\beta_0^2 + \sigma_{Yj}^2 + \tau_0^2)^3}\Big] \\
      =& \mathbb{E} \Big[\frac{(e_j - \beta_0 \epsilon_j) [\epsilon_j
        (\sigma_{Xj}^2 \beta_0^2 - \sigma_{Yj}^2 - \tau_0^2) - 2 e_j
        \sigma_{Xj}^2 \beta_0]}{(\sigma_{Xj}^2\beta_0^2 + \sigma_{Yj}^2 + \tau_0^2)^3}\Big] \\
      =& \frac{-2 (\sigma_{Yj}^2 + \tau_0^2) \sigma_{Xj}^2 \beta_0 -
        \beta_0 \sigma_{Xj}^2 (\sigma_{Xj}^2 \beta_0^2 - \sigma_{Yj}^2 - \tau_0^2)}{(\sigma_{Xj}^2\beta_0^2 + \sigma_{Yj}^2 + \tau_0^2)^3} \\
      =& \frac{- \sigma_{Xj}^2 \beta_0}{(\sigma_{Xj}^2\beta_0^2 + \sigma_{Yj}^2 + \tau_0^2)^2}.
    \end{split}
  \]
  Furthermore,
  \[
    \frac{\partial}{\partial \beta}
    \psi_{2j}(\beta_0,\tau_0)
    =  \sigma_{Xj}^2\frac{D_j - E_j}{(\sigma_{Xj}^2 \beta_0^2 + \sigma_{Yj}^2 + \tau_0^2)^4}
  \]
  where
  \[
    D_j = [2 (\hat{\Gamma}_j - \beta_0
    \hat{\gamma}_j) (-\hat{\gamma}_j) - 2 \sigma_{Xj}^2 \beta_0] (\tau_0^2 + \sigma_{Yj}^2 +
    \sigma_{Xj}^2 \beta_0^2)^2,
  \]
  \[
    E_j =
    [(\hat{\Gamma}_j - \beta_0
    \hat{\gamma}_j)^2 - (\tau_0^2 + \sigma_{Yj}^2 + \sigma_{Xj}^2
    \beta_0^2)] \cdot 2 (\tau_0^2 + \sigma_{Yj}^2 +
    \sigma_{Xj}^2 \beta_0^2) \sigma_{Xj}^2 \cdot 2\beta_0.
  \]
  It is not hard to see that both $D_j$ and $E_j$ have mean
  $0$. Finally,
  \[
    \begin{split}
      & \frac{\partial}{\partial \tau_0^2}
      \psi_{2j}(\beta_0,\tau_0) \\
      =& -\frac{\sigma_{Xj}^2}{(\sigma_{Xj}^2 \beta_0^2 + \sigma_{Yj}^2 + \tau_0^2)^2} -
      \frac{2 \sigma_{Xj}^2 [(\hat{\Gamma}_j - \beta_0
        \hat{\gamma}_j)^2 - (\sigma_{Xj}^2
        \beta_0^2 + \sigma_{Yj}^2 + \tau_0^2)]}{(\sigma_{Xj}^2 \beta_0^2 + \sigma_{Yj}^2 + \tau_0^2)^3}
    \end{split}
  \]
  It is easy to see that
  \[
    \mathbb{E}\Big[\frac{\partial}{\partial \tau^2}
    \psi_{2j}(\beta_0,\tau_0)
    \Big] = -\frac{\sigma_{Xj}^2}{(\sigma_{Xj}^2 \beta_0^2 + \sigma_{Yj}^2
      + \tau_0^2)^2}.
  \]
  In summary, we have proved that $\mathbb{E}[\nabla \bm
  \psi(\beta_0,\tau_0^2)] = - \tilde{\bm V}_2$.
\end{proof}

It is useful to write down the order of $\tilde{\bm V}_1$ and
$\tilde{\bm V}_2$:
\begin{equation} \label{eq:V-tilde-rate}
  \begin{split}
    \tilde{\bm V}_1 &=
    \begin{pmatrix}
      \Theta(n)
      & 0 \\
      0 & \Theta (n) \\
    \end{pmatrix},~
    \tilde{\bm V}_2 =
    \begin{pmatrix}
      \Theta(n)
      & \Theta(n^2) \\
      0 & \Theta(n^2) \\
    \end{pmatrix}.
  \end{split}
\end{equation}

Similar to the proof of \Cref{thm:clt-simple}, consider the
Taylor expansion (let $\bm{\theta} = (\beta,\tau^2)$)
\begin{equation} \label{eq:taylor-expansion-overdispersed}
  \bm{0} = \bm{\psi}(\hat{\bm{\theta}}) =
  \bm{\psi}(\bm{\theta}_0) + \nabla \psi(\bm{\theta}_0)
  (\hat{\bm{\theta}} - \bm{\theta}_0) +\frac{1}{2}
  \begin{pmatrix}
    (\hat{\bm{\theta}} - \bm{\theta}_0)^T \partial^2
    {\psi}_1(\tilde{\bm{\theta}}) (\hat{\bm{\theta}} - \bm{\theta}_0) \\
    (\hat{\bm{\theta}} - \bm{\theta}_0)^T \partial^2
    {\psi}_2(\tilde{\bm{\theta}}) (\hat{\bm{\theta}} - \bm{\theta}_0).
  \end{pmatrix}
  % O(\|\hat{\bm{\theta}} -
  % \bm{\theta}_0\|_2^2) \cdot O(\|\partial^2 \bm{\psi}(\tilde{\bm{\theta}})\|).
\end{equation}
By the consistency of $(\hat{\beta},p\hat{\tau}^2)$ and three Lemmas listed
after this paragraph, the third term on the right hand side is
negligible compared to the second term. The central limit theorem
\eqref{eq:clt-overdispersed} can then be proven by the same
arguments (normalizing by $\tilde{\bm{V}}_1$ and $\tilde{\bm{V}}_2$
and using Slutsky's lemma)
as in the beginning of the proof of \Cref{thm:clt-simple}.

\begin{lemma} \label{lem:5}
  $(\tilde{\bm{V}}_1)^{-1/2} \bm{\psi}(\beta_0,\tau_0^2) \overset{d}{\to} \mathrm{N}(\bm{0}, \bm{I}_2)$.
\end{lemma}
\begin{lemma} \label{lem:6}
  $(\tilde{\bm V}_2)^{-1} \nabla \bm{\psi}(\beta_0,\tau_0^2)
  \overset{p}{\to} - \bm I_2$.
\end{lemma}
\begin{lemma} \label{lem:7}
  Denote $\partial^2 \psi_i(\beta,\tau^2)$ to be all the
  second-order partial derivatives of $\psi_i(\beta,\tau^2)$,
  $i=1,2$. For a neighborhood $\mathcal{N}$ of
  $(\beta_0,p\tau_0^2)$ and $l=0,1,2$,
  \[
    \begin{split}
    \sup_{(\beta,\tau^2) \in \mathcal{N}}
    \Big|\frac{\partial^2}{\partial \beta^{2-l} (\partial \tau^2)^{l}} \psi_1(\beta,\tau^2)\Big| &=
    O_p(n^{l+1}),~\text{and}\\
    \sup_{(\beta,\tau^2) \in \mathcal{N}}
    \Big|\frac{\partial^2}{\partial \beta^{2-l} (\partial \tau^2)^{l}} \psi_2(\beta,\tau^2)\Big| &=
    O_p(n^{l+1}).
        \end{split}
  \]
\end{lemma}

Next we prove \Cref{lem:5,lem:6,lem:7}. Let
$\bm \psi_j(\beta,\tau^2) =
(\psi_{1j}(\beta,\tau^2),\psi_{2j}(\beta,\tau^2))$ for $j \in
[p]$. Since $\bm \psi_j,j\in [p]$ are
mutually independent and $p \to \infty$, it suffices to verify the following Lyapunov condition
\citep{bentkus2005lyapunov}
\[
  \sum_{j=1}^p\mathbb{E}\big[\|\tilde{\bm V}_1^{-1/2} \bm
  \psi_j(\beta_0,\tau_0^2) \|^3 \big] \to 0
\]
to prove \Cref{lem:5}. Because $\tilde{\bm V}_1$ is diagonal, it
suffices to verify this for each coordinate of $\bm \psi_j$. Similar
to the proof of \Cref{lem:1}, we can show that
\begin{equation} \label{eq:lyapunov}
  \sum_{j=1}^p\mathbb{E}[|\psi_{1j}(\beta_0, \tau_0^2)|^3] = O(p + \sqrt{n}\|\bm
  \gamma\|_1 + n \|\bm \gamma\|_2^2 + n^{3/2} \|\bm \gamma\|_3^3).
\end{equation}
Therefore, using $\|\bm \gamma\|_3/\|\bm \gamma\|_2 \to 0$, we obtain
\[
  \frac{\sum_{j=1}^p\mathbb{E}[|\psi_{1j}(\beta_0)|^3]}{(\tilde{\bm
      V}_1)_{11}^{3/2}} = O\Big(\frac{p + \sqrt{n}\|\bm
  \gamma\|_1 + n \|\bm \gamma\|_2^2 + n^{3/2} \|\bm \gamma\|_3^3}{n^{3/2}}\Big) \to 0.
\]

For $\psi_2$, since the third moment of a $\chi^2_1$
distribution exists,
\[
  \mathbb{E}[|\psi_{2j}(\beta_0,\tau_0^2)|^3] =
  O\Big(\frac{\sigma_{Xj}^6}{(\sigma_{Xj}^2 \beta_0^2 + \sigma_{Yj}^2 +
    \tau_0^2)^3}\Big) = O(1).
\]
% On the other hand,
% \[
%   (\tilde{\bm V}_1)_{22} =
%   \sum_{j=1}^p\frac{2(\sigma_{Xj}^2)^2}{(\sigma_{Xj}^2 \beta_0^2 +
%   \sigma_{Yj}^2 + \tau_0^2)^2} = \Theta\big(p (n \tau_n^2)^{-2}\big)
% \]
Thus
\[
  \frac{\sum_{j=1}^p\mathbb{E}[|\psi_{2j}(\beta_0)|^3]}{(\tilde{\bm
      V}_1)_{22}^{3/2}} = O\Big( \frac{p}{n^{3/2}} \Big) = O(n^{-1/2}) \to 0.
\]
This completes our proof of \Cref{lem:5}.

For \Cref{lem:6,lem:7}, it remains to bound the variance of $\nabla
\psi$ and $\partial^2 \psi$. Notice that, similar to the proof of
\Cref{lem:2,lem:3}, $\psi_{1j}(\beta,\tau^2)$ is a
homogeneous quadratic polynomial of $(\tilde{\gamma}_j, \tilde{e}_j,
\tilde{\epsilon}_j) = (\gamma_j/\sqrt{n},e_j/\sqrt{n},\epsilon_j/\sqrt{n})$:
\begin{equation*} \label{eq:psi-1j-2}
  \begin{split}
    \psi_{1j}(\beta) =& \frac{\big[(\beta_0 - \beta)
    \tilde{\gamma}_j + \tilde{e}_j -
    \beta \tilde{\epsilon}_j\big]}{[n \sigma_{Xj}^2
      \beta^2 + n (\sigma_{Yj}^2 + \tau^2)]^2} \cdot \\
    &\cdot \big[ (n\sigma_{Xj}^2 \beta \beta_0 + n
    (\sigma_{Yj}^2 + \tau^2))\tilde{ \gamma}_j + (n \sigma_{Xj}^2 \beta) \tilde{e}_j +
    (n \sigma_{Yj}^2) \tilde\epsilon_j\big].
  \end{split}
\end{equation*}
Therefore, its derivatives with respect to $\beta$ and $\tau^2$
remain to be homogeneous quadratic polynomials. As in the proof of
\Cref{lem:2}, this suggests that, for $(\beta,p\tau^2) \in
\mathcal{B}$, (recall that $\|\bm \gamma\|_4 \le \|\bm \gamma\|_3 \ll
\|\bm \gamma\|_2$)
\begin{equation} \label{eq:psi-1-beta}
  \begin{split}
  \mathrm{Var}\Big(\frac{\partial}{\partial \beta}
  \psi_1(\beta,\tau^2)\Big) &\le \mathbb{E}\Big[\Big(\frac{\partial}{\partial \beta}
  \psi_1(\beta,\tau^2)\Big)^2\Big] \\ &= O\big(n^2 \|\bm\gamma\|_4^4 + n \|\bm
  \gamma\|_2^2 + p \big) = o((\tilde{\bm V}_2)_{11}^2).
  \end{split}
\end{equation}
Therefore
$
(\tilde{\bm
  V}_2)_{11}^{-1}(\partial / \partial
\beta) \psi_1(\beta_0,\tau_0^2) \overset{p}{\to} -1,
$
where $(\bm{V})^{-1}_{ij}$ means the reciprocal of the $(i,j)$-th
entry of $\bm V$. Similarly,
\begin{equation} \label{eq:psi-1-tau}
  \mathrm{Var}\Big((\partial / \partial \tau^2)
  \psi_1(\beta,\tau^2)\Big) = O\big(n[n^2 \|\bm\gamma\|_4^4 + n \|\bm
  \gamma\|_2^2 + p] \big) = o((\tilde{\bm V}_2)_{12}^2).
\end{equation}
The extra $n$ comes from differentiating with respect to
$\tau^2 = O(1/p) = O(1/n)$. So $((\tilde{\bm
  V}_2)_{12})^{-1}(\partial / \partial \tau^2) \psi_1(\beta,\tau^2)
\overset{p}{\to} -1$.

For $\psi_2$, we have
\[
  \psi_{2j}(\beta, \tau^2) = \frac{\sigma_{Xj}^2}{\sigma_{Xj}^2 \beta_0^2 + \sigma_{Yj}^2 +
    \tau_0^2} \bigg[\frac{\big[(\beta_0 - \beta)
    \tilde{\gamma}_j + \tilde{e}_j -
    \beta \tilde{\epsilon}_j\big]^2}{n \sigma_{Xj}^2
    \beta^2 + n (\sigma_{Yj}^2 + \tau^2)} - 1\bigg].
\]
Using the same argument,
\begin{equation} \label{eq:psi-2-beta}
  \begin{split}
    \mathrm{Var}\Big((\partial / \partial \beta)
    \psi_2(\beta,\tau^2)\Big) = O\big(n^2 \|\bm\gamma\|_4^4 + n \|\bm
    \gamma\|_2^2 + p \big) =  o((\tilde{\bm V}_2)_{11}),
  \end{split}
\end{equation}
\begin{equation} \label{eq:psi-2-tau}
  \begin{split}
    \mathrm{Var}\Big((\partial / \partial \tau^2) \psi_2(\beta,\tau^2)\Big)
    &\le O\big(n[n^2 \|\bm\gamma\|_4^4 + n \|\bm
    \gamma\|_2^2 + p] \big) = o((\tilde{\bm V}_2)_{22}^2).
  \end{split}
\end{equation}
Therefore $((\tilde{\bm
  V}_2)_{22})^{-1}(\partial / \partial \tau^2) \psi_2(\beta,\tau^2)
\overset{p}{\to} -1$. We cannot claim the same conclusion for
$(\partial / \partial \tau^2) \psi_2(\beta,\tau^2)$ because
$(\tilde{\bm V}_2)_{21} = 0$. Nevertheless, the above results are
already enough to verify \Cref{lem:6}, because
\small
\[
  \begin{split}
    (\tilde{\bm V}_2)^{-1} \nabla \bm \psi(\beta_0,\tau_0^2) =&
    \begin{pmatrix}
      (\tilde{\bm V}_2)_{11}^{-1} & - (\tilde{\bm V}_2)_{11}^{-1}
      (\tilde{\bm V}_2)_{22}^{-1} (\tilde{\bm V}_2)_{12} \\
      0 & (\tilde{\bm V}_2)_{22}^{-1}
    \end{pmatrix}
    \begin{pmatrix}
      \frac{\partial}{\partial \beta} \psi_1 &   \frac{\partial}{\partial
        \tau^2} \psi_1 \\
      \frac{\partial}{\partial \beta} \psi_2 &   \frac{\partial}{\partial \tau^2} \psi_2 \\
    \end{pmatrix} \\
    =&
    \begin{pmatrix}
      (\tilde{\bm V}_2)_{11}^{-1} \frac{\partial}{\partial \beta} \psi_1 - L
      \frac{\partial}{\partial \beta} \psi_2 &   (\tilde{\bm V}_2)_{11}^{-1} \frac{\partial}{\partial \tau^2} \psi_1 - L
      \frac{\partial}{\partial \tau^2} \psi_2 \\
      (\tilde{\bm V}_2)_{22}^{-1} \frac{\partial}{\partial \beta} \psi_2 &   (\tilde{\bm V}_2)_{22}^{-1} \frac{\partial}{\partial \tau^2} \psi_2 \\
    \end{pmatrix},
  \end{split}
\]
\normalsize
where $L = (\tilde{\bm V}_2)_{11}^{-1}
(\tilde{\bm V}_2)_{22}^{-1} (\tilde{\bm V}_2)_{12} =
\Theta(n^{-1})$. Using
\cref{eq:psi-1-beta,eq:psi-1-tau,eq:psi-2-beta,eq:psi-2-tau}, it is
straightforward to verify that the right hand side converges to $\bm I_2$
in probability.

Finally, \Cref{lem:7} can
be proven similarly to \Cref{lem:3} using the rate of the variances
established above as they also extend to the second-order derivative
of $\psi_2$.

\subsection{Proof of \Cref{prop:raps-local-id}}
\label{sec:proof-crefpr-local}

It is easy to show $\mathbb{E}[\psi_2^{(\rho)}(\beta_0,\tau^2_0)] = 0$
by using $t_j(\beta_0,\tau_0^2) \sim \mathrm{N}(0, 1)$. For
$\psi_1^{(\rho)}$, since
\begin{equation} \label{eq:uj}
  u_j(\beta,\tau^2) = - \frac{\partial}{\partial \beta} t_j(\beta,\tau^2)
  = \frac{\hat\Gamma_j \sigma_{Xj}^2 \beta + \hat\gamma_j (\sigma_{Yj}^2
    + \tau^2)}{(\sigma_{Xj}^2 \beta^2 + \sigma_{Yj}^2 + \tau^2)^{3/2}},
\end{equation}
it is straightforward to verify that
\[
  \mathbb{E}\big[t_j(\beta_0,\tau_0^2) \cdot u _j(\beta_0,\tau_0^2)\big]
  = \mathbb{E}\bigg[\frac{(\hat{\Gamma}_j - \beta_0
    \hat{\gamma}_j)\big[\hat\Gamma_j \sigma_{Xj}^2 \beta_0 + \hat\gamma_j (\sigma_{Yj}^2
    + \tau_0^2)\big]}{(\sigma_{Xj}^2 \beta_0^2 + \sigma_{Yj}^2 +
    \tau_0^2)^2}\bigg] = 0.
\]
Since $t_j(\beta_0,\tau_0^2)$ and $u_j(\beta_0,\tau_0^2)$ are linear
transformations of jointly normal random variables, this implies that
$t_j(\beta_0,\tau_0^2) \independent u_j(\beta_0,\tau_0^2)$. Therefore
\[
  \mathbb{E}[\psi_{1j}^{(\rho)}(\beta_0,\tau_0^2)] \propto
  \mathbb{E}[\rho'(t_j(\beta_0,\tau_0^2))\cdot u_j(\beta_0,\tau_0^2)] = 0.
\]

By \Cref{lem:8} below, $\mathrm{E}[\nabla \bm
\psi^{(\rho)}] = -\tilde{\bm V}_2^{(\rho)}$ has full rank because
$\delta,c_3 > 0$.

\subsection{Proof of \Cref{thm:clt-raps}}
\label{sec:proof-crefthm:clt-ra}

Similar to the proof of \Cref{thm:clt-overdispersed}, we first show
$\tilde{\bm V}_1^{(\rho)}$ is the variance of $\bm \psi^{(\rho)}$ and
$\tilde{\bm V}_2^{(\rho)}$ is the expectation of $-\nabla \bm
\psi^{(\rho)}$ at the true parameter $(\beta,\tau^2) =
(\beta_0,\tau_0^2)$.

\begin{lemma} \label{lem:8}
  $\mathrm{Var}\big(\bm\psi^{(\rho)}(\beta_0,\tau_0^2)\big) = \tilde{\bm
    V}_1^{(\rho)}$, $\mathbb{E}\big[\nabla
  \bm\psi^{(\rho)}(\beta_0,\tau_0^2) \big] = -\tilde{\bm V}_2^{(\rho)}$.
\end{lemma}
\begin{proof}[Proof of \Cref{lem:8}]
  Let $\psi_{1j}^{(\rho)}$ and $\psi_{2j}^{(\rho)}$ be the $j$-th
  summand in \eqref{eq:raps-1} and \eqref{eq:raps-2}.   We will use
  the shorthand notation $t_{j0} = t_j(\beta_0,\tau_0^2)$ and $u_{j0}
  = u_j(\beta_0,\tau_0^2)$.

  Because $\rho'$
  is an odd function and $t_{j0} \sim \mathrm{N}(0,
  1)$, we have $\mathbb{E}[\rho'(t_{j0})] = 0$. Using
  $t_{j0} \independent u_{j0}$ (see
  \Cref{sec:proof-crefpr-local}), we obtain
  \[
    \begin{split}
      \mathrm{Var}\big(\psi^{(\rho)}_{1j}(\beta_0,\tau_0^2)\big) &= \mathrm{Var}\big(\rho'(t_{j0}\big)
      u_{j0})
      = c_1 \mathbb{E}\big[u_{j0}^2\big].
      % &= c_1 \frac{\mathbb{E}\big[(\hat\Gamma_j \sigma_{Xj}^2 \beta_0 + \hat\gamma_j (\sigma_{Yj}^2
      % + \tau_0^2))^2\big]}{(\sigma_{Xj}^2 \beta_0^2 + \sigma_{Yj}^2 +
      % \tau_0^2)^3} \\
      % &= c_1 \frac{\big[\Gamma_j (\sigma_{Xj}^2 \beta_0) + \gamma_j
      % (\tau_0^2 + \sigma{Yj}^2)\big]^2
      % + (\tau_0^2 + \sigma{Yj}^2) \sigma_{Xj}^2 (\tau_0^2 + \sigma{Yj}^2 + \sigma_{Xj}^2
      % \beta_0^2)}{(\tau_0^2 + \sigma{Yj}^2 + \sigma_{Xj}^2
      % \beta_0^2)^3} \\
      % &= c_1 \frac{\gamma_j^2
      % (\tau_0^2 + \sigma{Yj}^2) + \Gamma_j^2\sigma_{Xj}^2 + (\tau_0^2 + \sigma{Yj}^2) \sigma_{Xj}^2}{(\tau_0^2 + \sigma{Yj}^2 + \sigma_{Xj}^2
      % \beta_0^2)^2} \\
      % &= c_1 (\tilde{\bm V}_1)_{11}.\\
    \end{split}
  \]
  If we let $\rho(r) = r^2/2$ be the $l_2$-loss (so $c_1 = 1$), we
  recover the APS so $\mathbb{E}[u_{j0}] = (\tilde{\bm
    V}_1)_{11}$. Thus
  $\mathrm{Var}\big(\psi^{(\rho)}_{1j}(\beta_0,\tau_0^2)\big) = c_1
  (\tilde{\bm V}_1)_{11}$.

  The covariance of $\psi_{1j}^{(\rho)}$ and $\psi_{2j}^{(\rho)}$ is
  \[
    \begin{split}
      &\mathrm{Cov}\big(\psi_{1j}^{(\rho)}(\beta_0,\tau_0^2),
      \psi_{2j}^{(\rho)}(\beta_0,\tau_0^2)\big) \\
      \propto &
      \mathbb{E}\Big\{\rho'(t_{j0}) u_{j0} \cdot
      \big[t_{j0} \rho'(t_{j0}) -
      \delta\big]\Big\} \\
      = & \mathbb{E}[u_{j0}] \cdot \mathbb{E} \Big\{
      \rho'(t_{j0}) \cdot
      \big[t_{j0} \rho'(t_{j0}) -
      \delta \big] \Big\} \\
      = & 0.
    \end{split}
  \]
  The last expectation is $0$ because $\rho'$ is an odd function.

  The variance of $\psi_{2j}^{(\rho)}$ is
  \[
    \begin{split}
      \mathrm{Var}\big( \psi_{2j}^{(\rho)}(\beta_0,\tau_0^2) \big) &=
      \frac{(\sigma_{Xj}^2)^2}{(\sigma_{Xj}^2 \beta_0^2 + \sigma_{Yj}^2 +
        \tau_0^2)^2} \mathrm{Var}\big(t_{j0}
      \rho'(t_{j0})\big) \\
      &= c_2(\tilde{\bm V}_1)_{22}.
    \end{split}
  \]

  Next we turn to the derivatives of $\bm \psi^{(\rho)}$. First, by
  the chain rule,
  \[
    \frac{\partial}{\partial \beta} \psi^{(\rho)}_{1j}(\beta,\tau^2) =
    -\rho''(t_j(\beta,\tau^2)) u_j(\beta,\tau^2)^2 +
    \rho'(t_j(\beta,\tau^2)) \frac{\partial}{\partial \beta} u_j(\beta,\tau^2).
  \]
  The expectation of the first term at $(\beta_0,\tau_0^2)$ is
  \[
    \mathbb{E}[-\rho''(t_{j0}) u_{j0}^2] = -
    \mathbb{E}[\rho''(t_{j0})] \mathbb{E}[u_{j0}^2] =
    \delta (\tilde{\bm V}_2)_{11},
  \]
  where we have used the identity $\mathbb{E}[\rho''(R)] =
  \mathbb{E}[R \rho'(R)]$ for $R \sim \mathrm{N}(0,1)$, which can be
  proved by integration by parts and the fact that $\phi'(x) = -x \phi(x)$.
  The second term requires more calculations:
  \[
    \begin{split}
      &\mathbb{E}\Big[[\rho'(t_{j0}) \cdot
      \Big(\frac{\partial}{\partial \beta}
      u_j(\beta,\tau^2)\Big)\Big|_{(\beta,\tau^2) =
        (\beta_0,\tau_0^2)}\Big] \\
      =& \mathbb{E}\bigg[ \rho' (  t_{j0} ) \cdot
      \frac{\partial}{\partial \beta} \Big(\hat{\gamma}_j (\tau_0^2 + \sigma_{j2}^2) + \hat{\Gamma}_j
      \sigma_{j1}^2 \beta\Big)\Big|_{\beta = \beta_0} \cdot \frac{1}{(\sigma_{Xj}^2 \beta_0^2 + \sigma_{Yj}^2 +
        \tau_0^2)^{3/2}}  \bigg] \\
      &+ \mathbb{E}\bigg[ \rho' (t_{j0}) \cdot
      (\hat{\gamma}_j (\tau_0^2 + \sigma_{j2}^2) + \hat{\Gamma}_j
      \sigma_{j1}^2 \beta_0) \cdot \frac{\partial}{\partial \beta} \Big( \frac{1}{(\sigma_{Xj}^2 \beta_0^2 + \sigma_{Yj}^2 +
        \tau_0^2)^{3/2}} \Big) \Big|_{\beta = \beta_0}  \bigg] \\
      =& \mathbb{E}\bigg[ \rho' (t_{j0}) \cdot
      \frac{\partial}{\partial \beta} \Big( \hat{\gamma}_j(\tau_0^2 + \sigma_{j2}^2) + \hat{\Gamma}_j
      \sigma_{j1}^2 \beta\Big)\Big|_{\beta = \beta_0} \cdot \frac{1}{(\sigma_{Xj}^2 \beta_0^2 + \sigma_{Yj}^2 +
        \tau_0^2)^{3/2}} \bigg] \\
      =& \mathbb{E}\bigg[ \rho' (t_{j0}) \cdot
      \hat{\Gamma}_j \sigma_{j1}^2 \cdot \frac{1}{(\sigma_{Xj}^2 \beta_0^2 + \sigma_{Yj}^2 +
        \tau_0^2)^{3/2}}  \bigg] \\
      =&\frac{\sigma_{Xj}^2}{(\sigma_{Xj}^2 \beta_0^2 + \sigma_{Yj}^2 +
        \tau_0^2)^{3/2}} \mathbb{E}[\rho'(t_{j0}) (\Gamma_j + e_j)] \\
      =& \frac{\sigma_{Xj}^2 \sqrt{\sigma_{Yj}^2 + \tau_0^2}}{(\sigma_{Xj}^2 \beta_0^2 + \sigma_{Yj}^2 +
        \tau_0^2)^{3/2}} \mathbb{E}\bigg[\rho'(t_{j0}) \cdot
      \frac{e_j}{\sqrt{\tau_0^2 + \sigma_{j2}^2}}\bigg].
    \end{split}
  \]
  The second equality above is because $t_{j0} \independent u_{j0}$ and
  $\mathbb{E}[\rho'(t_{j0})] = 0$. Notice that $t_{j0}$ and
  $e_j\big/\sqrt{\tau_0^2 + \sigma_{j2}^2}$ are marginally distributed as
  the standard normal and
  \[
    \mathrm{Cov}\bigg(t_{j0},  \frac{e_j}{\sqrt{\sigma_{Yj}^2 +
        \tau_0^2}}\bigg) = \Big(\frac{\sigma_{Yj}^2 +
      \tau_0^2}{\sigma_{Xj}^2 \beta_0^2 + \sigma_{Yj}^2 +
      \tau_0^2}\Big)^{1/2}.
  \]
  It is not difficult to verify that if $R_1,R_2$ are $\mathrm{N}(0,1)$
  marginally and $\mathrm{Cov}(R_1,R_2) = \lambda$, then $\mathbb{E}[\rho'(R_1)
  R_2] = \lambda \delta$. Thus
  \begin{equation} \label{eq:cov-term-2}
    \mathbb{E}\Big[[\rho'(t_{j0}) \cdot
    \frac{\partial}{\partial \beta}
    u_{j0}\Big] = \delta \cdot
    \frac{\sigma_{Xj}^2(\sigma_{Yj}^2 + \tau_0^2)}{(\sigma_{Xj}^2 \beta_0^2 + \sigma_{Yj}^2 +
      \tau_0^2)^2} = \delta [(\tilde{\bm V}_1)_{11} - (\tilde{\bm V}_2)_{11}].
  \end{equation}
  To summarize,
  \[
    \begin{split}
      \mathbb{E}\Big[ \frac{\partial}{\partial \beta}
      \psi_{1j}^{(\rho)}(\beta_0,\tau_0^2) \Big]
      &= - \delta (\tilde{\bm
        V}_1)_{11} + \delta [(\tilde{\bm V}_1)_{11} - (\tilde{\bm
        V}_2)_{11}]
      = -\delta (\tilde{\bm V}_2)_{11}.
    \end{split}
  \]

  The other first-order derivative of $\psi_{1j}^{(\rho)}$ is
  \[
    \begin{split}
      \mathbb{E}\bigg[\frac{\partial}{\partial \tau^2} \psi_{1j}^{(\rho)}(\beta_0,\tau_0^2)\bigg]
      &= \mathbb{E} \bigg[\rho''(t_{j0}) \Big(\frac{\partial}{\partial \tau^2} t_{j0}\Big) u_{j0} +
      \rho'(t_{j0}) \Big(\frac{\partial}{\partial \tau^2} u_{j0}\Big)
      \bigg].\\
    \end{split}
  \]
  Using
  \begin{equation} \label{eq:t-tau}
    \frac{\partial}{\partial \tau^2} t_j(\beta,\tau^2) =  - \frac{t_j(\beta,\tau^2)}{2(\sigma_{Xj}^2 \beta_0^2 + \sigma_{Yj}^2 + \tau_0^2)}
  \end{equation}
  and the independent of $t_{j0}$, it is straightforward to show the
  first term has mean $0$. For the second term,
  \[
    \begin{split}
      \mathbb{E} \bigg[\rho'(t_{j0}) \Big(\frac{\partial}{\partial \tau^2} u_{j0}\Big)
      \bigg]
      &= \mathbb{E}\bigg[ \rho' (  t_{j0} ) \cdot
      \hat{\gamma}_j \frac{1}{(\sigma_{Xj}^2 \beta_0^2 + \sigma_{Yj}^2 +
        \tau_0^2)^{3/2}}  \bigg] \\
    \end{split}
  \]
  Similar to the derivation of \eqref{eq:cov-term-2}, one can show that
  \[
    \begin{split}
      \mathbb{E} \bigg[\rho'(t_{j0}) \Big(\frac{\partial}{\partial \tau^2} u_{j0}\Big)
      \bigg]
      &= \frac{- \delta\beta_0 \sigma_{Xj}^2}{\sigma_{Xj}^2 \beta_0^2 + \sigma_{Yj}^2 +
        \tau_0^2} = - \delta \cdot (\tilde{\bm V}_2)_{12}.
    \end{split}
  \]

  Finally we consider the derivatives of $\psi_{2j}^{(\rho)}$. Using
  \eqref{eq:t-tau}, we have
  \[
    \begin{split}
      \mathbb{E}\bigg[\frac{\partial}{\partial \tau^2} \psi^{(\rho)}_{2j}(\beta_0,\tau_0^2)\bigg] &=
      \mathbb{E}\bigg[\sigma_{Xj}^2\frac{-t_{j0}
        \rho'(t_{j0})/2 - t_{j0}^2 \rho''(t_{j0})/2 - (t_{j0} \rho'(t_{j0})
        - \delta)}{(\sigma_{Xj}^2 \beta_0^2 + \sigma_{Yj}^2 +
        \tau_0^2)^2}\bigg] \\
      &= -\frac{\delta+ c_3}{2} \cdot \frac{\sigma_{Xj}^2}{(\sigma_{Xj}^2 \beta_0^2 + \sigma_{Yj}^2 +
        \tau_0^2)^2}
    \end{split}
  \]
  Hence
  \[
    \mathbb{E}\bigg[\frac{\partial}{\partial \tau^2}
    \psi^{(\rho)}_{2}(\beta_0,\tau_0^2)\bigg] = - [(\delta + c_3)/2]
    (\tilde{\bm V}_2)_{22}.
  \]
  The last partial derivative is $(\partial/\partial \beta)
  \psi_2^{(\rho)}$. Its expectation at $(\beta_0,\tau_0^2)$ is
  \[
    \begin{split}
      &\mathbb{E}\Big[\frac{\partial}{\partial \beta}
      \psi^{(\rho)}_{2j}(\beta_0,\tau_0^2)\Big] \\
      =& \mathbb{E}\bigg[\frac{\frac{\partial}{\partial \beta}
        \Big[t_{j0} \rho'(t_{j0}) - \delta\Big]}{(\sigma_{Xj}^2 \beta_0^2 + \sigma_{Yj}^2 +
        \tau_0^2)^2}\bigg]\\
      &+ \mathbb{E}\bigg[\frac{\partial}{\partial \beta} \Big[\frac{1}{(\sigma_{Xj}^2 \beta^2 + \sigma_{Yj}^2 +
        \tau_0^2)^2}\Big]\Big|_{\beta = \beta_0} \cdot \Big[t_{j0} \rho'(t_{j0}) - \delta\Big] \bigg] \\
      =& \mathbb{E}\bigg[\frac{\frac{\partial}{\partial \beta}
        \Big[t_{j0} \rho'(t_{j0}) - \delta\Big]}{(\sigma_{Xj}^2 \beta_0^2 + \sigma_{Yj}^2 +
        \tau_0^2)^2}\bigg] \\
      % =& \mathbb{E}\bigg[\frac{\frac{\partial}{\partial \beta} t_{j0}\rho'(t_{j0})}{(\sigma_{Xj}^2 \beta_0^2 + \sigma_{Yj}^2 +
      % \tau_0^2)^2}\bigg] \\
      =& \mathbb{E}\bigg[\frac{- u_{j0}\rho'(t_{j0}) - t_{j0} \rho''(t_{j0})
        u_{j0}}{(\sigma_{Xj}^2 \beta_0^2 + \sigma_{Yj}^2 +
        \tau_0^2)^2}\bigg] \\
      =& 0.
    \end{split}
  \]
  The last equation is due to the independence of $t_{j0}$ and $u_{j0}$ and the fact that $\rho'(r)$
  and $r \rho''(r)$ are odd functions of $r$.
\end{proof}

To prove asymptotic normality of the RAPS estimator, we just need to
verify \Cref{lem:5,lem:6,lem:7} with $\bm \psi$ replaced by $\bm
\psi^{(\rho)}$ and $\tilde{\bm V}_1$, $\tilde{\bm V}_2$ replaced by
$\tilde{\bm V}_1^{(\rho)}$, $\tilde{\bm V}_2^{(\rho)}$. This requires
verifying the Lyapunov condition for the central limit theorem in
\Cref{lem:5} and bounding the derivatives of $\bm \psi^{(\rho)}$ to
prove \Cref{lem:6,lem:7}.

It is useful to notice that the rates of $\tilde{\bm V}_1$ and
$\tilde{\bm V}_2$ in \eqref{eq:V-tilde-rate} still apply to
$\tilde{\bm V}_1^{(\rho)}$ and $\tilde{\bm V}_2^{(\rho)}$. First,
using the boundedness of $\rho'$, we have
\[
  \sum_{j=1}^p
  \mathbb{E}\big[|\psi_{1j}^{(\rho)}(\beta_0,\tau_0^2)|^3\big] =
  \sum_{j=1}^p \mathbb{E}[|\rho'(t_{j0})|^3] \cdot
  \mathbb{E}[|u_{j0}|^3] =
  O\Big(\sum_{j=1}^p\mathbb{E}[|u_{j0}|^3]\Big).
\]
We can rewrite \eqref{eq:uj} as
\[
  u_{j0} = \frac{(\gamma_j\sqrt{n}) (\sigma_{Xj}^2 \beta_0 + \sigma_{Yj}^2
    + \tau_0^2) + (e_j\sqrt{n})(\sigma_{Xj}^2 \beta_0) + (\epsilon_j \sqrt{n})(\sigma_{Yj}^2
    + \tau_0^2)}{(\sigma_{Xj}^2 \beta_0^2 + \sigma_{Yj}^2 + \tau_0^2)^{3/2}\sqrt{n}}.
\]
So $u_{j0}$ is a linear combination of $\gamma_j\sqrt{n}$,
$e_j\sqrt{n}$, $\epsilon_j\sqrt{n}$. Using the same argument in
\Cref{sec:proof-crefthm:clt-ov}, for any positive integer $k$,
\begin{equation}
  \label{eq:uj-moment-bound}
  \mathbb{E}[|u_{j0}|^k] = O\Big(\sum_{l=0}^k \big(\sqrt{n}|\gamma_j| +
  1\big)^l\Big) = O\Big(\sum_{l=0}^k (\sqrt{n})^l|\gamma_j|^l\Big).
\end{equation}
From this it is easy to verify \cref{eq:lyapunov} still
holds for $\psi_{1j}^{(\rho)}(\beta_0,\tau_0^2)$, $j=1,2,\dotsc$, which implies that
\[
  \frac{\sum_{j=1}^p
    \mathbb{E}\big[\big|\psi_{1j}^{(\rho)}(\beta_0,\tau_0^2)\big|^3\big]}{(\tilde{\bm
      V}_1^{(\rho)})_{11}^{3/2}} \to 0.
\]
Furthermore,
\[
  \sum_{j=1}^p
  \mathbb{E}\big[|\psi_{2j}^{(\rho)}(\beta_0,\tau_0^2)|^3\big] =
  \sum_{j=1}^p\frac{(\sigma_{Xj}^2)^3}{(\sigma_{Xj}^2 \beta_0^2 +
    \sigma_{Yj}^2 + \tau_0^2)^3} \cdot c_5
\]
where $c_5 = \mathrm{E}[|R \rho'(R) - \delta|^3]$ for $R \sim
\mathrm{N}(0,1)$. Thus
\[
  \sum_{j=1}^p
  \mathbb{E}\big[|\psi_{2j}^{(\rho)}(\beta_0,\tau_0^2)|^3\big] = O(p)
  \ll (\tilde{\bm V}_1^{(\rho)})_{22}^{3/2}.
\]
To summarize, we have verified the Lyapunov condition for $\bm
\psi_j^{(\rho)}$. Consequently, the central limit theorem
$(\tilde{\bm{V}}_1^{(\rho)})^{-1/2}
\bm{\psi}^{(\rho)}(\beta_0,\tau_0^2) \overset{d}{\to}
\mathrm{N}(\bm{0}, \bm{I}_2)$ holds.

Next we restablish the variance bounds, namely
\cref{eq:psi-1-beta,eq:psi-1-tau,eq:psi-2-beta,eq:psi-2-tau}, for $\bm
\psi^{(\rho)}$. Similar to \Cref{sec:proof-crefthm:clt-ov}, we extend
\eqref{eq:uj-moment-bound}, the bound on the moments of $u_j$, to the
derivatives of $u_j$: for $(\beta,p\tau^2) \in \mathcal{B}$,
\begin{equation} \label{eq:uj-deriv-moment-bound}
  \mathbb{E}\bigg[\Big|\frac{\partial^{l_1+l_2}}{\partial
    \beta^{l_1} (\partial \tau^2)^{l_2}}u_j(\beta,\tau^2)\Big|^k\bigg]
  = O\Big(n^{l_2}\sum_{l=0}^k (\sqrt{n})^l|\gamma_j|^l\Big).
\end{equation}
Similarly,
\begin{equation} \label{eq:tj-moment-bound}
  \mathbb{E}\big[\big|t_j(\beta,\tau^2)\big|^k\big]
  = O\Big(\sum_{l=0}^k (\sqrt{n})^l|\gamma_j|^l\Big).
\end{equation}

Consider a partial derivative of $\psi^{(\rho)}_{1j}$:
\[
  \frac{\partial^{l_1+l_2}}{\partial
    \beta^{l_1} (\partial \tau^2)^{l_2}} \psi^{(\rho)}_{1j}(\beta,\tau^2) = \frac{\partial^{l_1+l_2}}{\partial
    \beta^{l_1} (\partial \tau^2)^{l_2}} \rho'(t_j(\beta,\tau^2)) u_j(\beta,\tau^2).
\]
By \cref{eq:uj,eq:t-tau}, It is a polynomial of derivatives (up to $(l_1+l_2+1)$-th
order) of $\rho(t_j(\beta,\tau^2))$, $t_j(\beta,\tau^2)$,
$u_j(\beta,\tau^2)$, and derivatives of $u_j(\beta,\tau^2)$, for which
we all have moment bounds. We will use the shorthand notation $t_j =
t_j(\beta,\tau^2)$ and $u_j = u_j(\beta,\tau^2)$ below. In particular,
\[
  \frac{\partial}{\partial \beta} \psi^{(\rho)}_{1j}(\beta,\tau^2) =
  -\rho''(t_j) u_j^2 +
  \rho'(t_j) \frac{\partial}{\partial \beta}u_j.
\]
Using the boundedness of $\rho'$ and $\rho''$ and \cref{eq:uj-deriv-moment-bound}, we have
\[
  \begin{split}
    \mathrm{Var}\Big(\frac{\partial}{\partial \beta}
    \psi^{(\rho)}_{1}\Big) &\le
    \mathrm{E}\Big[\Big(\frac{\partial}{\partial \beta}
    \psi^{(\rho)}_{1}\Big)^2\Big]\\
    &= O\bigg( \sum_{j=1}^p
    u_j^4 + \Big(\frac{\partial}{\partial \beta}
    u_j\Big)^2 \bigg) \\
    &= O\Big(\sum_{l=0}^4 (\sqrt{n})^l \|\bm \gamma\|_l^l\Big) =
    o\big((\tilde{\bm V}^{(\rho)}_2)_{11}^2\big)
  \end{split}
\]
Similarly,
\[
  \frac{\partial}{\partial \tau^2} \psi^{(\rho)}_{1j} =
  -\rho''\big(t_j\big) \frac{t_j u_j}{2(\sigma_{Xj}^2 \beta_0^2 + \sigma_{Yj}^2 + \tau_0^2)} +
  \rho'\big(t_j\big) \frac{\partial}{\partial \tau^2}u_j.
\]
Using the Cauchy-Schwarz inequality, we obtain
\[
  \begin{split}
    \mathrm{Var}\Big(\frac{\partial}{\partial \tau^2}
    \psi^{(\rho)}_{1}\Big) &\le
    \mathrm{E}\Big[\Big(\frac{\partial}{\partial \tau^2}
    \psi^{(\rho)}_{1}\Big)^2\Big]\\
    &= O\bigg( \mathbb{E} \Big[ \sum_{j=1}^p \frac{t_j^4 +
      u_j^4}{(1/n)^2} + \Big(\frac{\partial}{\partial \tau^2}
    u_j\Big)^2 \Big] \bigg) \\
    &= O\Big(n^2\sum_{l=0}^4 (\sqrt{n})^l \|\bm \gamma\|_l^l + n
    \sum_{l=0}^2 (\sqrt{n})^l \|\bm \gamma\|_l^l\Big) =
    o\big((\tilde{\bm V}^{(\rho)}_2)_{12}^2\big).
  \end{split}
\]

Next we consider the derivatives of $\psi_2^{(\rho)}$:
\[
  \begin{split}
    \frac{\partial}{\partial \beta} \psi^{(\rho)}_{2j} =&
    \frac{\sigma_{Xj}^2}{\sigma_{Xj}^2 \beta^2 + \sigma_{Yj}^2 +
      \tau^2} \cdot \Big[
    -u_j \rho'\big(t_j\big) -
    t_j \rho''\big(t_j\big)
    u_j\Big] - \\
    &- \frac{\sigma_{Xj}^4 \beta}{(\sigma_{Xj}^2 \beta^2 + \sigma_{Yj}^2 +
      \tau^2)^2} \cdot [t_j \rho'(t_j) - \delta],
  \end{split}
\]
thus, again using the Cauchy-Schwarz inequality,
\[
  \begin{split}
    \mathrm{E}\Big[\Big(\frac{\partial}{\partial \beta}
    \psi^{(\rho)}_{2}\Big)^2\Big] &= O\bigg( \mathbb{E}\Big[\sum_{j=1}^p
    u_j^2 + (t_j^4 + u_j^4) + t_j^2\Big]\bigg) \\
    &= O\Big( \sum_{l=0}^4 (\sqrt{n})^l \|\bm
    \gamma\|_l^l\Big) = o\big((\tilde{\bm V}^{(\rho)}_2)_{22}\big).
  \end{split}
\]
Finally,
\[
  \begin{split}
    \frac{\partial}{\partial \tau^2} \psi^{(\rho)}_{2j} =&
    \frac{\sigma_{Xj}^2}{\sigma_{Xj}^2 \beta^2 + \sigma_{Yj}^2 +
      \tau^2} \cdot \Big[ - \frac{t_j}{2(\sigma_{Xj}^2 \beta^2 + \sigma_{Yj}^2 +
      \tau^2)} (\rho'(t_j) - t_j \rho''(t_j))\Big] - \\
    &- \frac{\sigma_{Xj}^2}{(\sigma_{Xj}^2 \beta^2 + \sigma_{Yj}^2 +
      \tau^2)^2} \cdot [t_j \rho'(t_j) - \delta].
  \end{split}
\]
Thus
\[
  \begin{split}
    \mathrm{E}\Big[\Big(\frac{\partial}{\partial \tau^2}
    \psi^{(\rho)}_{2}\Big)^2\Big] &= O\bigg(\frac{1}{1/n} \mathbb{E}\Big[\sum_{j=1}^p
    t_j^2 + t_j^4\Big]\bigg) \\
    &= O\Big(n \sum_{l=0}^4 (\sqrt{n})^l \|\bm
    \gamma\|_l^l\Big) \\
    &= o\big((\tilde{\bm V}^{(\rho)}_2)_{22}^2\big).
  \end{split}
\]

In summary, we have re-established \cref{eq:psi-1-beta,eq:psi-1-tau,eq:psi-2-beta,eq:psi-2-tau} for $\bm
\psi^{(\rho)}$. Therefore \Cref{lem:6} still holds for $\bm
\psi^{(\rho)}$ with $\tilde{\bm V}_2$ replaced by $\tilde{\bm V}_2^{(\rho)}$.

Finally we prove \Cref{lem:7} for the RAPS $\bm \psi^{(\rho)}$. Notice
that, for $(\beta,p\tau^2) \in \mathcal{B}$, $t_j(\beta,\tau^2) =
O_p(\sqrt{n}|\gamma_j| + 1)$ and $u_j(\beta,\tau^2) =
O_p(\sqrt{n}|\gamma_j| + 1)$. These rates also hold for the partial
derivatives of $t_j$ and $u_j$ with respect to $\beta$. Therefore, by
the boundedness of $\rho'$, $\rho''$ and $\rho'''$,
\[
\frac{\partial^2}{\partial \beta^2} \psi_{1j}^{(\rho)}(\beta,\tau^2) =
\rho'''(t_j) u_j^3 - 3 \rho''(t_j) u_j \frac{\partial}{\partial \beta}
u_j + \rho'(t_j) \frac{\partial}{\partial \beta} u_j = O_p\big((\sqrt{n}|\gamma_j| + 1)^3\big).
\]
Hence
\[
\frac{\partial^2}{\partial \beta^2} \psi_1^{(\rho)}(\beta,\tau^2) =
\sum_{j=1}^p \frac{\partial^2}{\partial \beta^2}
\psi_{1j}^{(\rho)}(\beta,\tau^2) = O_p(p + \sqrt{n} \|\bm\gamma\|_1 +
n \|\bm \gamma\|_2^2 + n^{3/2}\|\bm\gamma\|_3^3).
\]
Using the assumption that $\|\bm \gamma\|_3^3 = O(1/\sqrt{p})$, it is
easy to show that the right hand side is $O_p(n) = O_p\big((\tilde{\bm
  V}_2^{(\rho)})_{11}\big)$. Rates of the other partial derivatives
can be proved analogously and we omit further detail.

%%% Local Variables:
%%% mode: latex
%%% TeX-master: t
%%% End:

\end{document}